\documentclass[
  aps,
  prx,
  reprint,
  amsmath,
  amssymb,
  floatfix
]{revtex4-2}

\usepackage[utf8]{inputenc}
\usepackage[T1]{fontenc}

\usepackage{graphicx}
\usepackage{epstopdf}    
\usepackage{subcaption}  
\usepackage{comment}
\usepackage{textcomp}
\usepackage[colorlinks=true, linkcolor=blue, citecolor=blue, urlcolor=blue]{hyperref}
\begin{document}

\title{Topology to Elasticity Mapping in Thermal Glass}
\author{Avinash Kumar Jha}
\affiliation{Department of Physics, Indian Institute of Technology Roorkee, Roorkee, Uttarakhand 247667, India}

\begin{abstract}
Building on our previous work on thermal nonaffine metrics and thermal strain fields, we examine how local steric geometry controls thermally driven rearrangements in an amorphous glass-forming solid. As a structural descriptor, we employ the topological steric ``\(\theta\)-mismatch'' order parameter, which quantifies local deviations of bond angles from the steric reference geometry and is used to distinguish low- and high-mismatch particle populations. We then compare their thermal nonaffine response, thermal deviatoric strain, and thermal volumetric strain over a broad temperature range at several state points. High-\(\Delta\theta\) sites preferentially exhibit larger thermal nonaffine motion and stronger thermal deviatoric strain, particularly in the deeply jammed regime, indicating that geometrically frustrated regions are preferentially associated with localized mechanical distortions. The thermal volumetric response exhibits a distinct mismatch-dependent crossover: low-\(\Delta\theta\) sites predominantly contribute to the small-strain response, whereas high-\(\Delta\theta\) sites contribute more strongly to the large-strain tail. The contrast between low- and high-mismatch populations is most pronounced in the deeply jammed regime, remains identifiable through the intermediate viscoelastic regime, and progressively weakens upon heating toward the liquid state. These results establish a direct connection between local steric frustration and thermally activated mechanical response, showing that thermal nonaffine metrics and thermal strain fields provide a physically transparent bridge between local structure and thermally driven rearrangements in amorphous glasses.
\end{abstract}

\maketitle

\section{Introduction}

Understanding how local structure controls thermally activated rearrangements is a central problem in the physics of disordered solids. In crystalline materials, symmetry and topology strongly constrain the onset and spatial organization of local motion; in amorphous solids, by contrast, the absence of long-range order makes it essential to identify local structural descriptors that correlate with mechanical response and dynamical heterogeneity. A broad range of structural indicators has been proposed in this context, including soft vibrational modes, machine-learned softness, and geometric or topological motifs associated with mobility \cite{manning2011vibrational,schoenholz2016structural,cubuk2015identifying,tong2018revealing,schoenholz2014understanding,schoenholz2016structural,rottler2014predicting,richard2020predicting}.

Complementary to purely structural measures, nonaffine diagnostics provide a direct description of local mechanical response. The nonaffine displacement measure \(D_{\min}^2\), originally introduced for strain-driven plasticity \cite{falk1998dynamics,shi2005strain,maloney2006amorphous,maloney2004universal,utter2008experimental,tsamados2009local,shang2014evolution,pinney2016structure}, has become a standard tool for identifying localized rearrangements under external deformation. In our previous work \cite{jha2026thermal,jha2026thermally}, we extended this idea to purely thermal trajectories and introduced thermal nonaffine metrics together with thermal strain fields, allowing local mechanical response to be characterized even in the absence of externally imposed shear. The present work builds on that framework and asks whether these thermally defined mechanical measures can be related directly to local geometric frustration.
For clarity, throughout this paper the temperature at which the mean-squared displacement enters its quiescent plateau coincides, within numerical resolution, with the temperature at which the thermal nonaffine field becomes very sparse and strongly attenuated. We refer to this common crossover, at each state point, as the solidification or mechanically locked temperature. In our earlier thermal-relaxation study, this same temperature scale marked the onset of the quiescent solid-like regime in the mean-squared displacement \cite{jha2026thermal}. In our follow-up thermal-nonaffine study, we showed that nonaffine fields remain clearly appreciable below the glass-transition temperature in an intermediate viscoelastic window, and that only deeper in the glass do they become strongly suppressed and mechanically locked \cite{jha2026thermally}. In this work, we therefore use the solidification temperature as the common reference point where the MSD is flat and the destabilizing thermal nonaffine field becomes very sparse.
To address this question, we use the steric ``\(\theta\)-mismatch'' order parameter introduced by Tong and Tanaka \cite{tong2018revealing} as a structural reference. This quantity is based on the local contact geometry of neighboring particles and measures deviations of instantaneous bond angles from the steric reference geometry. Because it depends only on local geometry, \(\Delta\theta\) provides a purely structural measure of frustration and does not require any dynamical information. It therefore offers a natural way to classify regions of a glass according to their local geometric environment.
Here we combine these two viewpoints and study whether regions of strong steric frustration are also the regions that sustain larger thermal nonaffine motion and stronger thermal strain. In particular, we partition particles into low- and high-\(\Delta\theta\) populations and compare their thermal nonaffine response, thermal deviatoric strain, and thermal volumetric strain over a broad range of temperatures and state points. This allows us to test, in a purely thermally driven setting, whether a geometric scalar can predict local mechanical response in an amorphous solid.
Our main result is that high-\(\Delta\theta\) sites preferentially host larger thermal nonaffine response and stronger thermal deviatoric strain in the deeply jammed solid, while thermal volumetric strain exhibits a robust mismatch-dependent crossover that distinguishes low- and high-mismatch populations across the solid and intermediate regimes. These correlations weaken progressively as the system approaches the liquid state. Together, the results establish a direct mapping from local steric frustration to thermally activated mechanical response, and show that thermal nonaffine metrics and thermal strain fields provide a compact and physically transparent description of rearrangements in amorphous solids.\\

The paper is organized as follows. Section~\ref{sec:methods} describes the system preparation, simulation protocol, and the characterization of the thermodynamic and thermal-mechanical temperature scales, including the mechanically locked or solidification line. Section~\ref{steric_exp} introduces the steric reference angle and examines the temperature-dependent mismatch distributions. Section~\ref{evol_per} describes the evolution of the mean per-particle mismatch with temperature, while Sec.~\ref{density_dep} analyzes the density dependence of the low- and high-mismatch populations. Section~\ref{sec:thermal_metrics} summarizes the mathematical formulation of the thermal nonaffine and thermal strain metrics used in the present analysis. Sections~\ref{therm_nonaff}--\ref{sec:volumetric_theta} then examine, respectively, the thermal nonaffine, thermal deviatoric-strain, and thermal volumetric-strain responses of low- and high-\(\theta\)-mismatch populations across temperature and density. Section~\ref{sec:spatial_correlation} provides a real-space comparison by mapping the spatial organization of the strongest thermal mechanical responses onto the steric-mismatch field. Section~\ref{sec:conclusions} summarizes the principal findings and their implications for the relation between local steric geometry and thermally driven mechanical response in amorphous solids.
\section{System Preparation, Simulation Protocol, and Thermal Mechanical Stability Characterization}
\label{sec:methods}

\begin{align}
    U_{jk}(r)&=4\epsilon\left[\left(\frac{\sigma_{jk}}{r_{jk}}\right)^{12}
    -\left(\frac{\sigma_{jk}}{r_{jk}}\right)^{6}
    +\frac{1}{4}\right],
    \qquad \frac{r_{jk}}{\sigma_{jk}} < 2^{1/6},
    \nonumber\\
    &=0 \qquad \mbox{otherwise.}
\end{align}

This is the Weeks--Chandler--Andersen (WCA) potential
\cite{tanaka2010critical,filion2011simulation,weeks1971role}. Here \(\epsilon\) sets the
energy scale, lengths are reported in units of the mean particle diameter \(\langle\sigma\rangle\), and temperature is expressed in units of
\(\epsilon/k_{\mathrm B}\), with \(k_{\mathrm B}=1\). The effective diameter of a pair of particles is $\sigma_{jk}=\frac{\sigma_j+\sigma_k}{2}$, where \(\sigma_j\) is the diameter of particle \(j\), and $r_{jk}=|\mathbf r_j-\mathbf r_k|$ is the center-to-center distance between particles \(j\) and \(k\).
Throughout this work, \(\rho\) denotes the reduced number density, $\rho=\frac{N}{L^2}$, where \(N\) is the total number of particles and \(L\) is the linear
dimension of the two-dimensional square simulation box. We perform NVT molecular-dynamics simulations using an isokinetic thermostat \cite{brown1984comparison}. The system contains \(N=2000\) particles and represents a two-dimensional polydisperse colloidal glass former (2DPC), with particle diameters sampled from a Gaussian distribution. The degree of polydispersity is defined as $\Delta=\frac{\sqrt{\langle\sigma^2\rangle-\langle\sigma\rangle^2}}{\langle\sigma\rangle}.$
We set the average particle diameter to \(\langle\sigma\rangle=1.0\) and the polydispersity to \(\Delta=0.11\). Simulations are performed at three densities, \(\rho=0.900\), \(0.950\), and \(1.000\).

For each density, the initial configuration is taken from a well-equilibrated high-temperature liquid and directly quenched to the desired temperature, followed by equilibration and production runs. The equilibration time for the supercooled and lower-temperature states is kept fixed at \(t_w=945000\). A total of 10--12 independent runs are used for the analysis. For the temporal analysis, configurations are sampled using the block-based log-linear saving protocol introduced in our previous work \cite{jha2026thermally}. Unless otherwise stated, all quantities reported in the present study are calculated from consecutive entries of this log-linear saved trajectory, which provides dense sampling at short time intervals while extending naturally to longer time separations. This protocol allows the thermal nonaffine response and the associated strain fields to be evaluated consistently across all temperatures and state points without imposing a single external time lag. The complete details of the configuration-saving scheme and the alternative fixed-lag protocol based on the structural relaxation time
\(\tau_\alpha\) are given in Ref.~\cite{jha2026thermally}

The characteristic temperature scales relevant to the present study are summarized in Table~\ref{tab:temperature}. The thermodynamic glass-transition temperature \(T_{g,\tau_\alpha}\) is obtained from a VFT fit of the alpha-relaxation time obtained from the self-intermediate scattering function (FSKT), while \(T_{g,D}\) is obtained from the diffusion coefficient \(D\). \(T_{\rm VFT}\) denotes the temperature obtained from the VFT fit, and \(T_{\rm onset}\) denotes the onset
temperature below which supercooled-liquid behavior begins. In addition to these thermodynamic temperature scales, we introduce the mechanically locked temperature \(T_{\rm ML}\), which is referred to throughout this manuscript as the \emph{solidification line}.

The solidification line is identified from the thermal nonaffine field obtained in our previous work \cite{jha2026thermally}. Specifically, \(T_{\rm ML}\) is the temperature at which the spatially resolved thermal nonaffine field undergoes a pronounced attenuation and becomes very sparse, signaling the crossover to a mechanically quiescent state. The values identified in this way are \(T_{\rm ML}=0.050\), \(0.232\), and \(0.251\) for \(\rho=0.900\), \(0.950\), and \(1.000\), respectively. Within numerical resolution, this same temperature scale coincides with the onset of the quiescent plateau in the mean-squared displacement identified in our earlier thermal-relaxation study \cite{jha2026thermal}. Thus, the solidification line is a mechanical criterion for the onset of the deeply arrested state and is distinct from the thermodynamic glass-transition temperature.

\begin{table}[t]
\centering
\caption{Density dependence of the characteristic temperatures for the 2DPC system. \(T_{g,\tau_\alpha}\) is the thermodynamic glass-transition temperature obtained from a VFT fit of the alpha-relaxation time obtained from the self-intermediate scattering function (FSKT), \(T_{g,D}\) is the glass-transition temperature obtained from the diffusion coefficient \(D\), \(T_{\rm VFT}\) is the temperature obtained from the VFT fit, and \(T_{\rm onset}\) is the onset temperature below which supercooled-liquid behavior begins. \(T_{\rm ML}\) denotes the mechanically locked temperature, hereafter referred to as the solidification line. It is identified from the thermal nonaffine field as the temperature at which the spatially resolved nonaffine field undergoes a pronounced attenuation and becomes very sparse, marking the crossover to mechanical quiescence \cite{jha2026thermally}. Within numerical resolution, this temperature
also coincides with the onset of the quiescent mean-squared-displacement regime identified in our earlier thermal-relaxation study \cite{jha2026thermal}.}
\label{tab:temperature}
\begin{tabular}{c c c c c c}
\hline
Density & \(T_{g,\tau_\alpha}\) & \(T_{g,D}\) & \(T_{\rm VFT}\) &
\(T_{\rm onset}\) & \(T_{\rm ML}\) \\
\hline
0.900 & 0.168 & 0.160 & 0.062 & 0.265 & 0.050 \\
0.950 & 0.343 & 0.331 & 0.178 & 0.679 & 0.232 \\
1.000 & 0.596 & 0.571 & 0.385 & 1.483 & 0.251 \\
\hline
\end{tabular}
\end{table}

The resulting temperature hierarchy is \(T_{\rm ML}<T_g\) at all state points. The interval $T_{\rm ML}<T<T_g$ corresponds to a mechanically active viscoelastic regime in which appreciable thermal nonaffine activity persists even below the thermodynamic glass-transition temperature \cite{jha2026thermally}. For $T\leq T_{\rm ML}$, the system enters the deeply arrested, mechanically locked regime associated with the solidification line, where the destabilizing thermal nonaffine field is drastically attenuated and spatially sparse, signaling the suppression of thermally driven mechanical rearrangements \cite{jha2026thermally}. Thus, \(T_{\rm ML}\), or equivalently the solidification line, provides a mechanical criterion distinct from the thermodynamic \(T_g\) for identifying the crossover to the mechanically stable, quiescent regime. These temperature scales define the deeply arrested, viscoelastic, and liquid-like regimes used throughout the
remainder of the manuscript.
\section{Steric Reference Angle and Temperature‐Dependent Mismatch Distributions}
\label{steric_exp}
\begin{figure}[htb!]
  \centering
  \includegraphics[width=0.55\textwidth]{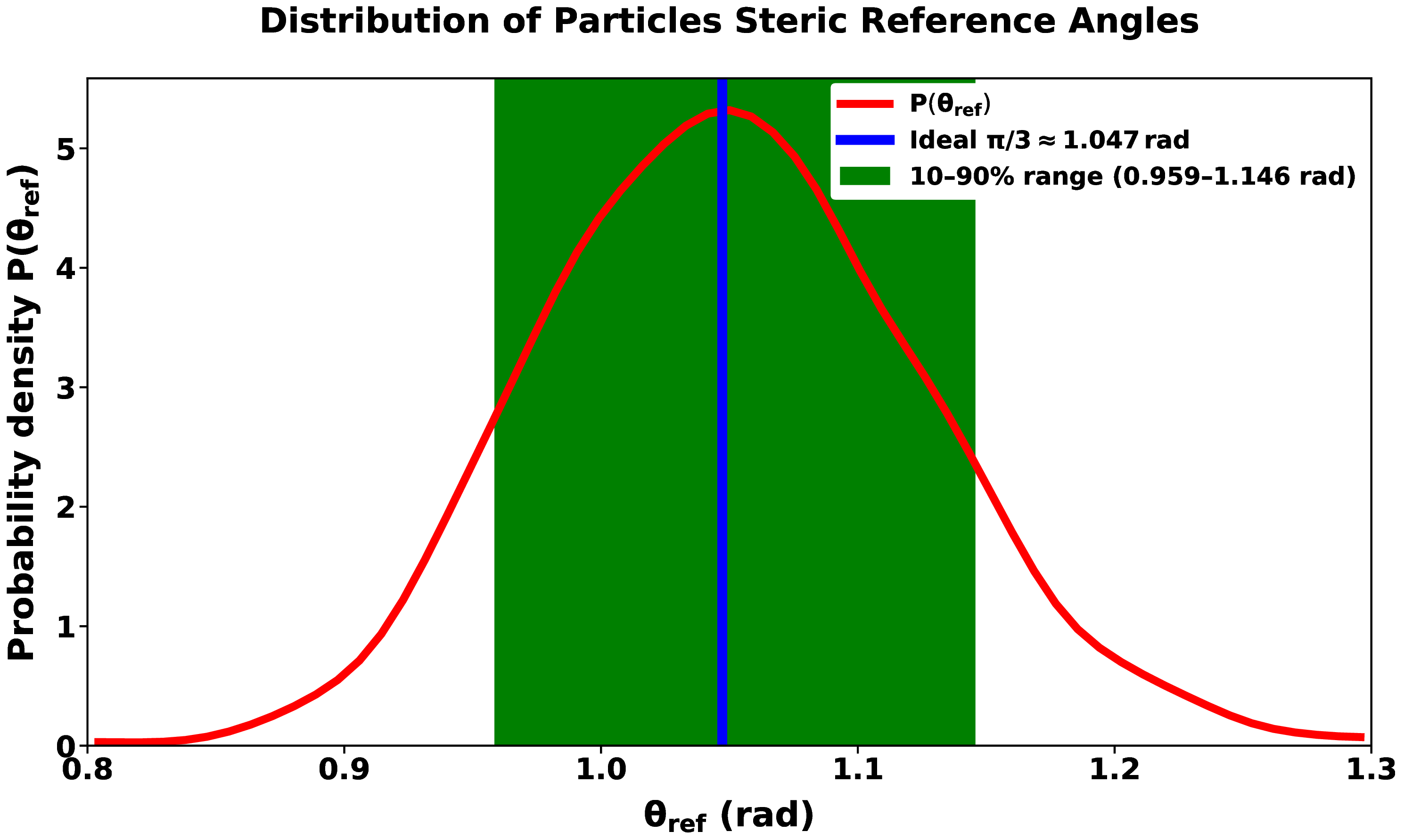}
   \caption{Probability density $P(\theta_{\rm ref})$ of per-particle steric-reference angles computed from particle diameters (red). The vertical blue line marks the ideal touching angle for monodisperse disks, $\pi/3\approx1.047\,$rad. The green shaded band indicates the 10--90\% interval of the per-particle averages (0.959--1.146\,rad); the data were produced with polydispersity $=0.11$, which gives the measured broadening around the monodisperse reference. This interval quantifies the typical deviation of local steric geometry from the ideal ``just-touch'' motif and serves as a convenient benchmark for topology mechanics comparisons.}
 \label{fig:theta1a}
\end{figure}

\begin{figure*}[t!]
  \centering
  \includegraphics[width=0.48\textwidth]{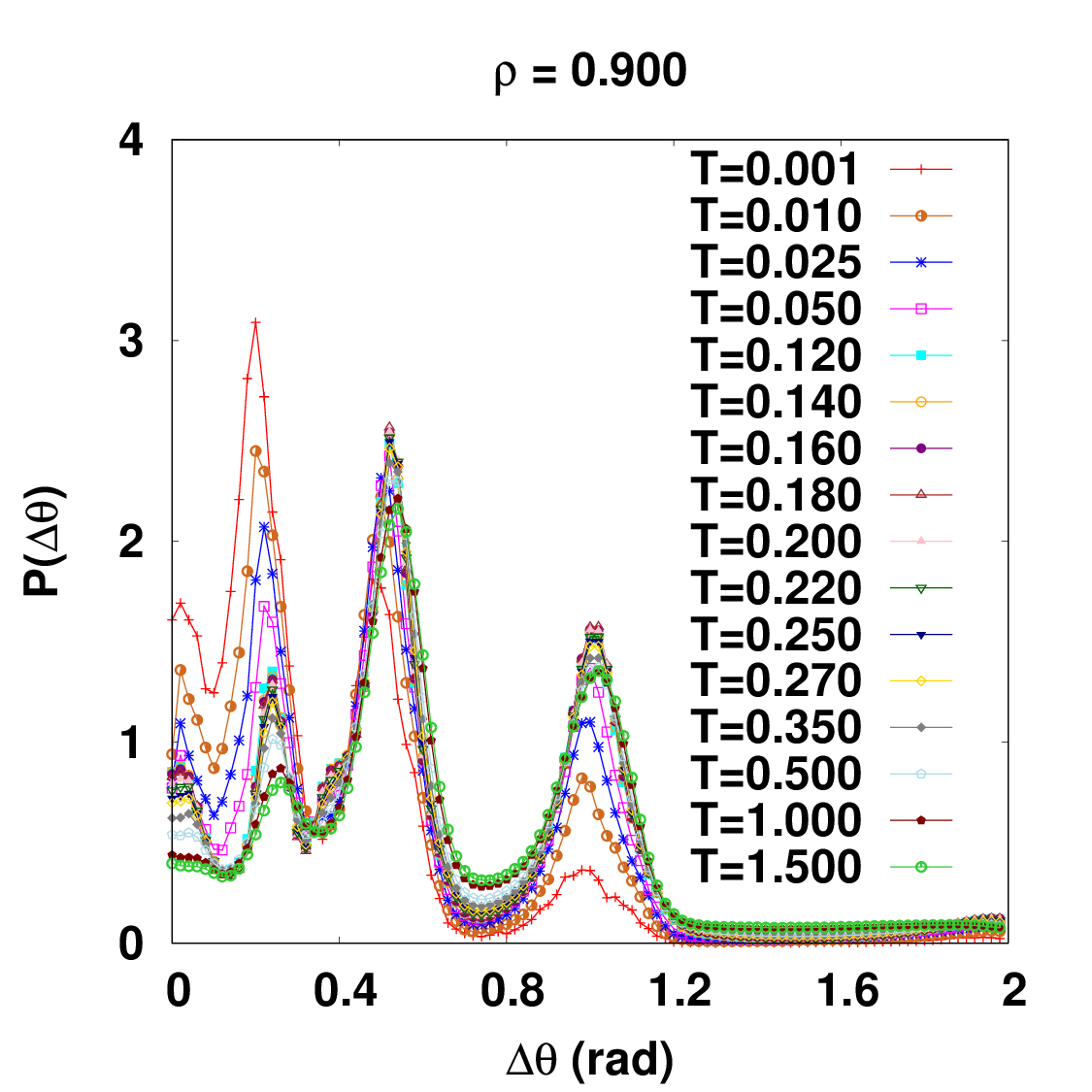}\hfill
  \includegraphics[width=0.48\textwidth]{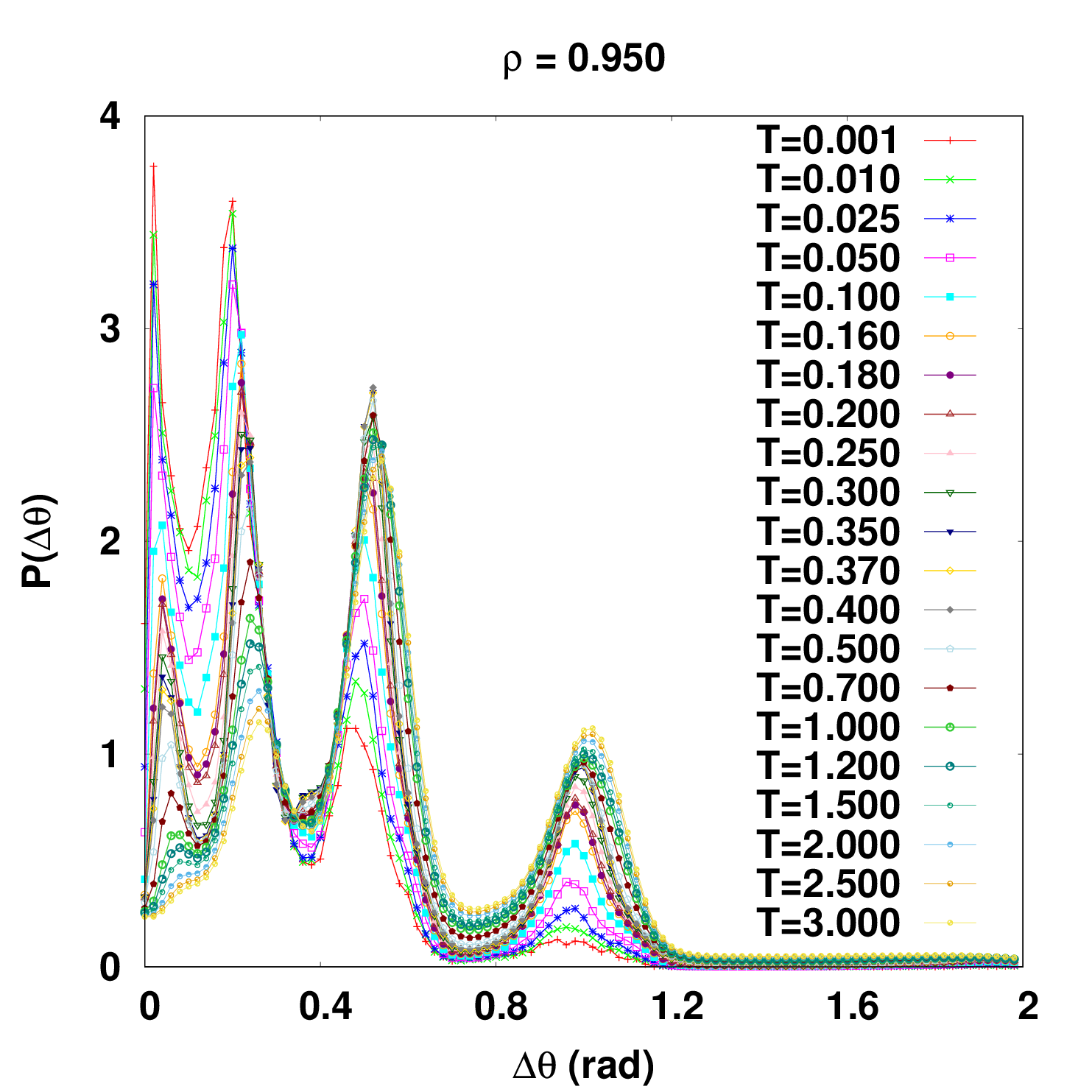}

  \vspace{0.5pt}
 
  \includegraphics[width=0.48\textwidth]{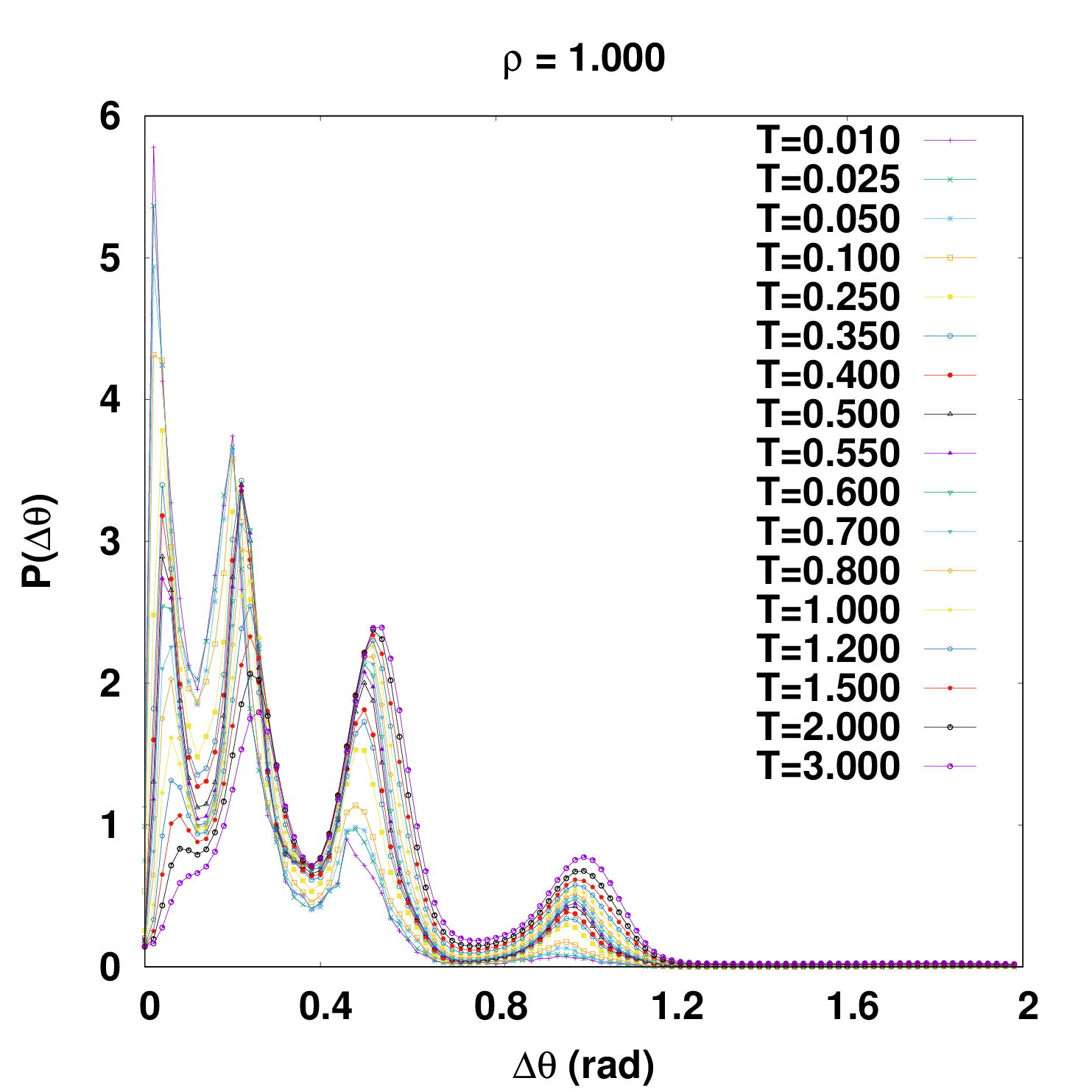}
 \caption{
 \small
    \parbox{\textwidth}{%
    \raggedright
 Temperature dependence of the per-particle steric mismatch distribution $P(\Delta\theta)$ at three densities. Each panel shows probability densities of the per-particle mismatch $\Delta\theta_i$ (rad) computed at a range of temperatures (legend; colours correspond to temperatures listed).  Panels are (a) $\rho=0.900$, (b) $\rho=0.950$, and (c) $\rho=1.000$.  The horizontal axis is the mismatch angle $\Delta\theta$ (radians), defined as the average absolute deviation of instantaneous bond angles from the steric reference angles, and the vertical axis is the probability density $P(\Delta\theta)$. Three characteristic peaks appear in the distributions (small-, mid-, and large-$\Delta\theta$), reflecting the dominant local steric motifs produced by polydispersity; the large-angle peak near the monodisperse touching value ($\sim\pi/3$) corresponds to locally near-touching triplets, while the smaller-angle peaks reflect alternate contact geometries created by size disorder.  At the lowest temperatures(deeply solid regime), the three peaks are most sharply resolved (distinct structural motifs are well preserved). In physical terms, this means the following: in the cold solid, the \emph{low}--$\Delta\theta$ (small-angle) band becomes a tall, narrow feature; most particles sit close to their steric, “just-touch” geometry and show only small vibrational deviations, whereas the \emph{high}--$\Delta\theta$ (large-angle) band is strongly suppressed (few particles populate strongly frustrated or bond-broken motifs).  This separation of peaks indicates that different local motifs exist as distinct populations in the solid and are not washed out by thermal motion. As temperature increases, the peaks broaden and their relative heights change, indicating thermal activation and local rearrangements that wash out fine steric detail. The overall peak near $\pi/3$ remains robust across temperature but widens with increasing $T$. Three systematic, density-dependent observations are evident and are important for subsequent analysis.  (i) For the low-$\Delta\theta$ (small-angle) motif, the lowest-temperature curves show the largest peak amplitude while the highest-temperature curves show the smallest; the small-angle motif is therefore most sharply populated in the cold solid.  (ii) As density increases, the small-angle peak develops substructure: at $\rho=0.950$ and especially at $\rho=1.000$, the low-angle feature splits into two nearby subpeaks (indicating multiple closely related small-angle motifs that are resolved only under stronger steric constraints).  (iii) The peak amplitudes evolve oppositely with density: the height of the small-angle (lowest $\Delta\theta$) peak increases with density, whereas the height of the large-angle (highest $\Delta\theta$) peak decreases with density. 
\label{fig:theta2a}
  }
}
\end{figure*}

To characterize the local steric environment, we employ the steric ``\(\theta\)-mismatch'' measure introduced by Tong and Tanaka \cite{tong2018revealing}. For each central particle \(i\) with \(n_i\) contact neighbors, we first compute the mismatch angle for every pair \((j,k)\) of those neighbors,
\begin{equation}
\label{m1}
\Delta\theta_{i,j,k}
\;=\;
\bigl|\theta_{\mathrm{measured}}^{\,i,j,k}-\theta_{\mathrm{ref}}^{\,i,j,k}\bigr|,
\qquad 1\le j<k\le n_i
\end{equation}

and then define the particle’s total mismatch as the sum over all such neighbor‐pairs,

\begin{equation}
\label{m2}
\Delta\theta_i \;=\; \frac{1}{n_i}
\sum_{1\le j<k\le n_i}\Delta\theta_{i,j,k}\,.
\end{equation}
This ensures \(\Delta\theta_i\) captures every local bond‐angle deviation around particle \(i\).
In disordered, repulsive‐sphere packings the “just‐touching” geometry between any three mutually contacting particles \(i\), \(j\), and \(k\) defines a purely steric reference angle, \(\theta_{\rm ref}\), against which all instantaneous bond angles \(\theta_{\mathrm{measured}}\) may be compared.  If particle diameters are \(d_i\), \(d_j\), and \(d_k\), then in the unstressed, contact configuration, the interatomic distances are simply  
\[
  r_{ij} = \tfrac12(d_i + d_j),\quad
  r_{ik} = \tfrac12(d_i + d_k),\quad
  r_{jk} = \tfrac12(d_j + d_k),
\]  
and by the Law of Cosines the ideal, zero‐strain angle at \(i\) is  
\begin{equation}
\theta_{\rm ref}
    = \arccos\!\Bigl(\frac{r_{ij}^2 + r_{ik}^2 - r_{jk}^2}{2\,r_{ij}\,r_{ik}}\Bigr).
    \label{m3}
\end{equation} 
In the monodisperse limit (\(d_i=d_j=d_k=D\)) this reduces to  
\begin{equation}
  \theta_{\rm ref} = \arccos\!\bigl(\tfrac12\bigr)=60^\circ,
  \label{m4}
\end{equation}
recovering the familiar hexagonal “six‐fold” bond‐angle; in a polydisperse assembly one obtains a narrow distribution of \(\theta_{\rm ref}\) values that nonetheless cluster tightly around \(60^\circ\), justifying its continued use as the canonical zero‐strain benchmark as shown in Fig.~\ref{fig:theta1a} which quantifies the polydispersity-induced spread of the steric reference geometry. Because \(\theta_{\rm ref}\) is computed solely from the particle diameters through the law of cosines, its distribution is a direct structural fingerprint of size disorder rather than a dynamical observable. The peak near the monodisperse touching angle \(\pi/3\) shows that the local steric reference remains centered close to the ideal triangular geometry, while the finite width of the distribution reflects the geometric broadening introduced by polydispersity. In this sense, Fig.~\ref{fig:theta1a} provides the zero-strain geometric baseline for the rest of the analysis.
Defining the \emph{mismatch angle} from equation~\eqref {m1}, we accumulate its frame-averaged probability density \(P(\Delta\theta)\) across a wide temperature sweep from the high-\(T\) normal liquid, down through the supercooled regime, across the glass transition \(T_g\), and into the deeply arrested solid.  Strikingly, \(P(\Delta\theta)\) consistently resolves four separate ``bands'' of mismatch (Fig.~\ref{fig:theta2a}). Importantly, the lowest-\(\Delta\theta\) region itself splits into two closely spaced sub-bands (i.e., two near-zero-angle peaks). The relative weight of the very first (leftmost) sub-peak is density dependent: at the lowest packing (\(\rho=0.900\)) this leftmost peak is relatively weak, whereas at \(\rho=0.950\) and at high  \(\rho=1.000\) the same sub-peak grows dramatically in amplitude. This behavior indicates that stronger steric confinement (higher density) resolves and stabilizes multiple closely related small-angle contact motifs that are only weakly populated at low density. The remaining mid- and large-\(\Delta\theta\) bands correspond to other steric motifs produced by polydispersity, and their relative populations shift systematically with both temperature and density.  We now describe these four features of Fig.~\ref{fig:theta2a} in turn.

\begin{enumerate}
 \item \textbf{Very-low mismatch band} (\(\Delta\theta\!\approx\!0.00\!-\!0.10\ \mathrm{rad}\)).  
A narrow, near-zero spike whose relative weight increases strongly with the packing fraction (weak at \(\rho=0.900\), appreciable at \(\rho=0.950\), dominant at \(\rho=1.000\)). This feature corresponds to contact triples whose instantaneous bond angles are essentially identical to their steric zero-strain reference (near-equilateral three-particle motifs): such motifs experience only tiny thermal vibrations and therefore contribute a sharply peaked, nearly elastic signature to \(P(\Delta\theta)\). Because this peak strengthens with density, it is naturally interpreted as the fingerprint of tightly caged, mechanically stable local arrangements that are increasingly favored as free volume is reduced. These particles typically have full contact neighborhoods (no loose or missing contacts) and tend to sit in the most locally overconstrained parts of the contact network; consequently, they participate predominantly in high-frequency, localized vibrational modes rather than in soft, rearranging degrees of freedom.

\item \textbf{Low mismatch band} (\(\Delta\theta\!\approx\!0.10\!-\!0.32\ \mathrm{rad}\)).  
A second small-angle peak immediately to the right of the very-low spike, present at all densities and showing modest thermal broadening on heating. This broader small-angle population reflects slightly distorted steric motifs, e.g., triangles where modest polydispersity, small elastic deformations, or contact geometry (non-ideal touching distances) shift the bond angle away from the perfect steric value. Physically, these configurations are still mechanically confined (they do not correspond to free rattlers), but possess a small amount of geometric frustration or residual strain that makes them more susceptible to anharmonic fluctuations and to participation in cooperative, low-amplitude rearrangements as temperature increases. In practice, these particles often have coordination similar to the very low group but show larger local free volume or slightly asymmetric neighbor size combinations, and therefore occupy an intermediate mechanical role between perfectly caged and fluid-like sites.

\item \textbf{Intermediate‐mismatch band} ($0.33\lesssim\Delta\theta\lesssim0.7\,\mathrm{rad}$).  
    Its height ordering \emph{reverses} relative to the low band: at low~$T$ its weight is suppressed, then increases into the supercooled regime, peaking before falling again at high~$T$.  
    This band captures cooperative, anharmonic distortions as particles begin to escape their cages.

\item \textbf{High‐mismatch band} ($\Delta\theta\gtrsim0.7\,\mathrm{rad}$).  
    Broad and lower in amplitude, most prominent in the liquid states and vanishing sharply upon glass formation.  
    These large‐angle excursions reflect bond breaking or highly frustrated local motifs, regions far from the zero‐strain reference where liquid‐like dynamics dominate.  
    Crucially, the precipitous drop in this band’s height as $T$ falls below the supercooled regime demonstrates its extreme sensitivity to dynamical arrest, underscoring the contrast with the first band’s persistence.
\end{enumerate}

\noindent
Taken together, the fact that four well‐defined mismatch bands persist from the deeply glassy to the high–temperature liquid regime reveals a clear hierarchy of local structural responses: small‐angle vibrations around the purely steric “just‐touch” geometry; moderate, cooperative distortions as particles begin to break free of their cages; and large, topologically frustrated excursions associated with bond breaking and rearrangement. Because our \(\Delta\theta\) is computed solely from the instantaneous “just‐touch” reference angles via the Law of Cosines, it naturally captures contributions from every conceivable local motif: 5-fold, 7-fold, and beyond in a genuinely disordered packing. In contrast to a perfect crystal, which would yield a single, narrow \(\Delta\theta\) peak at the canonical \(60^\circ\) zero-strain angle, the emergence of multiple peaks here serves as a direct fingerprint of amorphous disorder. Moreover, the enduring prominence of the very-low and low mismatch sub-peaks deep into the glass, together with the abrupt collapse of the high-mismatch band upon dynamical arrest, underscores both the robustness of purely vibrational fluctuations around the \(\sim60^\circ\) reference and the fragility of cooperative, large-angle distortions under cooling.

\begin{figure*}[t!]
  \centering
  \includegraphics[width=0.48\textwidth]{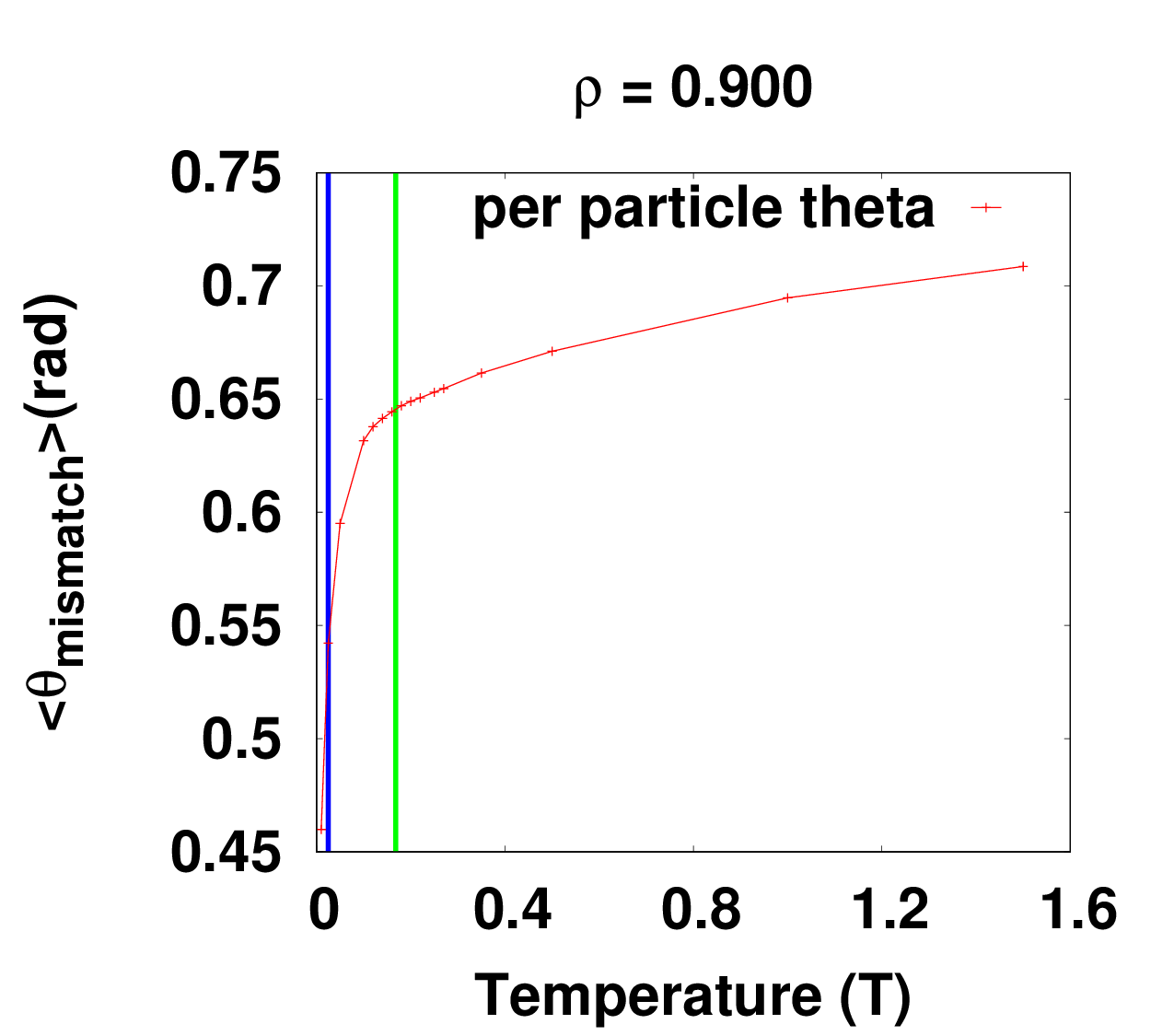}
  \includegraphics[width=0.48\textwidth]{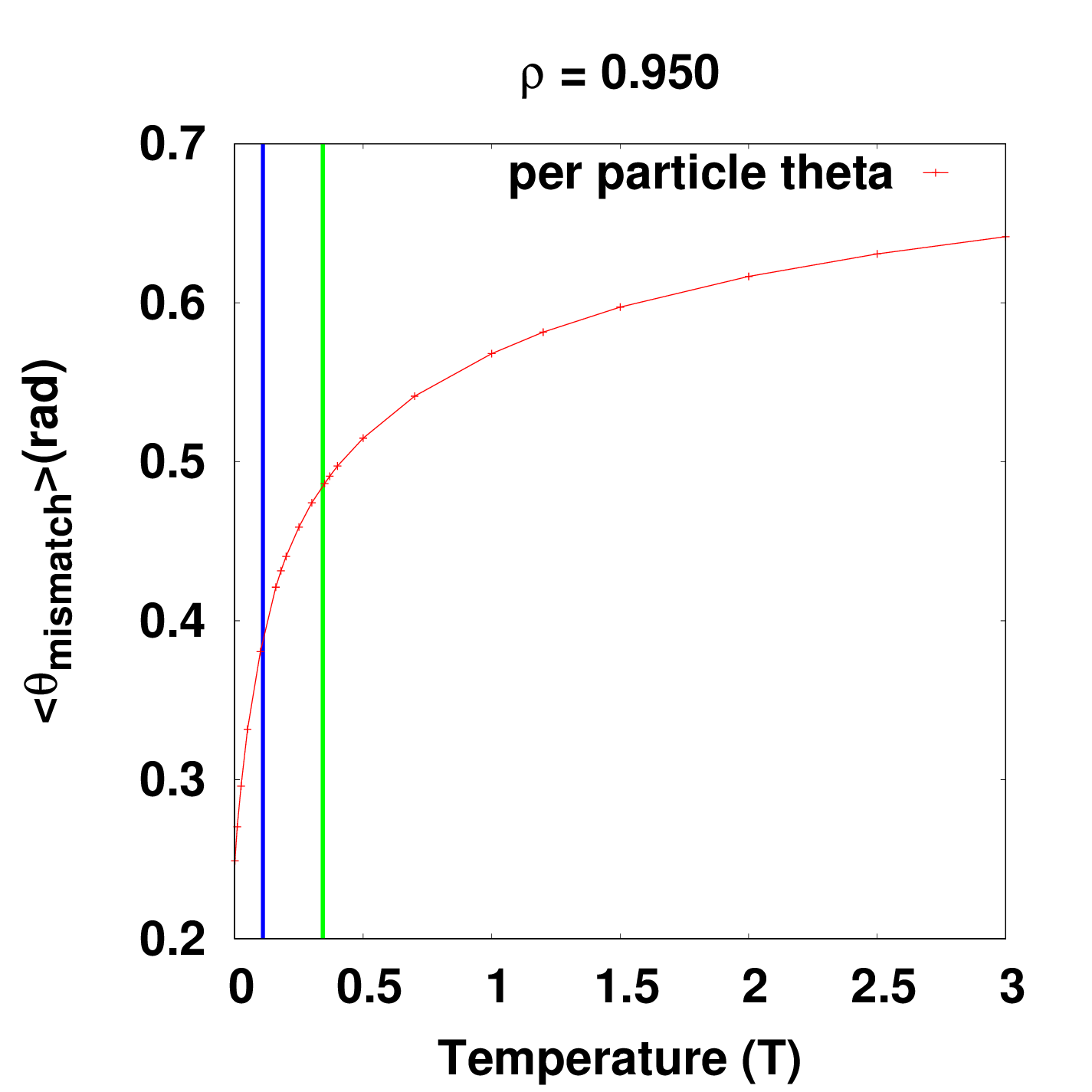}
  \includegraphics[width=0.48\textwidth]{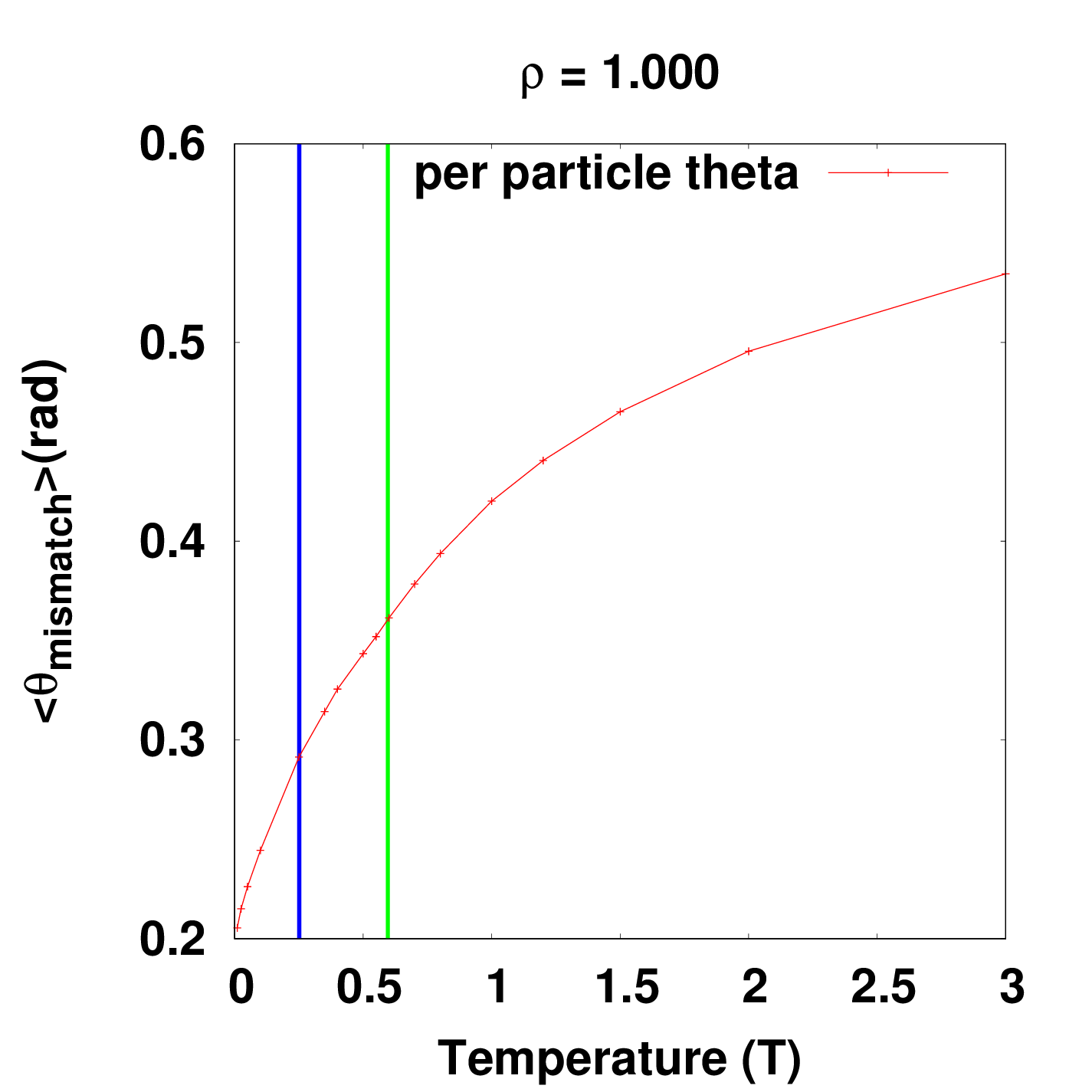}
  \caption{Temperature dependence of the mean steric mismatch. $\langle\theta_{\rm mismatch}\rangle$ (red curves) for three densities. Blue and green vertical lines mark the solidification (lower) and glass–transition (higher) temperatures, respectively. Below the solidification line particles remain tightly caged and $\langle\theta_{\rm mismatch}\rangle$ is small (only vibrational excursions); between the two markers modest, cooperative distortions produce a slow, density-dependent rise (the slope is shallowest at highest $\rho$); above $T_g$ particles escape cages and the mean mismatch grows rapidly toward a liquid plateau. Increasing density shifts both characteristic temperatures upward and compresses the angular span (approx.\ 0.45–0.72\,rad at $\rho=0.900$, 0.25–0.64\,rad at $\rho=0.950$, and 0.20–0.54\,rad at $\rho=1.000$), reflecting progressively stronger steric caging.}

  \label{fig:theta3a}
\end{figure*}

 \begin{figure*}[htb!]
  \centering
  \includegraphics[width=0.48\textwidth]{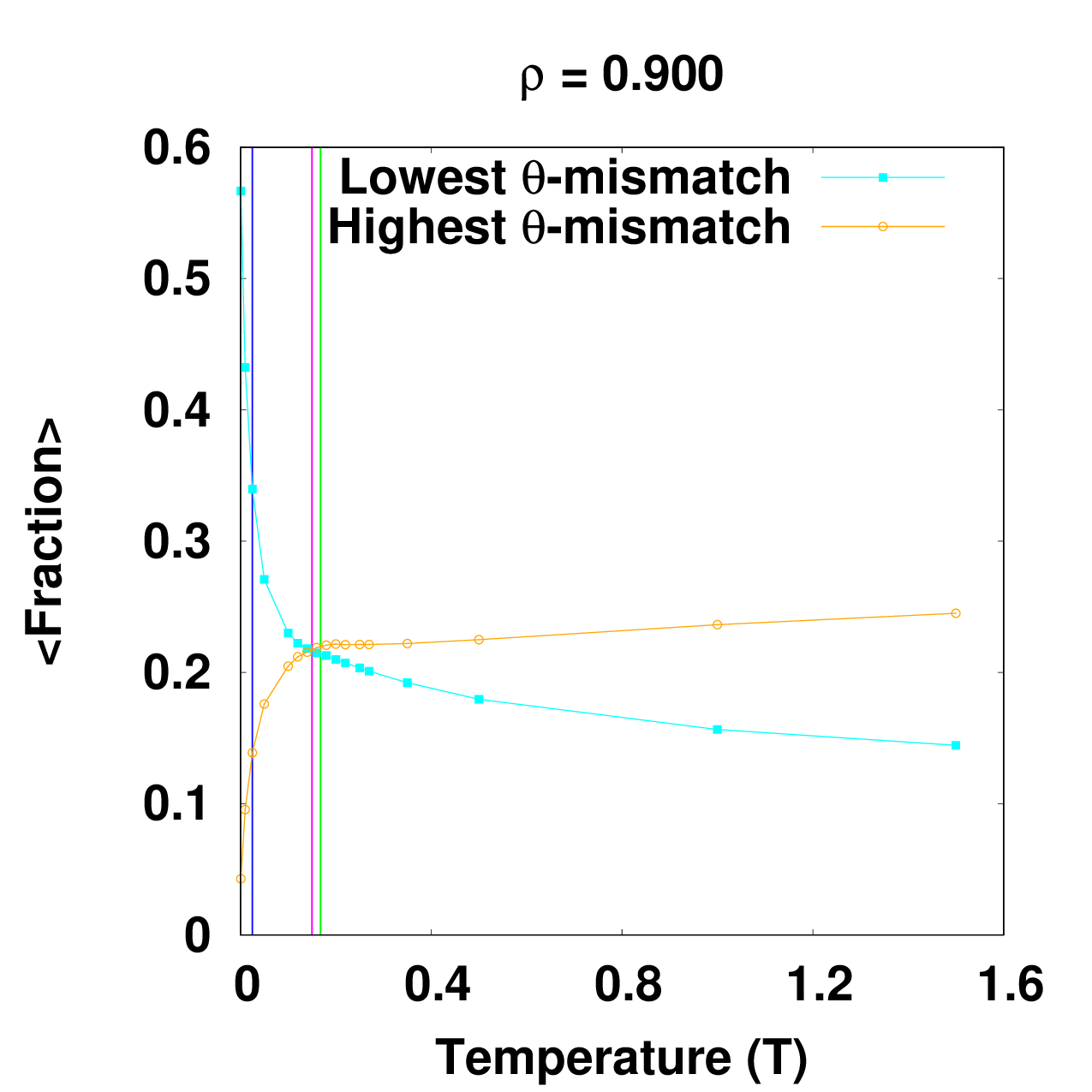}
  \includegraphics[width=0.48\textwidth]{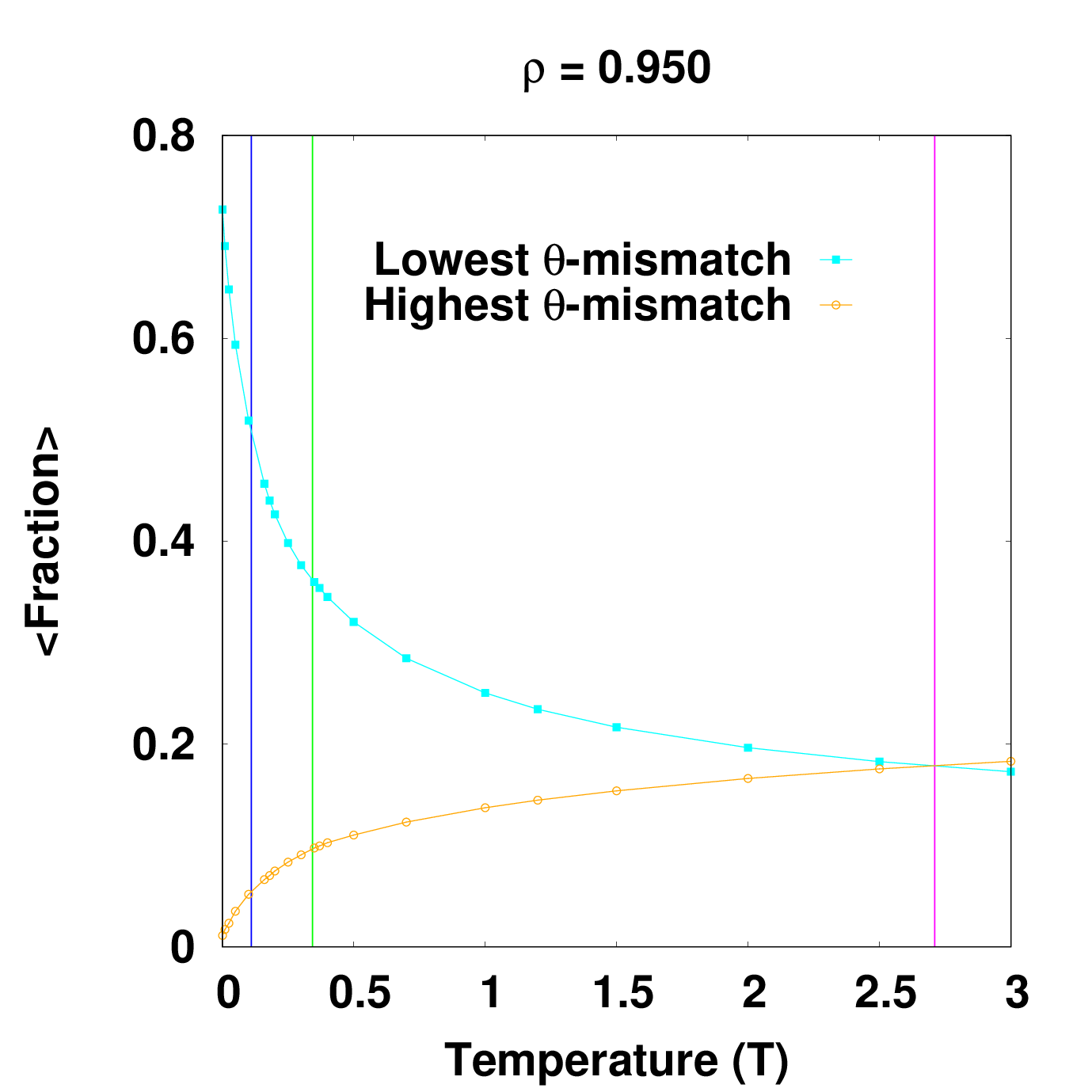}
  \includegraphics[width=0.48\textwidth]{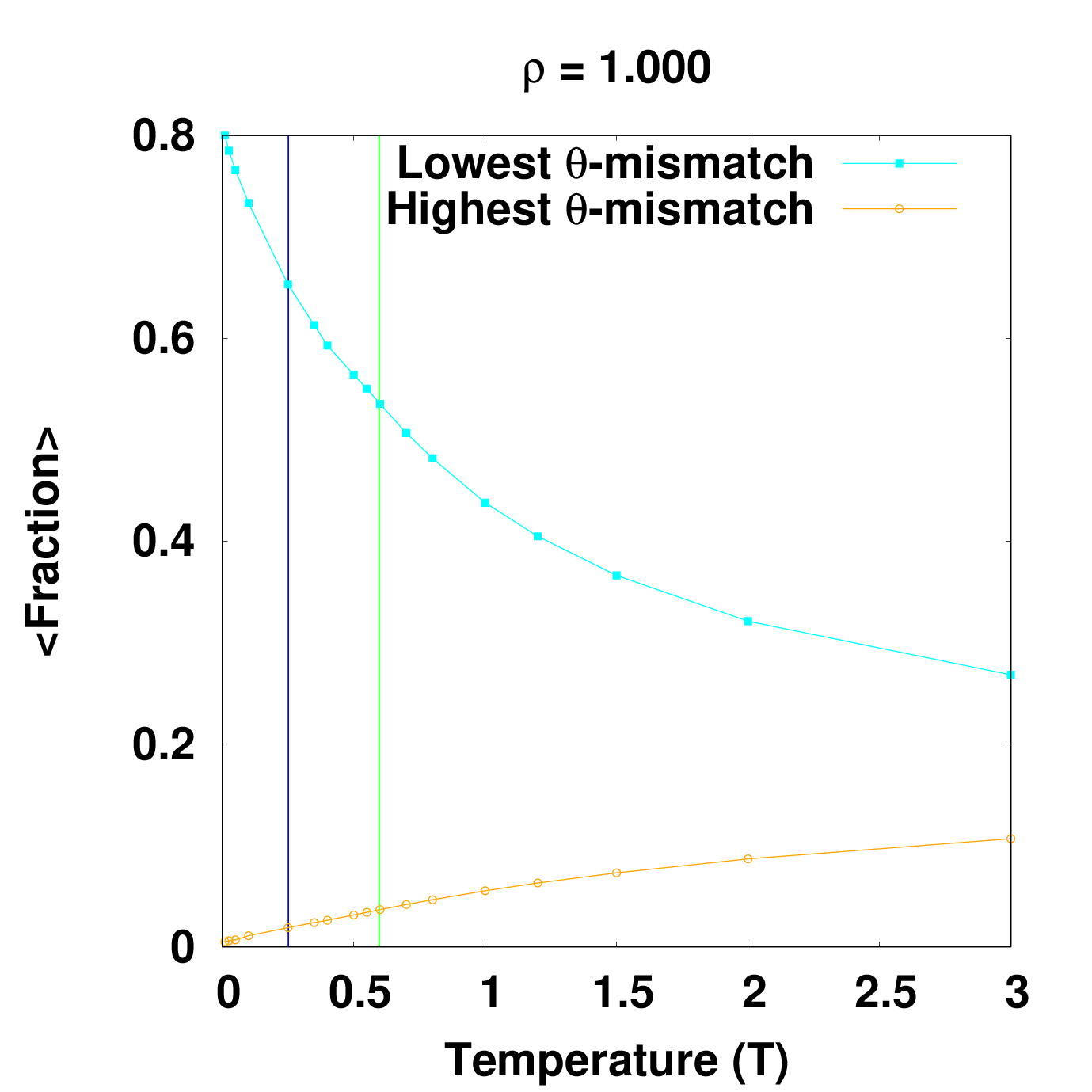}
 \caption{Frame-averaged fractions of low- and high-mismatch particles vs.\ temperature at three densities.  Each panel shows the fraction of particles with $\Delta\theta\le0.30\,$rad (``low-mismatch'', cyan) and $\Delta\theta\ge1.00\,$rad (``high-mismatch'', orange) for (top-left) $\rho=0.900$, (top-right) $\rho=0.950$, and (bottom) $\rho=1.000$.  The cutoffs were fixed by the first and third peaks of the full $\Delta\theta$ distributions from Fig.~\ref{fig:theta2a} and applied uniformly at all densities; vertical lines mark the same characteristic temperatures shown elsewhere (solidification, glass transition, etc.).  At $\rho=0.900$ the two fractions cross at $T\!\approx\!0.3$, indicating a crossover from vibration-dominated (low-$\Delta\theta$) behaviour to frequent large-angle/bond-breaking events; at $\rho=0.950$ the crossover shifts to $T\!\approx\!0.6$, reflecting stronger steric confinement, while at $\rho=1.000$ no crossover is observed up to $T=3.0$ (low-mismatch sites remain dominant).  Overall, increasing density delays the growth of high-mismatch populations and preserves vibrational (low-mismatch) geometry to higher temperatures.}
\label{fig:theta_frac}
\end{figure*}
\section{Evolution of the Per‐Particle Mismatch Angle with Temperature}
\label{evol_per}
At temperatures below the solidification line (blue vertical, as shown in Fig.~\ref{fig:theta3a}, particles remain tightly caged around their steric “just–touch” geometry, so the average mismatch angle 

$\langle\theta_{\rm mismatch}\rangle$ changes drastically around these temperatures across all densities and reflects only small vibrational excursions.  Heating into the intermediate regime between the solidification and glass–transition lines (green vertical) tiny cooperative distortions begin, but the growth of $\langle\theta_{\rm mismatch}\rangle$ remains modest until the glass transition. The slope in this regime is shallowest at the highest density, consistent with stronger steric caging. Once \(T\) exceeds the glass‐transition temperature \(T_{g}\), we see a much steeper rise; particles readily escape their cages, leading to rapid growth of the average bond‐angle distortion.

Beyond the glass transition, liquid‐like rearrangements dominate and drive  \(\langle\theta_{\rm mismatch}\rangle\)  rapidly upward toward a high‐temperature plateau.  As density increases, both the solidification and glass–transition temperatures shift upward and the overall angular span compresses (from roughly 0.45–0.72 \,rad at \(\rho=0.900\), to 0.25–0.64\,rad at \(\rho=0.950\), to 0.20–0.54 \,rad at \(\rho=1.000\)), revealing how vibrational, cooperative, and fully fluid distortions successively govern the local geometry upon heating. Finally, well above \(T_g\), each curve flattens toward its liquid‐state plateau, with the limiting \(\langle\theta_{\rm mismatch}\rangle\) decreasing from approximately 0.72\,rad at \(\rho = 0.900\) to 0.54\,rad at \(\rho = 1.000\).

\section{Density‐Dependence of Low‐ and High‐Mismatch Fractions}
\label{density_dep}
Fig.~\ref{fig:theta_frac} presents the frame-averaged fractions of particles with mismatch angle $\Delta\theta\le0.3\,$rad (``low-mismatch'') and $\Delta\theta\ge1.0\,$rad (``high-mismatch'') as functions of temperature for three packing densities, $\rho=0.900$, $0.950$, and $1.000$.  These cutoffs were fixed once anchored to the first and third peaks in the full $\Delta\theta$ distributions from Fig.~\ref{fig:theta2a} and then applied uniformly at all densities.

At $\rho=0.900$, the low- and high-mismatch fractions intersect at $T\approx0.3$, marking the crossover from vibration-dominated behavior to cooperative rearrangements: below this point, thermal vibrations around the steric ``just-touch'' angle prevail, whereas above it, bond-breaking excursions become more numerous.  Raising the density to $\rho=0.950$ shifts the intersection upward to $T\approx0.6$, reflecting stronger steric confinement that delays large-angle distortions.  At the highest density, $\rho=1.000$, the two curves remain distinct throughout the simulated range (up to $T=3.0$), indicating that even at elevated temperatures vibrational (low-$\Delta\theta$) fluctuations outnumber bond-breaking events.

In the deeply solid regime (left of the blue solidification line), almost all particles occupy the low-mismatch band, signaling that bond angles remain near the zero-strain reference.  Entering the viscoelastic window (between the blue and green lines) causes a sharp drop in the low-mismatch fraction as cooperative motions initiate, though the high-mismatch fraction remains small.  Beyond the glass transition (right of the green line), large-angle rearrangements become facile: the high-mismatch fraction grows rapidly, while the low-mismatch fraction continues its gradual decline.  At the highest temperatures, both fractions plateau: a persistent $\sim$20–25\% of particles still vibrate near zero strain, and a nonzero tail of highly mismatched bonds endures in the liquid.

Quantitatively, increasing $\rho$ both slows the decay of the low-mismatch fraction and retards the growth of the high-mismatch fraction, thereby shifting their crossover to higher $T$.  Denser packings raise the barrier for cooperative, large-angle excursions; vibrational fluctuations around the steric reference persist to higher temperatures, and bond-breaking distortions require progressively more thermal energy.  This monotonic density dependence provides a clear fingerprint of how packing fraction tunes the competition between localized vibrations and collective structural relaxation.  

\section{Mathematical Formulation of Thermal Nonaffine and Thermal Strain Metrics}
\label{sec:thermal_metrics}

Following our previous work \cite{jha2026thermal,jha2026thermally}, we characterize thermally driven local mechanical response using a thermal nonaffine displacement metric and the corresponding local thermal strain field. These quantities provide a particle-resolved description of the mechanical response generated solely by thermal fluctuations, without any externally imposed deformation. In the present work, they serve as the mechanical measures that we compare with the local steric ``\(\theta\)-mismatch'' in order to determine whether geometrically frustrated regions are preferentially associated with enhanced thermal nonaffine motion and local strain.

The thermal nonaffine metric is obtained from the nonaffine component of the particle displacement relative to a locally optimized affine transformation between two configurations separated by a finite thermal time interval. The resulting particle-resolved field quantifies the extent to which the local thermal motion cannot be described by an affine deformation. From the same local displacement-gradient field, we obtain the corresponding thermal strain tensor, which is decomposed into its deviatoric and volumetric components. The thermal deviatoric strain characterizes local shear-like distortions, whereas the thermal volumetric strain measures local dilational or compressional response.

We use these three thermal mechanical observables: the thermal nonaffine metric, thermal deviatoric strain, and thermal volumetric strain throughout the remainder of the manuscript. Their complete mathematical derivation and implementation, including the explicit equations and decomposition of the thermal strain field, have already been presented in detail in our previous studies \cite{jha2026thermal,jha2026thermally}; therefore, those derivations are not reproduced here. We use the previously established formulations directly in the present analysis and focus on their relationship with the local steric ``\(\theta\)-mismatch'' in the subsequent topology--mechanics comparison.
\section{Behavioiur of thermal nonaffine for low and high theta mismatch}
\label{therm_nonaff}
\subsection{Non‐Affine Displacement Distributions at \(\rho=0.900\)}
To probe the structural origin of particle rearrangements, we analyze the conditional probability distributions of non-affine displacements \( D^2_{\min} \) for particles classified by their local \(\theta\)-mismatch. Specifically, we compare particles with low and high \(\theta\)-mismatch at various temperatures spanning the solid, intermittent, supercooled, and normal liquid regimes.

In the solid regime (e.g., \( T = 0.010 \)), the distribution is sharply peaked at low \( D^2_{\min} \), where low \(\theta\)-mismatch particles dominate. However, at the larger \( D^2_{\min} \) values (around \( \sim 1.2 \)), the high \(\theta\)-mismatch particles become significantly more probable, indicating that even in the rigid solid, rare rearrangements are spatially localized in structurally disordered regions.

As temperature increases into the intermittent regime (\( 0.050 \leq T \leq 0.168 \)), this contrast becomes more pronounced: low \(\theta\)-mismatch particles continue to contribute mainly to small displacements, while high \(\theta\)-mismatch particles dominate the large displacement tail, highlighting enhanced dynamic heterogeneity.

At the glass transition temperature (\( T_g = 0.169 \)), the shape of the distribution begins to change, suggesting the onset of collective mobility. In the supercooled regime (\( 0.180 \leq T \leq 0.270 \)), both groups begin to exhibit broader displacement distributions, but high \(\theta\)-mismatch particles still preferentially undergo larger rearrangements, though the distinction becomes less sharp.

Finally, in the normal liquid regime (\( 0.350 \leq T \leq 0.500 \)), the conditional distributions for low and high \(\theta\)-mismatch particles overlap more substantially, indicating that structural mismatch becomes less predictive of dynamic behavior as thermal motion dominates.

This progression confirms that \(\theta\)-mismatch is a strong structural indicator of non-affine displacement in the deeply supercooled and glassy states, but loses discriminative power in the normal liquid regime.

\subsection{Non‐Affine Displacement Distributions at \(\rho=0.950\)}

\begin{figure*}[htb]
  \centering
  \begin{subfigure}[b]{0.32\textwidth}
    \includegraphics[width=\linewidth]{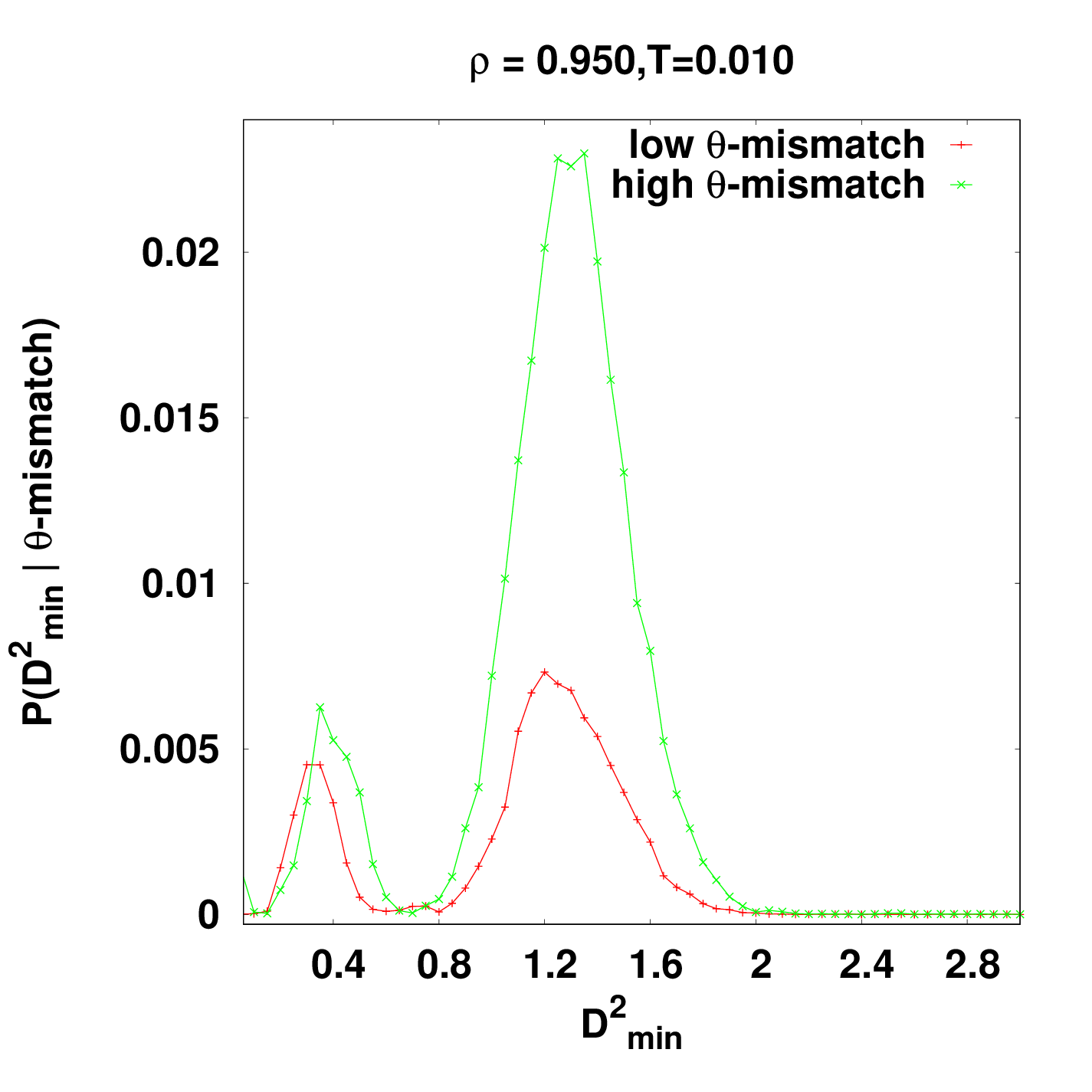}
    \caption{$T=0.010$}
  \end{subfigure}\hfill
  \begin{subfigure}[b]{0.32\textwidth}
    \includegraphics[width=\linewidth]{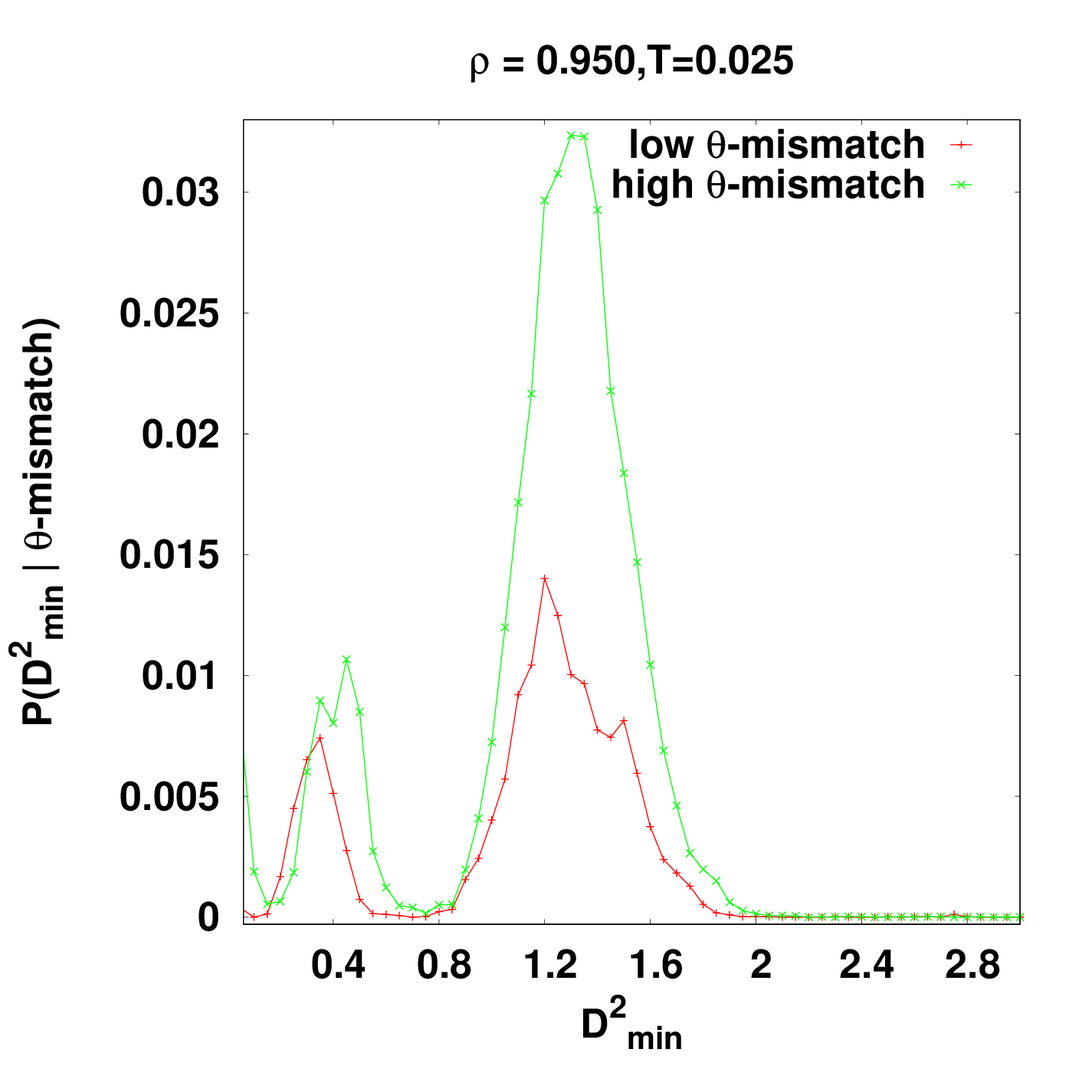}
    \caption{$T=0.025$}
  \end{subfigure}\hfill
   \begin{subfigure}[b]{0.32\textwidth}
    \includegraphics[width=\linewidth]{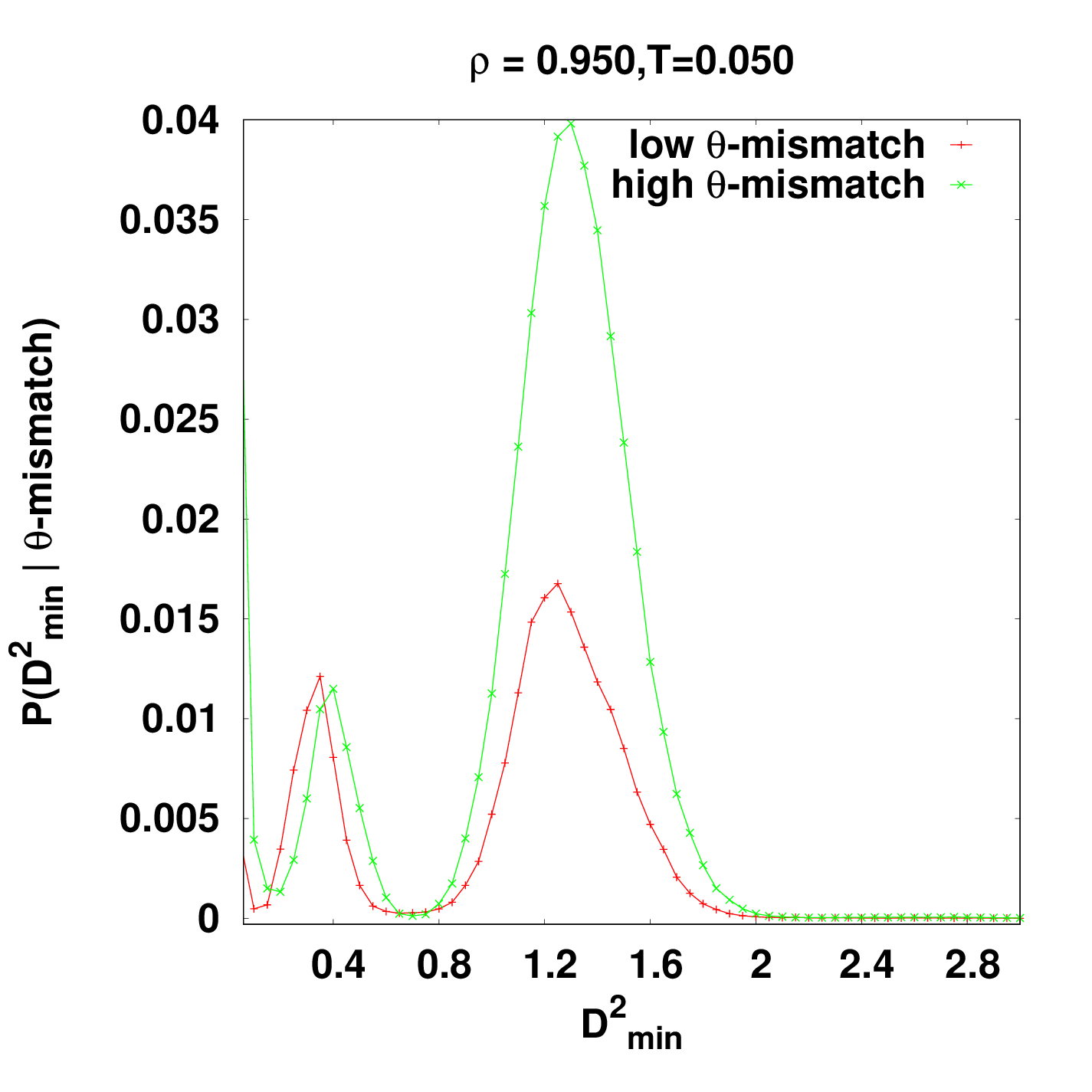}
    \caption{$T=0.050$}
  \end{subfigure}
  \vspace{0.5cm}
   \begin{subfigure}[b]{0.32\textwidth}
    \includegraphics[width=\linewidth]{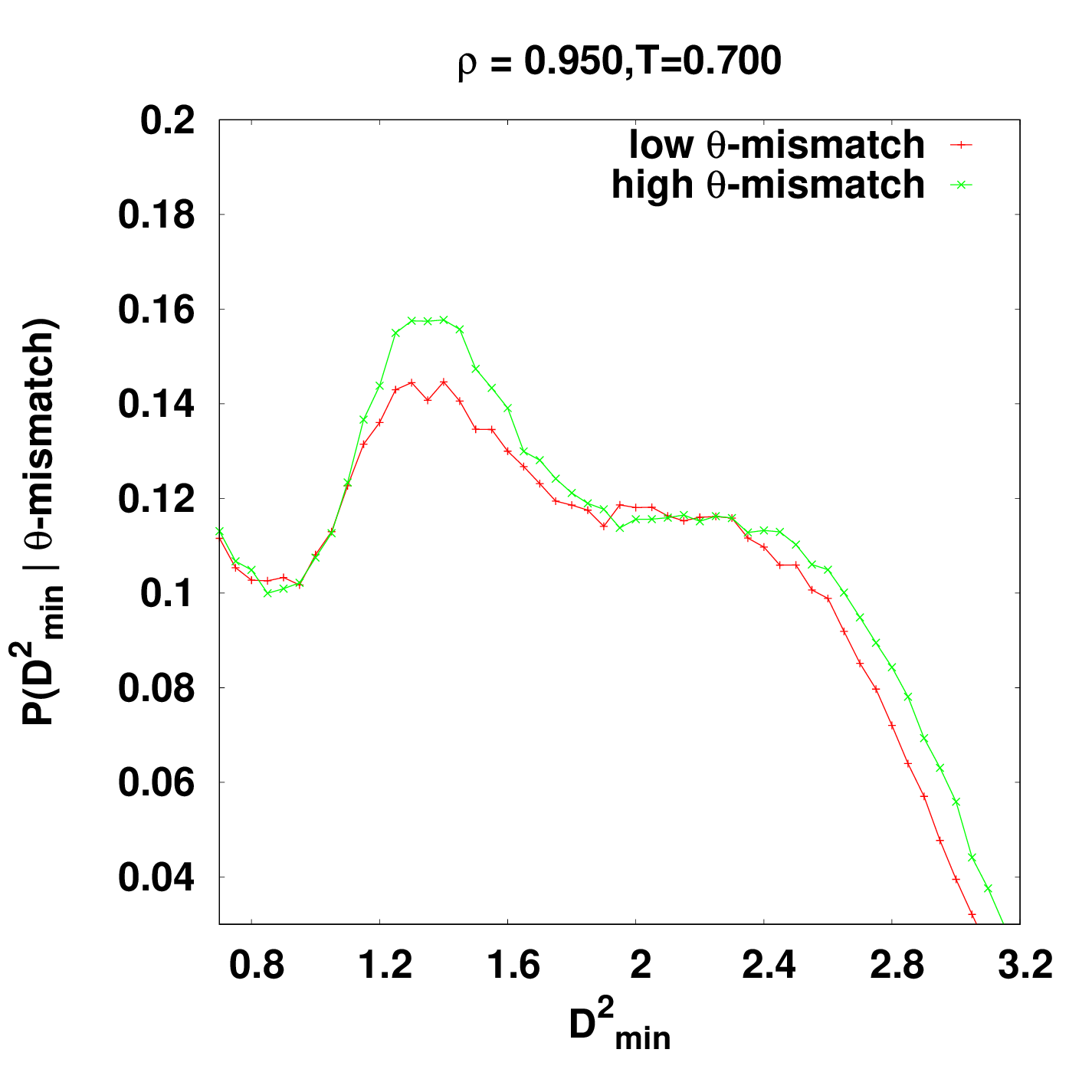}
    \caption{$T=0.700$}
  \end{subfigure}\hfill
  \begin{subfigure}[b]{0.32\textwidth}
    \includegraphics[width=\linewidth]{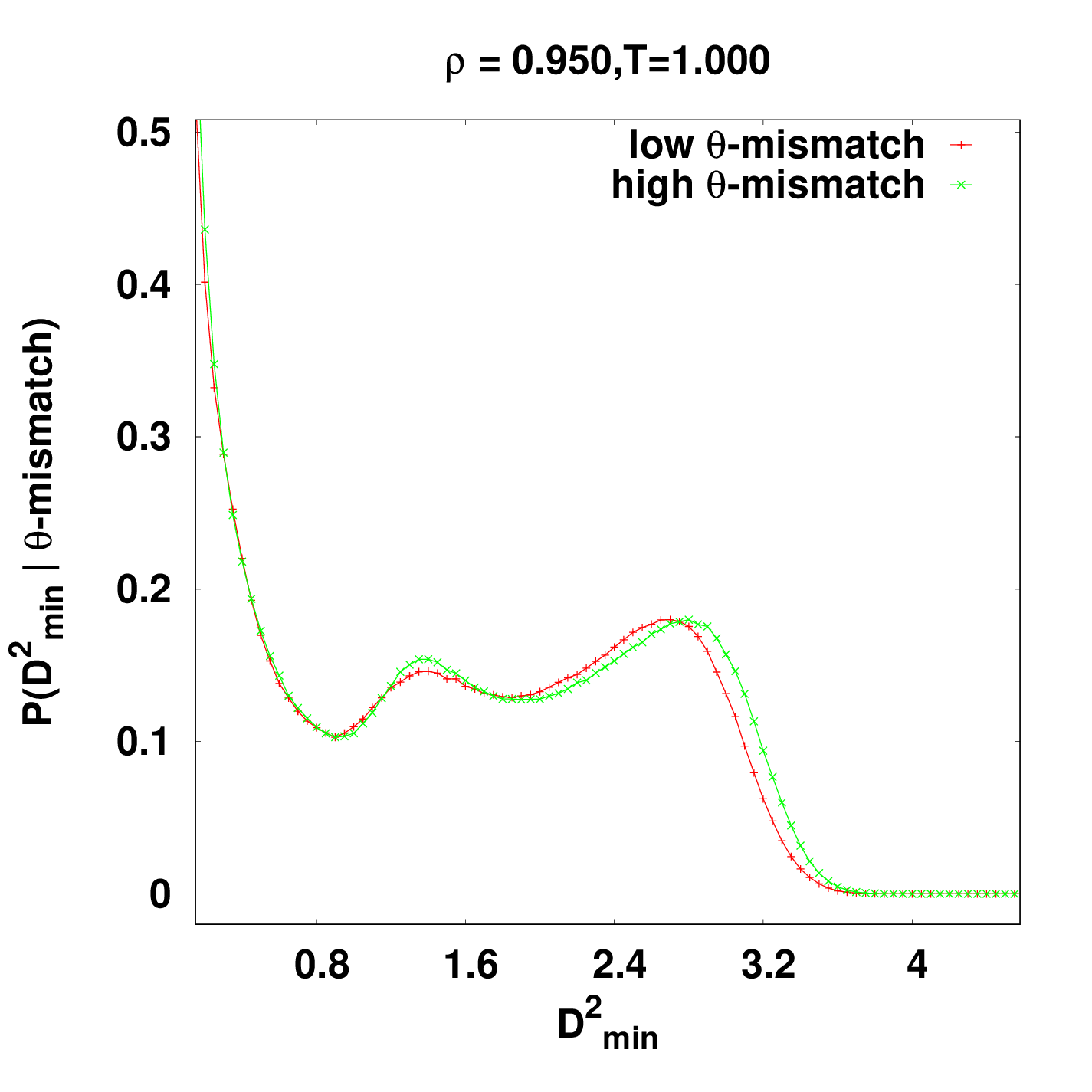}
    \caption{$T=1.000$}
  \end{subfigure}\hfill
  \begin{subfigure}[b]{0.32\textwidth}
    \includegraphics[width=\linewidth]{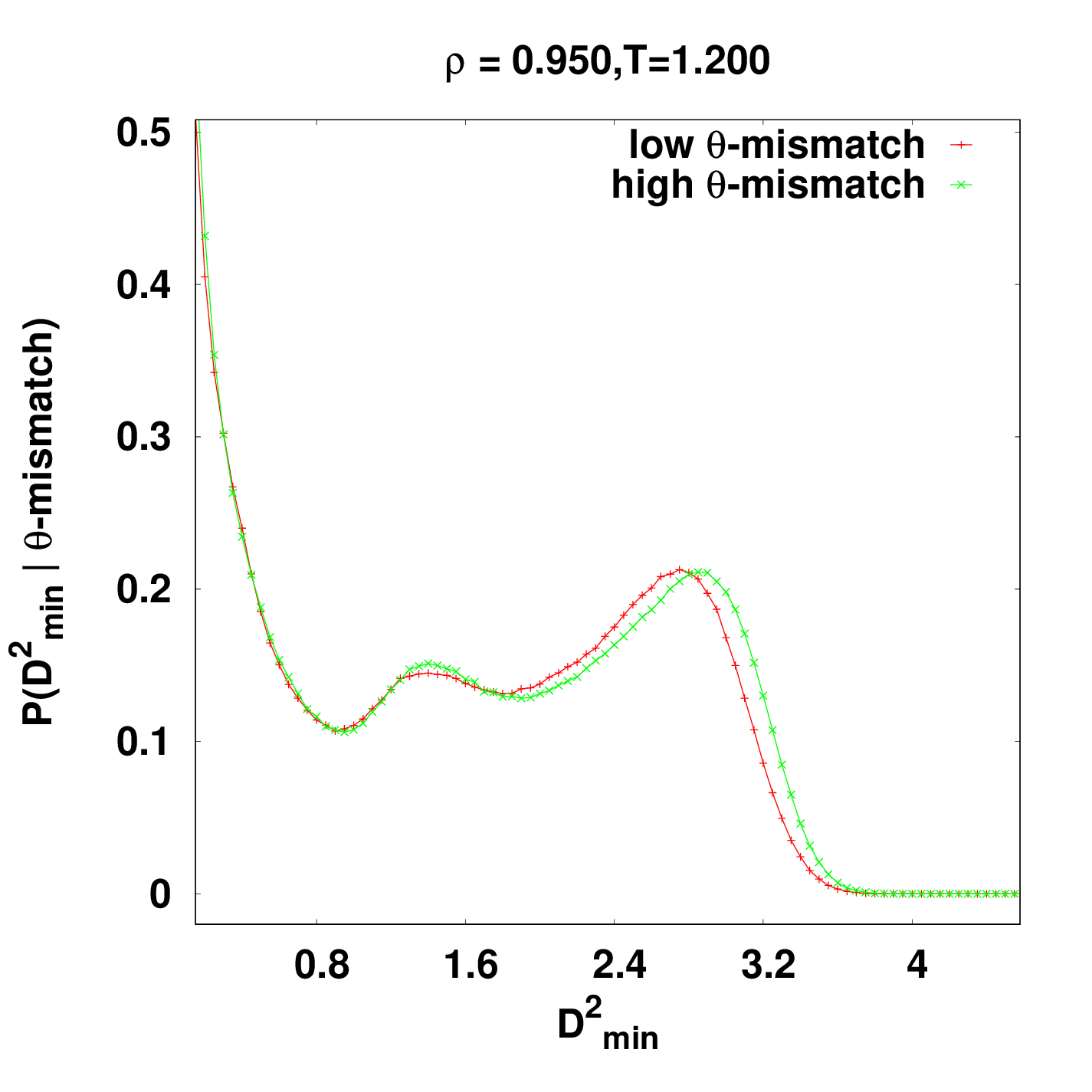}
    \caption{$T=1.200$}
  \end{subfigure}
 \caption{Conditional probability densities $P(D_{\min}^2\mid\Delta\theta)$ at $\rho=0.950$.
Curves compare the lowest 10\% in $\Delta\theta$ (``low--mismatch'', red) with the highest 10\% (``high--mismatch'', green) over a range of temperatures; vertical markers indicate $T_{\rm solid}=0.110$, $T_g=0.343$, and $T_{\rm onset}=0.679$. 
In the deeply jammed solid ($T\ll T_{\rm solid}$), low--mismatch particles are tightly localized at small $D_{\min}^2$ (peak $\sim0.5$) while high--mismatch sites at large $D_{\min}^2$ value. 
Entering the intermittent window ($T\sim0.1$--0.16), a distinct large--$D_{\min}^2$ peak ($\sim1.2$) grows for high--mismatch particles, whereas low--mismatch particles remain concentrated at small displacements, indicating that rearrangements nucleate in geometrically frustrated regions. 
In the supercooled regime ($T_g\lesssim T \lesssim T_{\rm onset}$), both distributions broaden and the high--mismatch PDF develops extended weight to $D_{\min}^2>2$, signifying enhanced dynamic heterogeneity. 
Finally, in the normal liquid ($T\gtrsim0.7$), the conditional PDFs converge (broad peaks near $D_{\min}^2\!\approx\!1.5$), showing that steric mismatch becomes a weaker predictor of nonaffine activity when thermal motion dominates.}
\label{fig:thetaa4}
\end{figure*}

Fig.~\ref{fig:thetaa4} presents the conditional probability density \(P\bigl(D_{\min}^2 \mid \theta\text{-mismatch}\bigr)\) for low‐ and high‐\(\theta\)–mismatch particles at \(\rho=0.950\), over temperatures spanning the solid, intermittent, supercooled, and normal‐liquid regimes.  The key thermodynamic markers are the solidification temperature \(T_{\rm solid}=0.110\), the cage‐escape onset \(T_{\rm onset}=0.679\), and the glass‐transition \(T_g=0.343\), indicated here for reference.

At \(T=0.010\ll T_{\rm solid}\) (deep solid), both populations are tightly localized at very small \(D_{\min}^2\), but the low‐\(\theta\)–mismatch particles (red) dominate this regime: their curve peaks sharply near \(D_{\min}^2\approx0.5\), while the high‐\(\theta\)–mismatch group (green) contributes only a modest tail, indicating rare, highly frustrated contacts even in the arrested solid.  

By \(T=0.025\) and \(0.050\) (still below \(T_{\rm solid}\)), the same pattern persists, but the high‐\(\theta\) peak begins to grow a secondary shoulder at \(D_{\min}^2\approx1.2\), presaging local bond‐breaking events in structurally mismatched regions.

Crossing into the intermittent regime (\(0.110 < T < 0.343\)), at \(T=0.100\) and \(T=0.160\), the distinction sharpens: low‐\(\theta\) particles remain confined to \(D_{\min}^2\lesssim0.8\), while high‐\(\theta\) particles overwhelmingly populate the larger‐displacement peak at \(D_{\min}^2\sim1.2\).  This underscores the onset of cooperative, cage‐escape motions preferentially in geometrically mismatched zones.

At the glass‐transition \(T_g=0.343\) (not shown) and into the supercooled regime (\(0.350 \le T \le 0.679\)), both distributions broaden.  The low‐\(\theta\) curve begins to develop a shoulder near \(D_{\min}^2\sim1.2\), while the high‐\(\theta\) distribution extends further into tail regions (\(D_{\min}^2>2.0\)), reflecting increased dynamic heterogeneity.

Finally, in the normal liquid (\(T\ge0.700\)), the two curves converge in shape: both low‐ and high‐\(\theta\) populations exhibit broad, overlapping peaks centered around \(D_{\min}^2\approx1.5\), indicating that thermal motion dominates over steric mismatch and dissolves the structural memory imprinted by \(\theta\)-mismatch.

Overall, at \(\rho=0.950\), \(\theta\)-mismatch remains a strong predictor of non‐affine displacements deep into the supercooled regime, but its discriminative power fades as \(T\) approaches the normal‐liquid limit.

\subsection{Nonaffine Response at \(\rho=1.000\)}
\label{sec:nonaffine_1000}

\begin{figure*}[ht!]
  \centering
  \begin{subfigure}[b]{0.32\textwidth}
    \includegraphics[width=\linewidth]{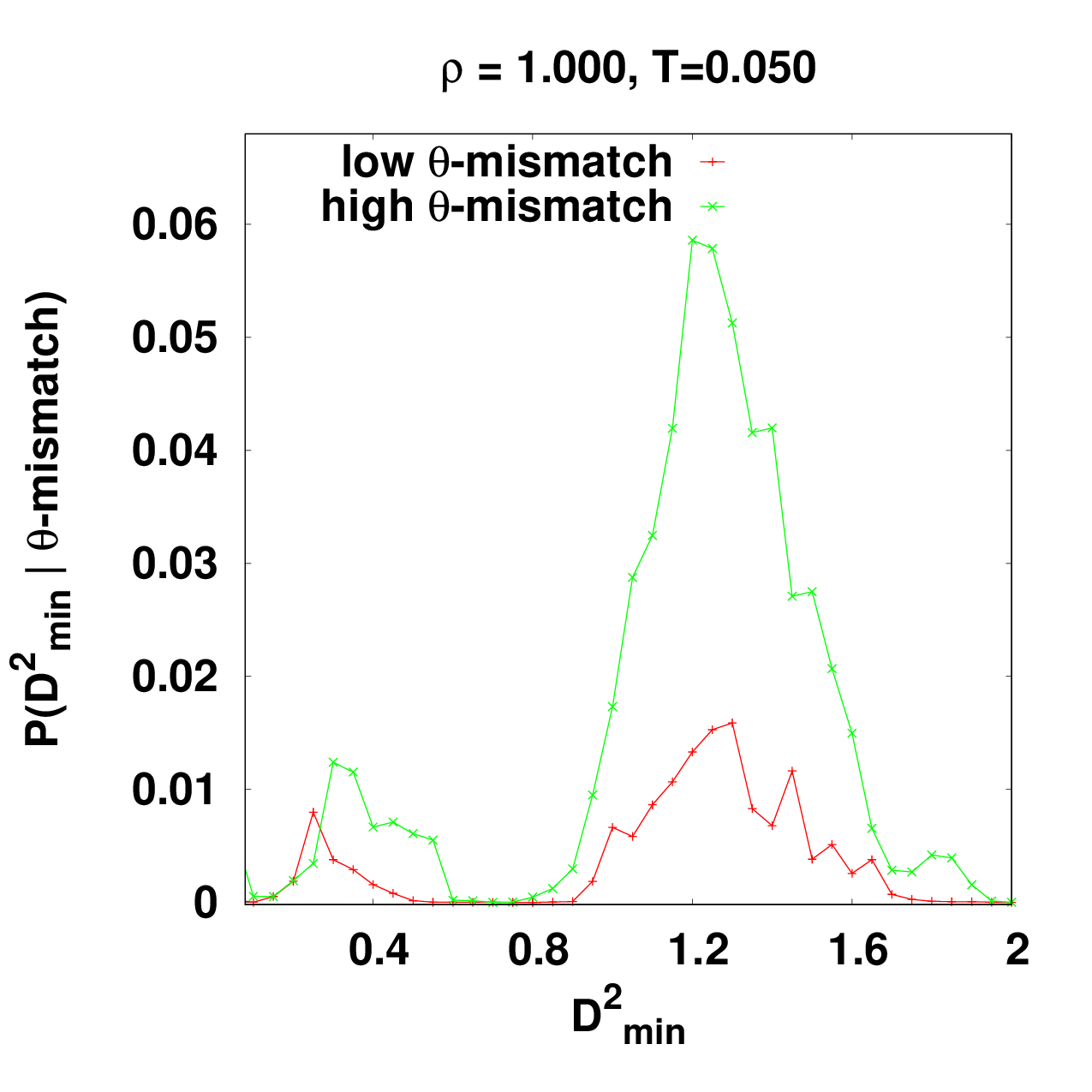}
    \caption{$T=0.050$}
  \end{subfigure}\hfill
   \begin{subfigure}[b]{0.32\textwidth}
    \includegraphics[width=\linewidth]{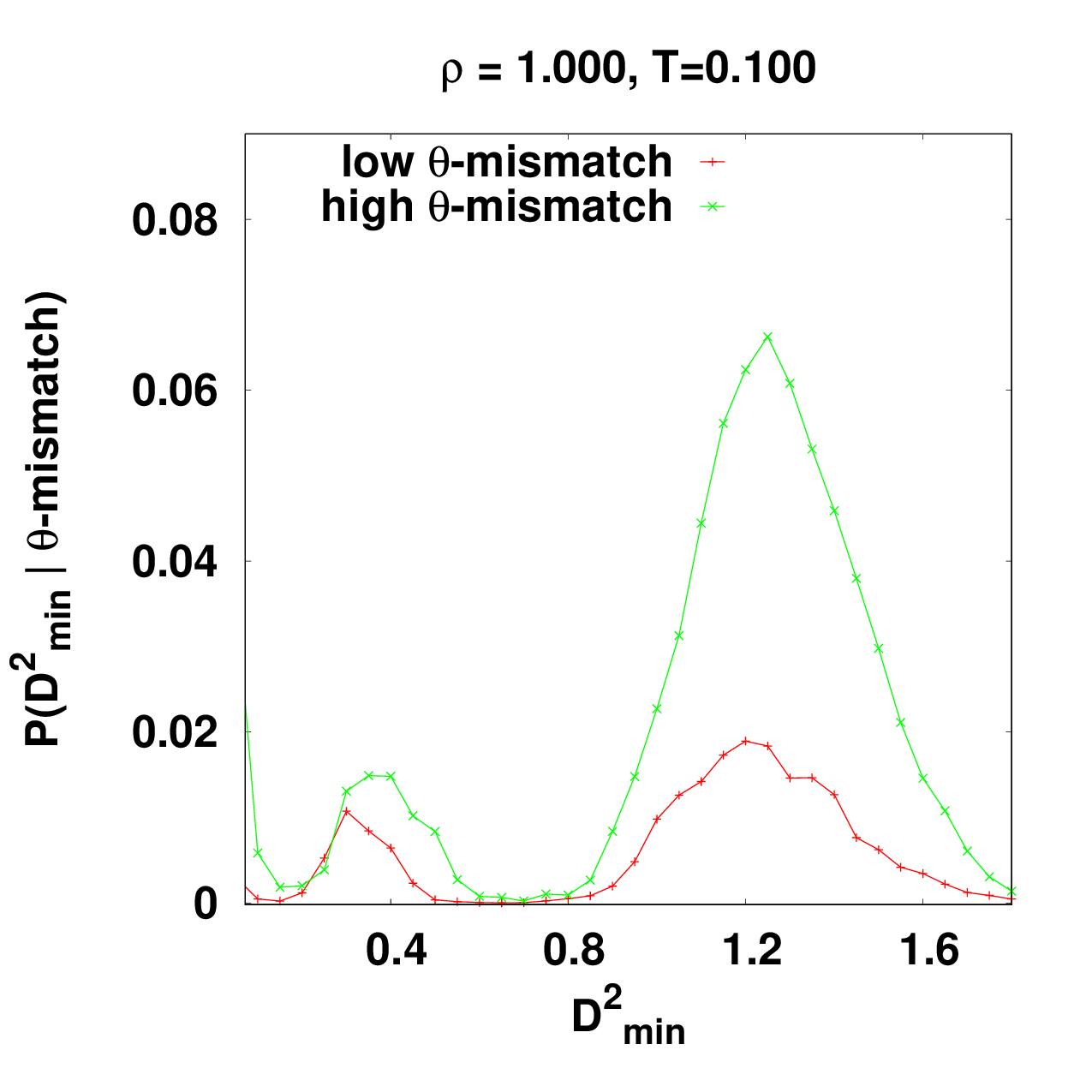}
    \caption{$T=0.100$}
  \end{subfigure}
  \begin{subfigure}[b]{0.32\textwidth}
    \includegraphics[width=\linewidth]{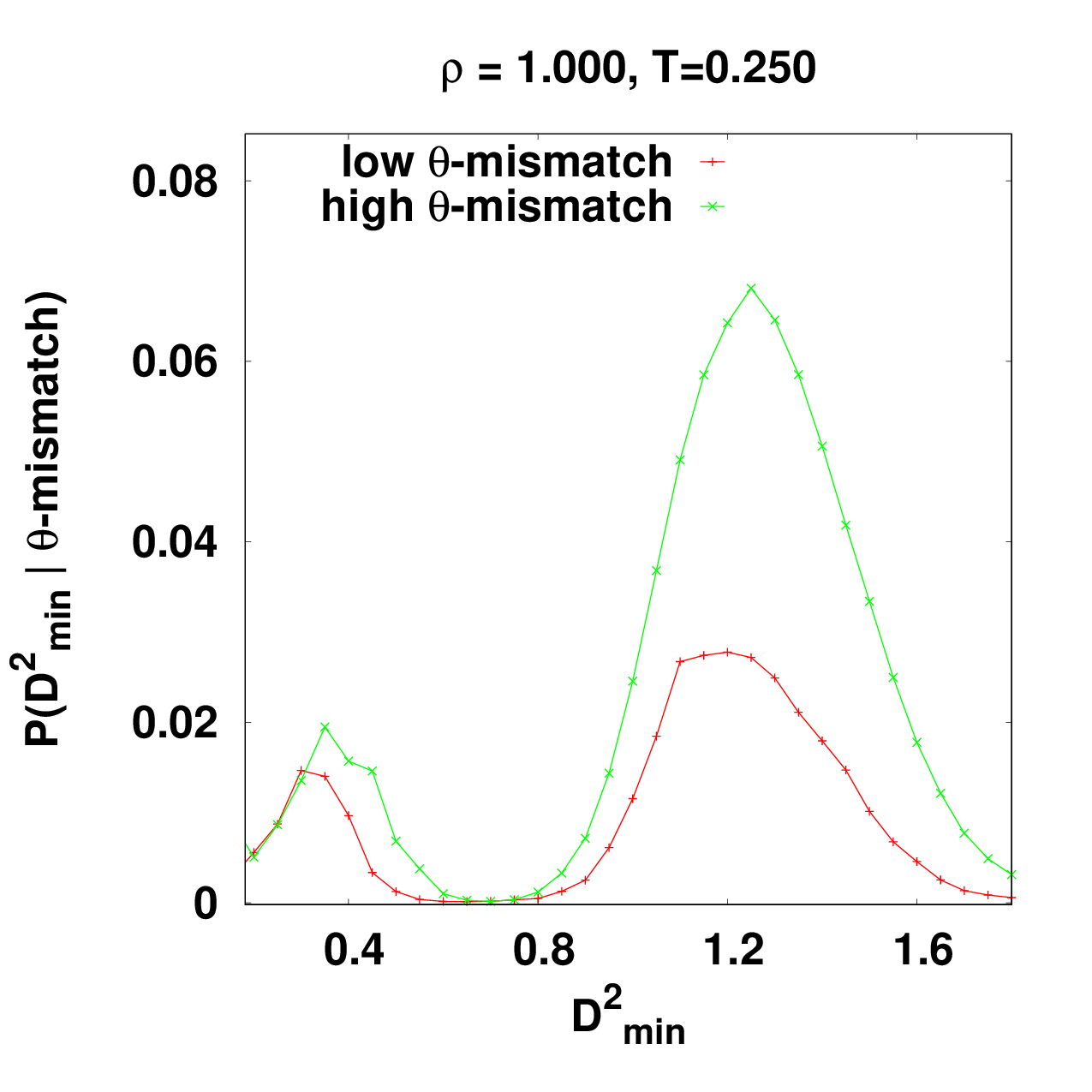}
    \caption{$T=0.250$}
  \end{subfigure}
  \vspace{0.5cm}
 \vspace{0.5em}
   \begin{subfigure}[b]{0.32\textwidth}
    \includegraphics[width=\linewidth]{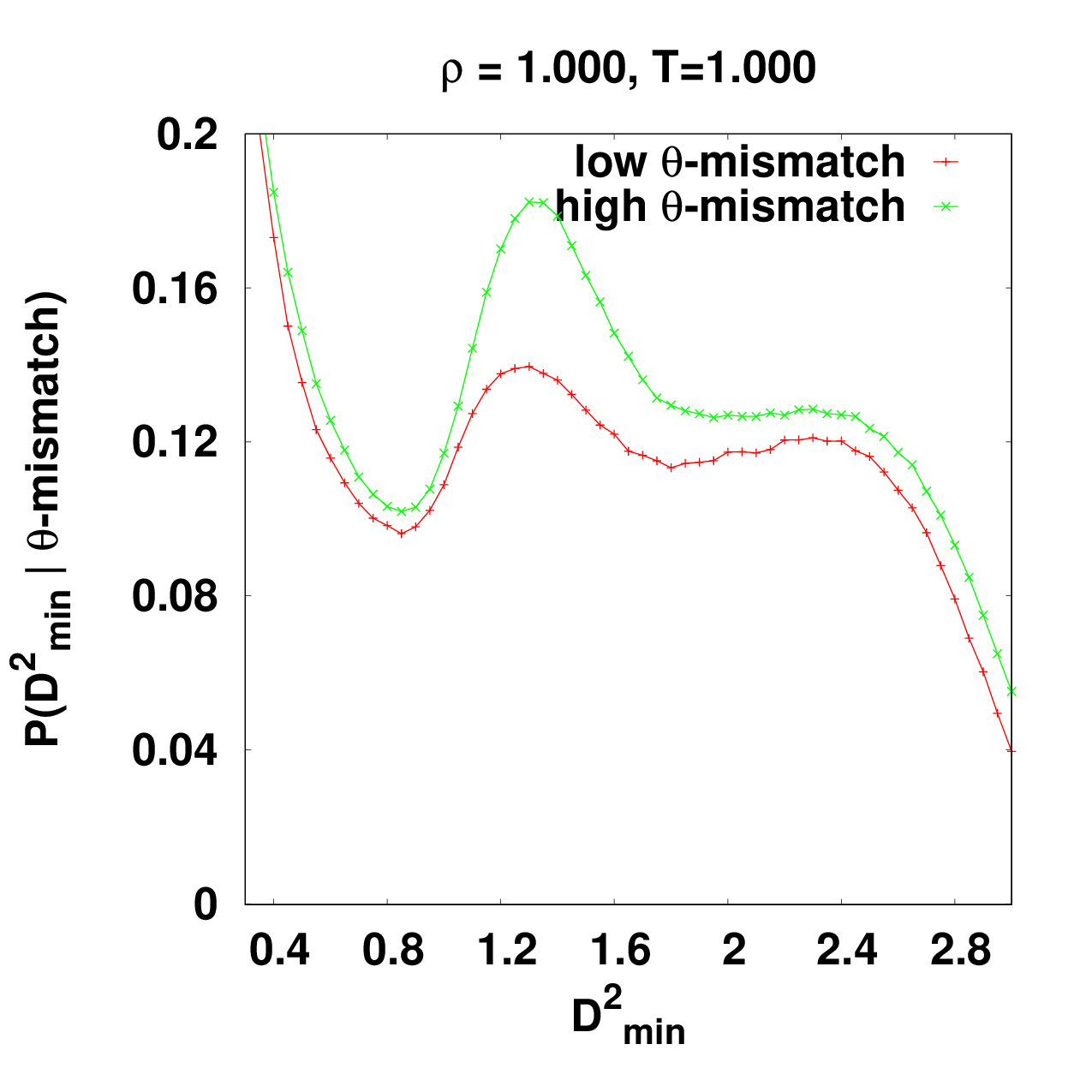}
    \caption{$T=1.000$}
  \end{subfigure}\hfill
   \begin{subfigure}[b]{0.32\textwidth}
    \includegraphics[width=\linewidth]{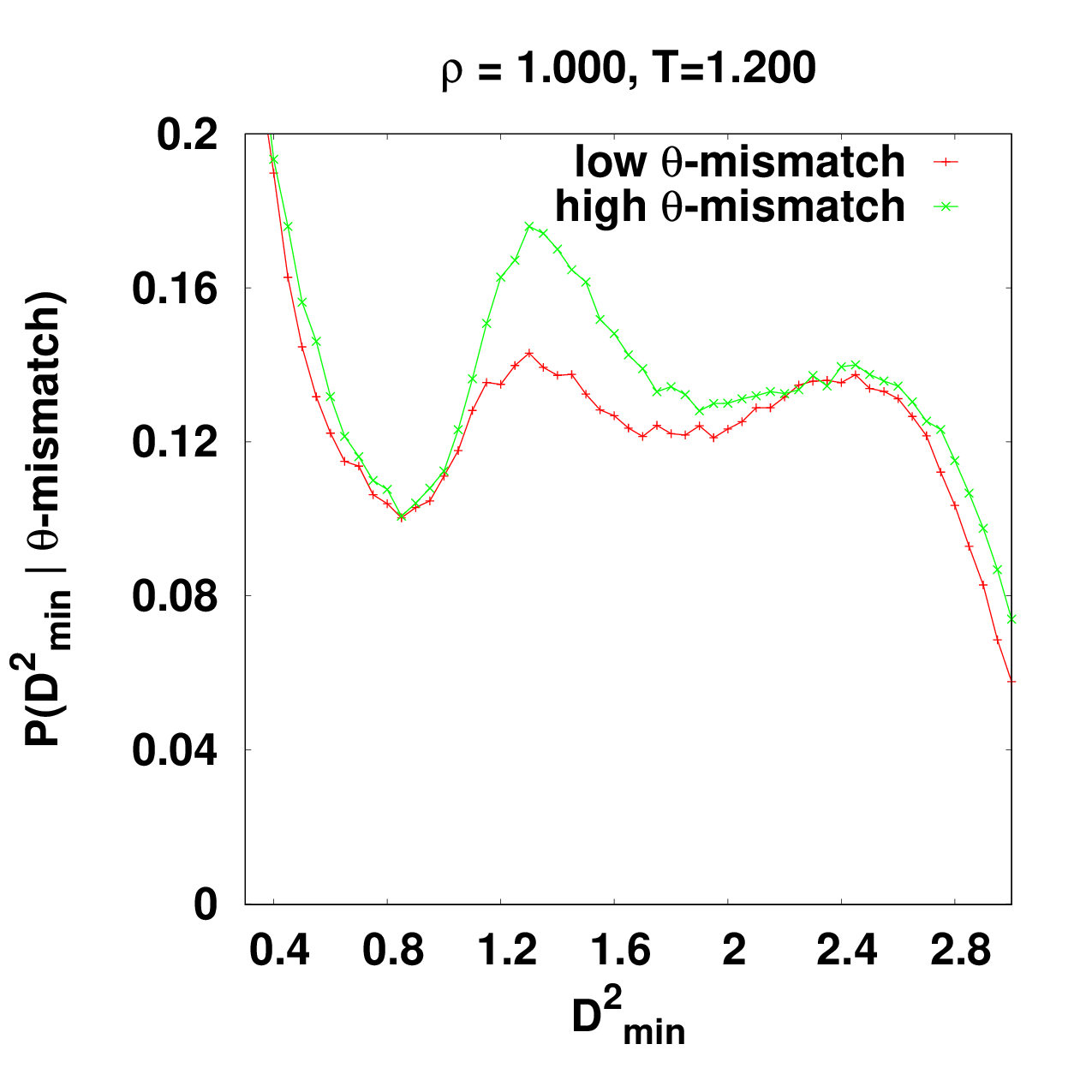}
    \caption{$T=1.200$}
  \end{subfigure}\hfill
  \begin{subfigure}[b]{0.32\textwidth}
    \includegraphics[width=\linewidth]{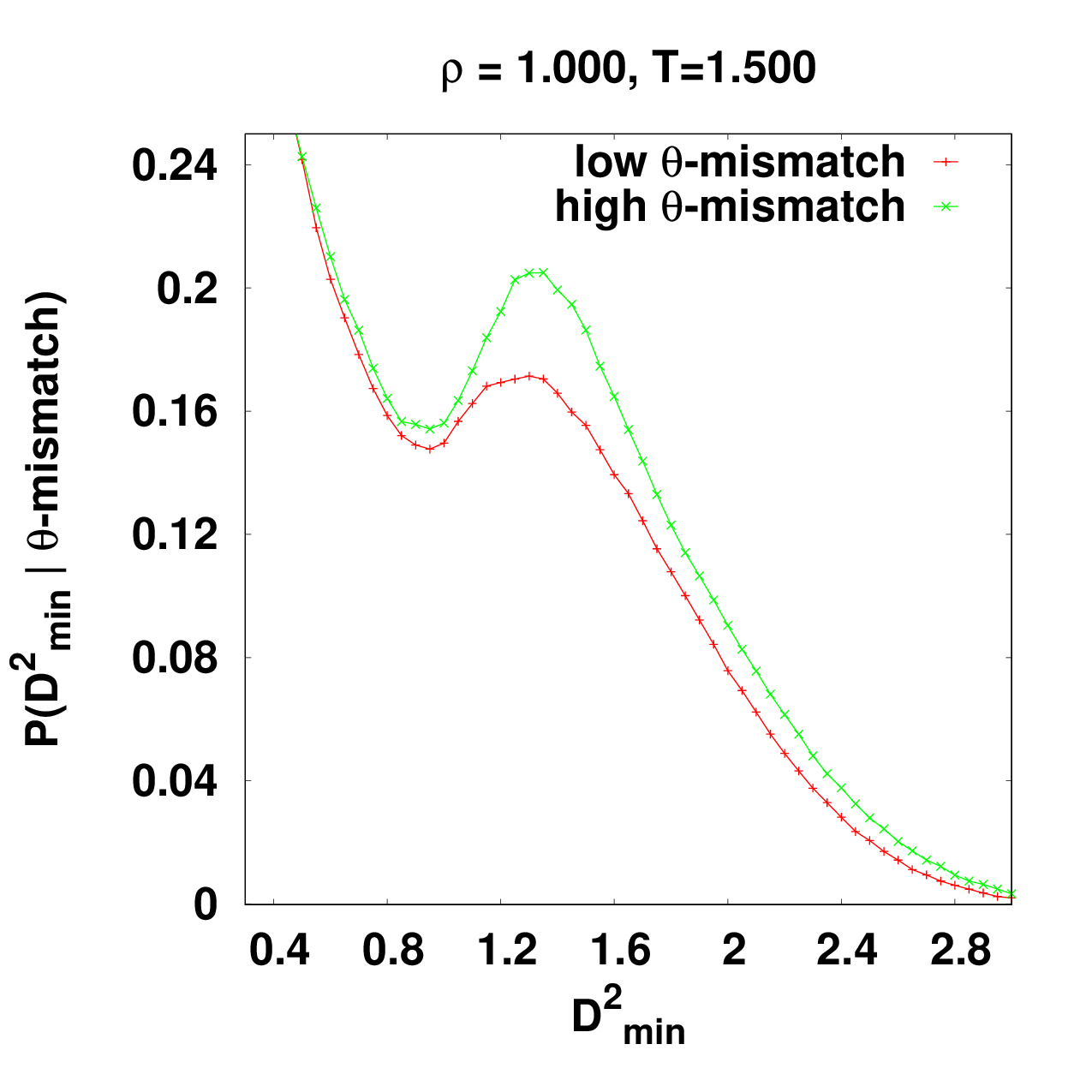}
    \caption{$T=1.500$}
  \end{subfigure}
  \caption{Conditional distributions $P(D_{\min}^2\mid\,$low-/high-$\Delta\theta)$ at $\rho=1.000$ (red: lowest 10\% $\Delta\theta$, green: highest 10\%).  In the deep solid ($T\ll T_{\rm solid}=0.251$) both groups peak at small $D_{\min}^2$, but high-$\Delta\theta$ sites show an enhanced shoulder at $D_{\min}^2\sim1.2$ indicating rare, frustration-driven hops.  In the viscoelastic window, the high-$\Delta\theta$ population dominates the large-$D_{\min}^2$ peak (cooperative cage escapes), while above $T_g$ both curves broaden and largely overlap as thermal motion erases structural memory.  Vertical markers show $T_{\rm solid}$, $T_g$, and $T_{\rm onset}$.}

  \label{fig:thetaa7}
\end{figure*}

At the highest density considered, \(\rho=1.000\), the strong steric confinement provides a stringent test of the relation between local geometric mismatch and thermally driven mechanical response. The characteristic temperatures at this state point are \(T_{\rm ML}=0.251\), \(T_g=0.596\), and \(T_{\rm onset}=1.483\). We examine the conditional probability density $P\!\left(D_{\min}^2\mid\Delta\theta\right)$
for the two topology-selected populations consisting of the \(10\%\) lowest- and highest-\(\Delta\theta\) particles. The corresponding distributions are shown in Fig.~\ref{fig:thetaa7}.
In the deeply arrested regime, \(T\leq T_{\rm ML}\), both populations are concentrated at small \(D_{\min}^2\), consistent with predominantly vibrational motion within tightly confined cages. The low-\(\Delta\theta\) population exhibits a narrow maximum around \(D_{\min}^2\simeq0.4$--$0.5\), whereas the high-\(\Delta\theta\) population has a much weaker contribution to this low-displacement peak but already develops a small shoulder near \(D_{\min}^2\simeq1.2\). Thus, even when the system as a whole is mechanically locked, the rare nonaffine events that survive are preferentially associated with geometrically frustrated regions. This distinction becomes substantially more pronounced on entering the viscoelastic regime, \(T_{\rm ML}<T<T_g\). The high-\(\Delta\theta\) distribution develops a pronounced peak near \(D_{\min}^2\simeq1.2\), while the low-\(\Delta\theta\) distribution remains concentrated at smaller displacements and develops only a weaker shoulder in the same region. The emergence of this second population of larger nonaffine displacements indicates that thermally activated cage escapes occur preferentially at sites with larger steric mismatch, while low-mismatch regions remain comparatively constrained.

Upon further heating through \(T_g\) into the supercooled regime, both conditional distributions broaden, indicating that thermally activated rearrangements increasingly involve a larger fraction of the system. The low-\(\Delta\theta\) population progressively acquires weight at larger \(D_{\min}^2\), whereas the high-\(\Delta\theta\) population develops a more extended large-\(D_{\min}^2\) tail, with values beyond \(D_{\min}^2\sim2\) becoming appreciable. The two populations therefore become less distinct with increasing temperature, although the high-mismatch sites continue to carry a relative excess of the largest nonaffine displacements. By \(T\sim1\), the two distributions have become qualitatively similar, indicating a substantial loss of the structural selectivity of \(\Delta\theta\), while a residual enhancement of the high-\(\Delta\theta\) tail remains.

Above the onset temperature, in the normal-liquid regime, the conditional distributions become broad and strongly overlapping, with characteristic displacements of order unity. Thermal motion now dominates over the local steric distinction, and the memory of the initial geometric mismatch is largely lost. The progressive evolution from a sharply separated low-displacement population and a weak high-\(D_{\min}^2\) tail in the mechanically locked regime, to a distinct high-\(D_{\min}^2\) population in the viscoelastic regime, and finally to overlapping distributions in the liquid demonstrates that the predictive role of steric mismatch is strongly temperature dependent. Most importantly, at \(\rho=1.000\), high-\(\Delta\theta\) regions emerge as the preferred locations of thermally driven nonaffine rearrangements over the low-temperature mechanically active regime, providing a direct link between local geometric frustration and thermal mechanical response.
\section{Comparative Summary of Thermal Nonaffine Response Across Density}
\label{sec:d2min_comparative}
Comparing the conditional thermal nonaffine distributions across \(\rho=0.900\), \(0.950\), and \(1.000\) reveals a common structural association together with a systematic density dependence. At all three state points, the low-temperature response is dominated by small nonaffine displacements, reflecting predominantly vibrational motion within the local cages, while the high-\(\Delta\theta\) population systematically contributes more strongly to the large-\(D_{\min}^2\) tail. This distinction is already visible in the deeply arrested regime, where rare nonaffine events remain preferentially localized at geometrically mismatched sites. Upon heating into the mechanically active viscoelastic regime, the separation becomes more pronounced as high-\(\Delta\theta\) regions develop enhanced weight at the larger-\(D_{\min}^2\) values associated with cage-escape events. The distinction subsequently weakens upon further heating because thermal activation progressively mobilizes the low-\(\Delta\theta\) population as well. The temperature range over which this structural selectivity persists, however, depends strongly on density: at \(\rho=0.900\) the two conditional distributions begin to overlap more substantially at comparatively low temperature, whereas at \(\rho=0.950\) the preferential large-displacement response of high-\(\Delta\theta\) sites persists deeper into the supercooled regime.
At \(\rho=1.000\), the strong steric confinement maintains the distinction over the widest temperature range, with high-\(\Delta\theta\) sites retaining an enhanced large-displacement tail even as the two distributions progressively converge at high temperature. Thus, increasing density not only shifts the characteristic temperature scales upward but also extends the temperature range over which local steric mismatch remains predictive of thermal nonaffine rearrangements. Across all state points, the overall trend establishes the central structure response relation of this analysis: regions of larger steric mismatch are preferential sites for thermally activated nonaffine rearrangements, while this structural selectivity is progressively erased as thermal motion becomes dominant in the liquid regime.

\begin{figure*}[ht!]
  \centering
  \begin{subfigure}[b]{0.32\textwidth}
    \includegraphics[width=\linewidth]{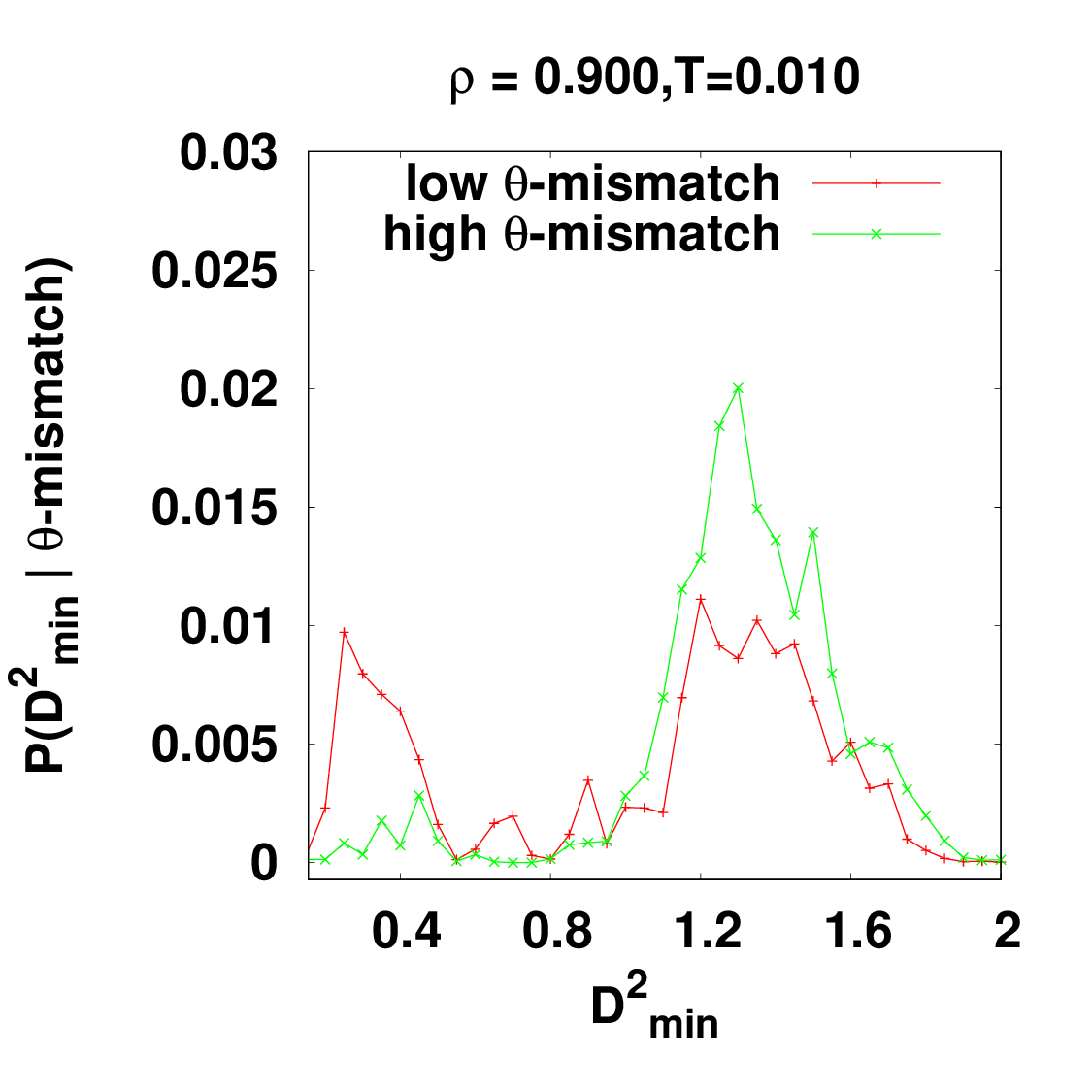}
    \caption{$T=0.010$}
  \end{subfigure}\hfill
  \begin{subfigure}[b]{0.32\textwidth}
    \includegraphics[width=\linewidth]{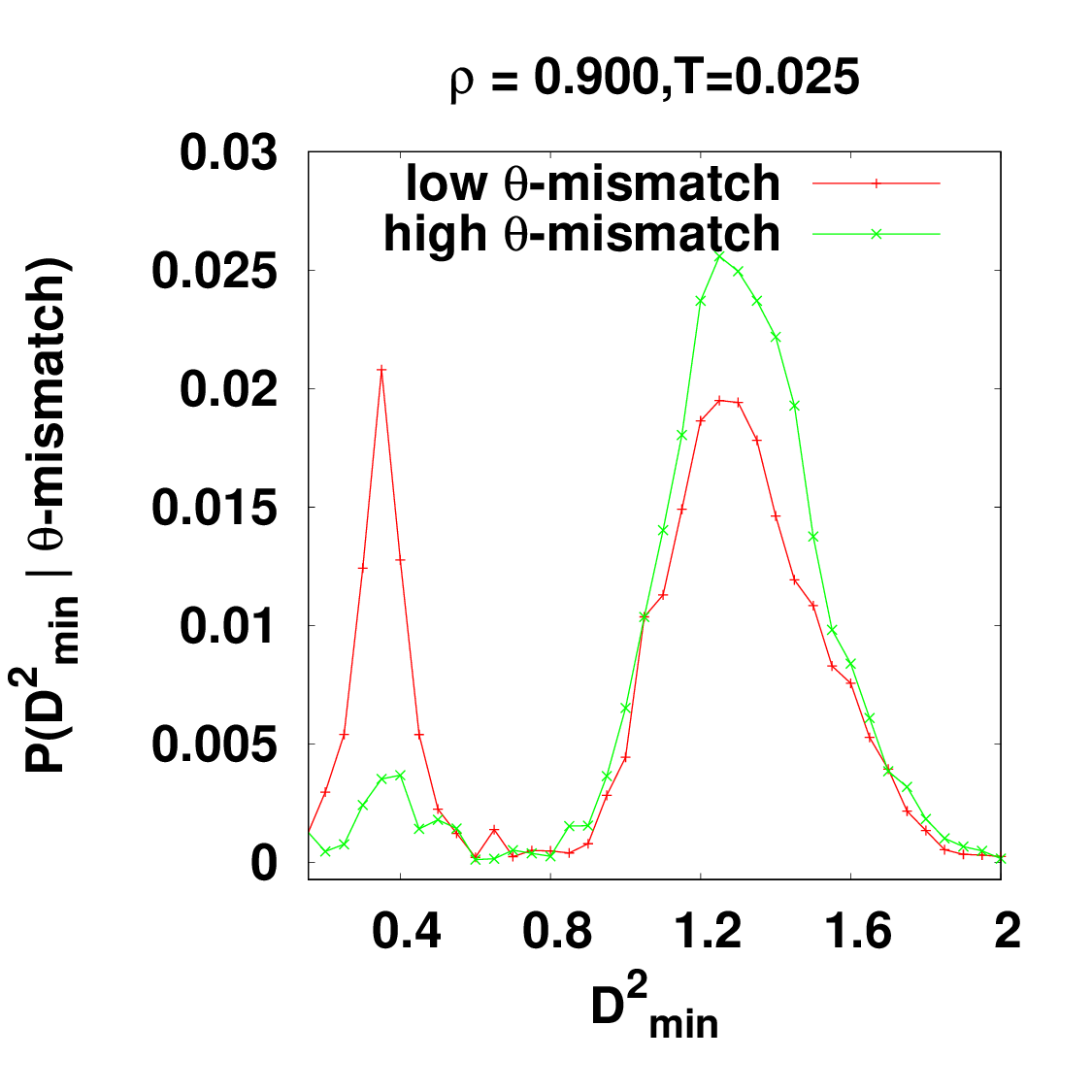}
    \caption{$T=0.025$}
  \end{subfigure}\hfill
  \begin{subfigure}[b]{0.32\textwidth}
    \includegraphics[width=\linewidth]{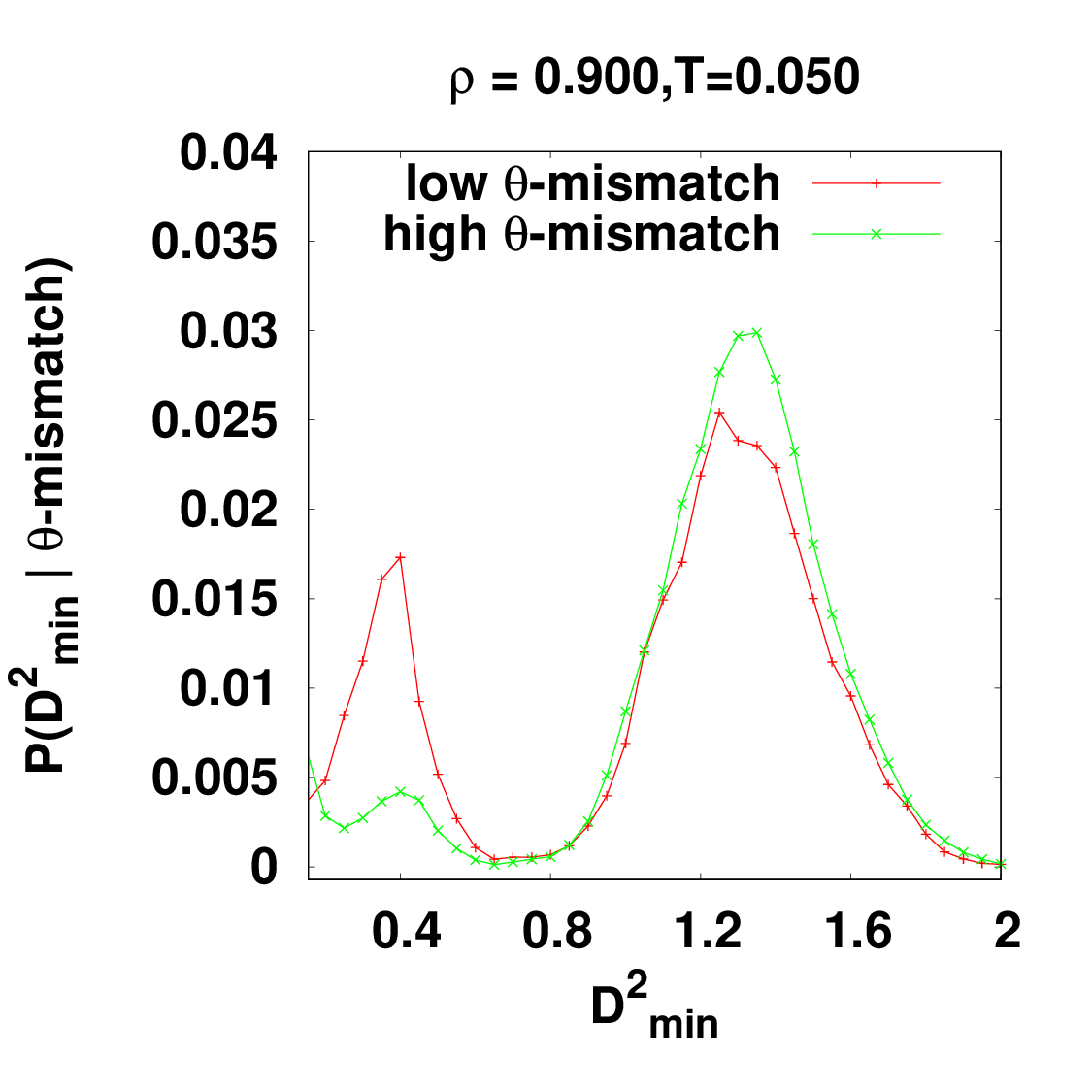}
    \caption{$T=0.050$}
  \end{subfigure}

  \vspace{0.5cm}

  \begin{subfigure}[b]{0.32\textwidth}
    \includegraphics[width=\linewidth]{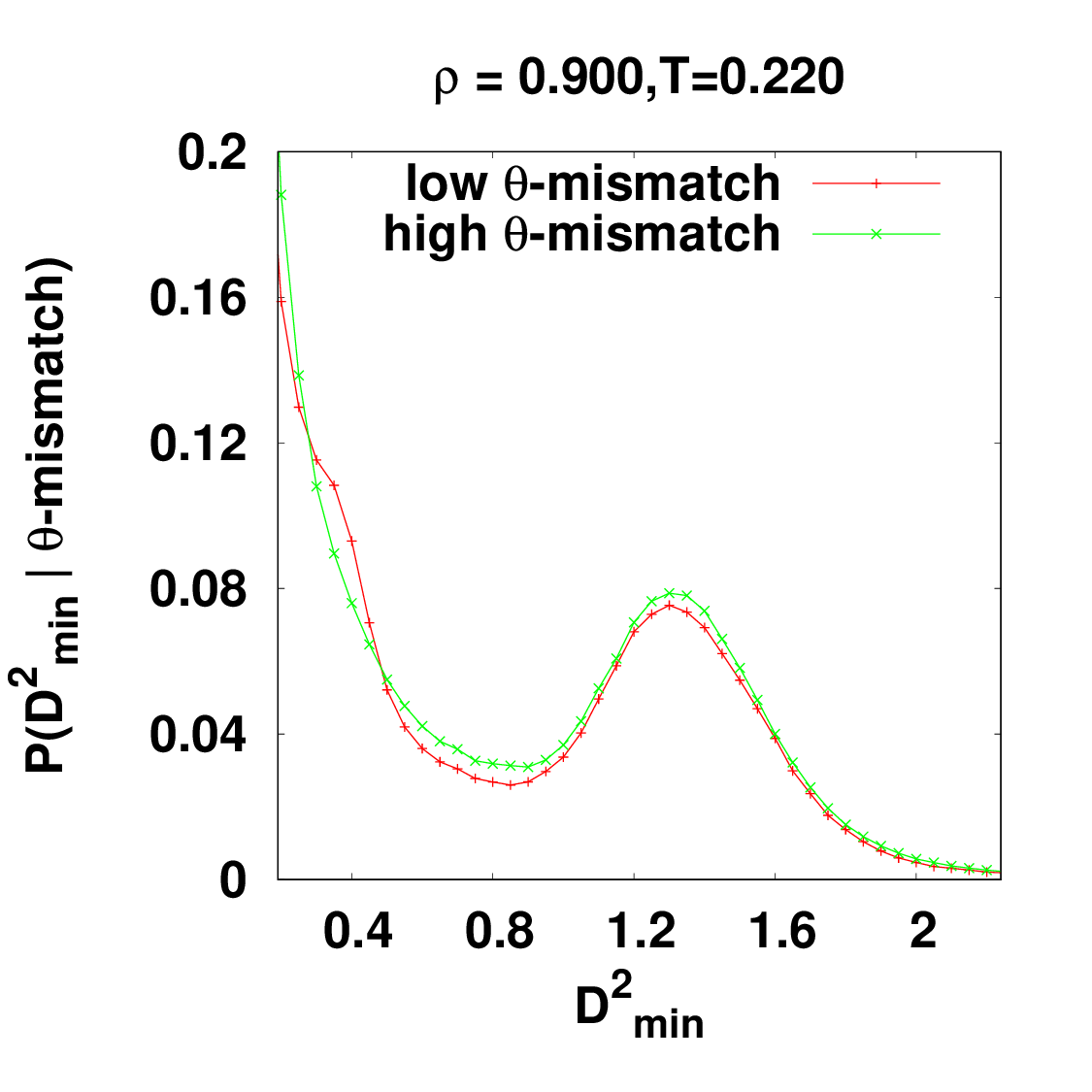}
    \caption{$T=0.220$}
  \end{subfigure}\hfill
  \begin{subfigure}[b]{0.32\textwidth}
    \includegraphics[width=\linewidth]{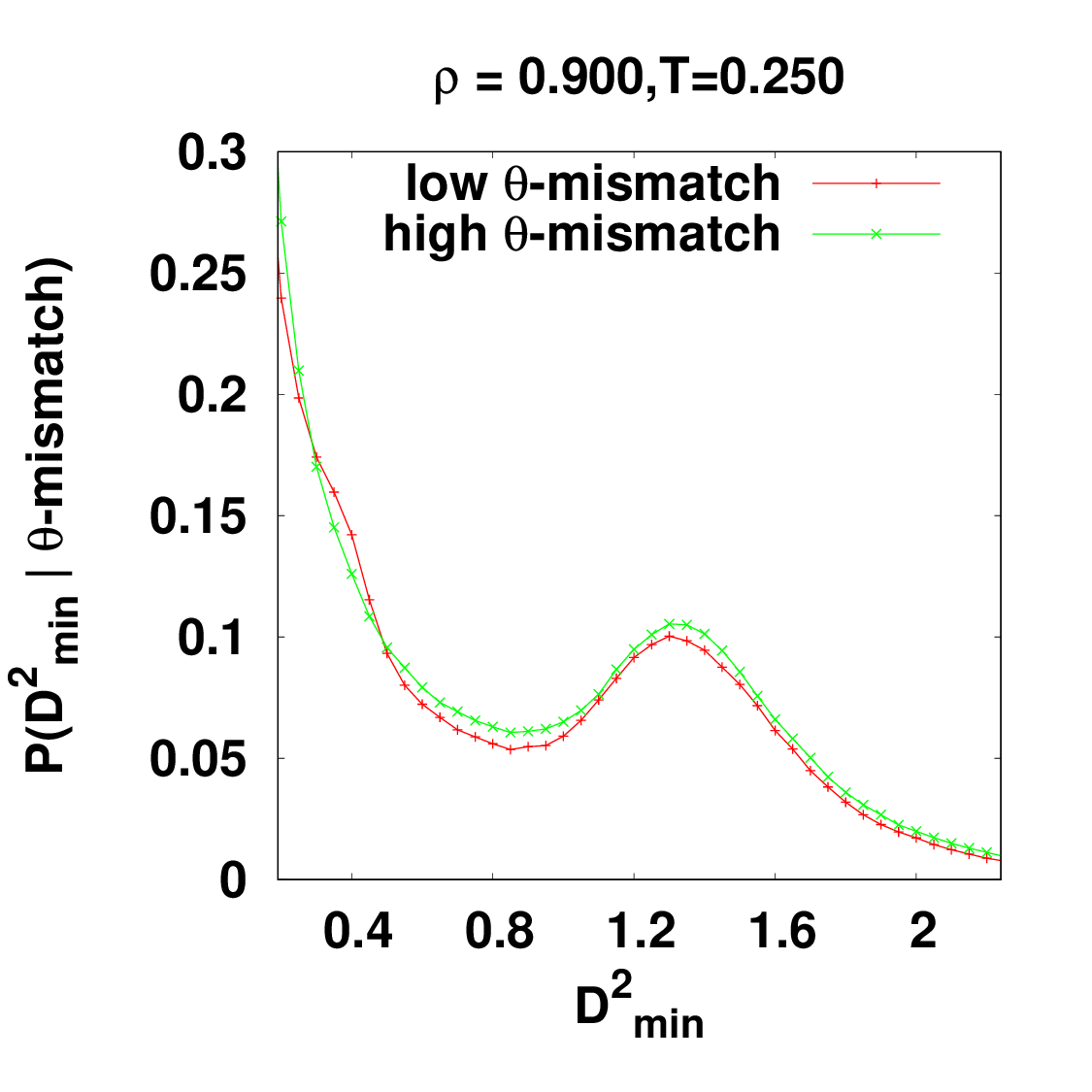}
    \caption{$T=0.250$}
  \end{subfigure}\hfill
  \begin{subfigure}[b]{0.32\textwidth}
    \includegraphics[width=\linewidth]{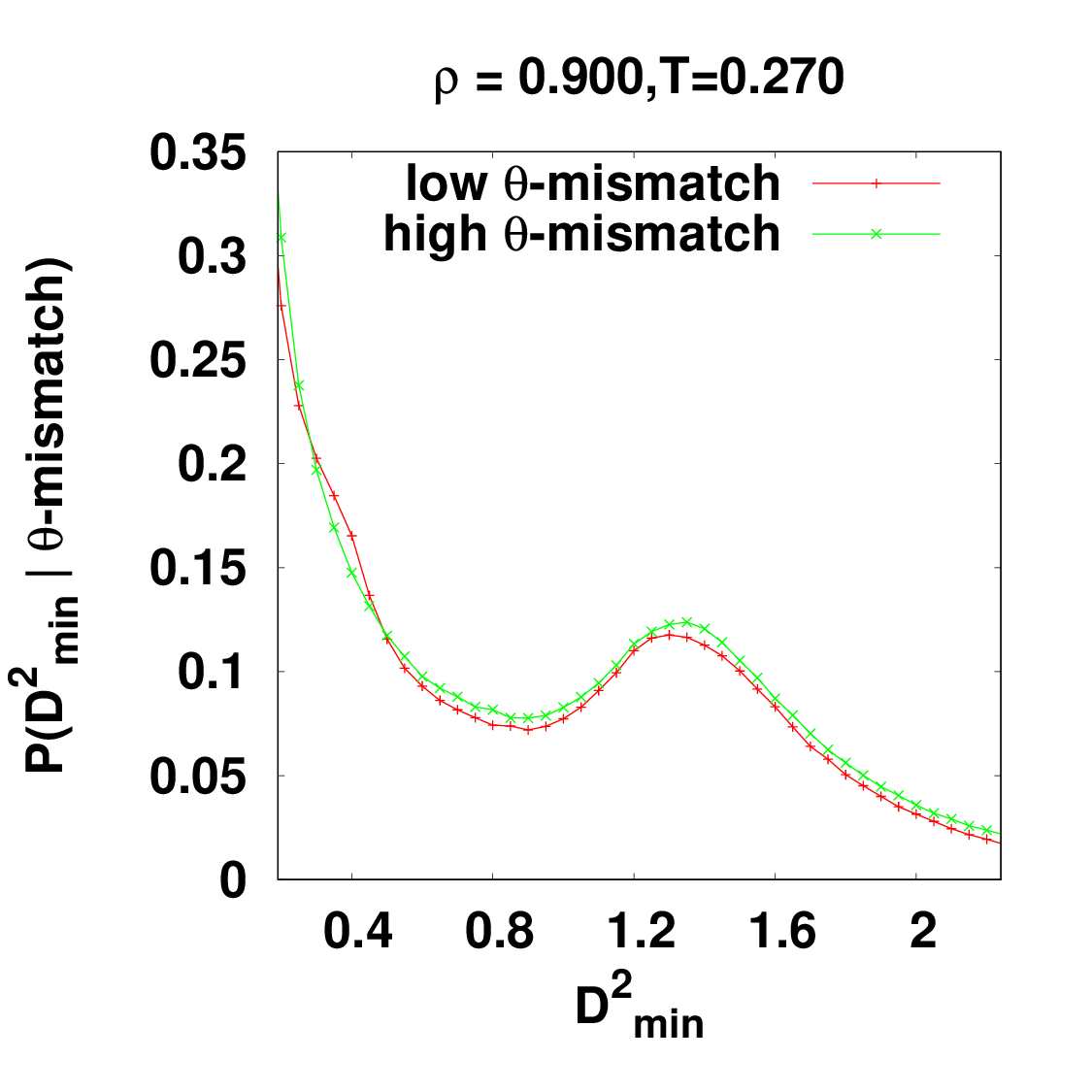}
    \caption{$T=0.270$}
  \end{subfigure}

 \caption{Conditional probability densities $P(D_{\min}^2\mid\Delta\theta)$ at $\rho=0.900$. 
Each curve compares the lowest 10\% in $\Delta\theta_i$ (``low--mismatch'', red) and the highest 10\% (``high--mismatch'', green) at a single temperature; panels (or traces) span the solid, intermittent, supercooled and normal-liquid regimes (example temperatures marked in the legend). 
In the deeply jammed solid, the distributions are sharply peaked at small $D_{\min}^2$ (low--mismatch sites), while the large-$D_{\min}^2$ is dominated by high--mismatch sites, rare, localized rearrangements concentrate in structurally frustrated regions. 
Entering the intermittent regime enhances this contrast (high--mismatch increasingly populates the large-displacement tail).  Around $T_g$ the distributions broaden and the separation weakens, and in the normal liquid the low/ high--mismatch PDFs substantially overlap, showing that steric mismatch becomes a poorer predictor of nonaffine activity when thermal motion dominates.}
\label{fig:thetaa1}
\end{figure*}

\begin{figure*}[ht!]
  \centering
  \begin{subfigure}[b]{0.32\textwidth}
    \includegraphics[width=\linewidth]{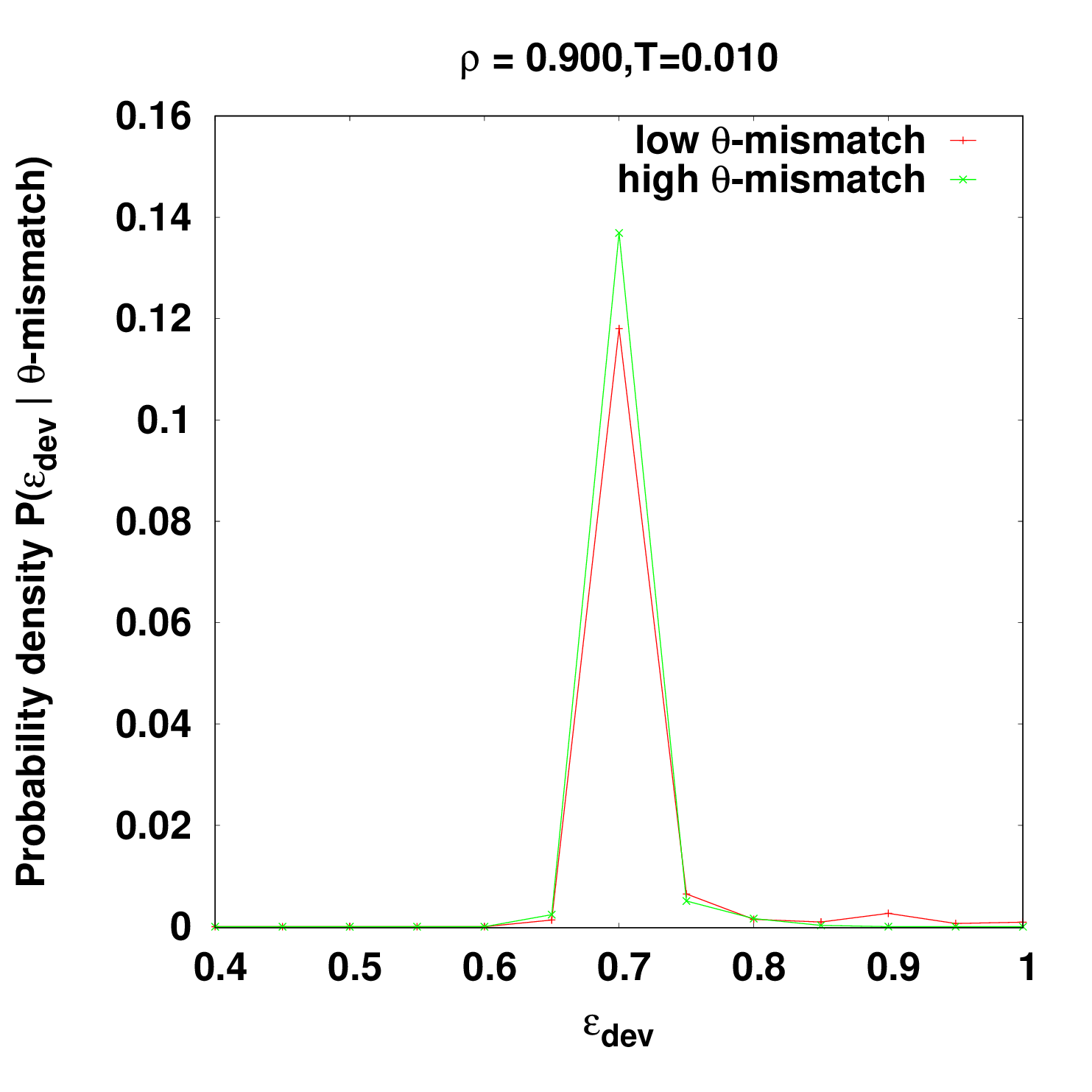}
    \caption{$T=0.010$}
  \end{subfigure}\hfill
  \begin{subfigure}[b]{0.32\textwidth}
    \includegraphics[width=\linewidth]{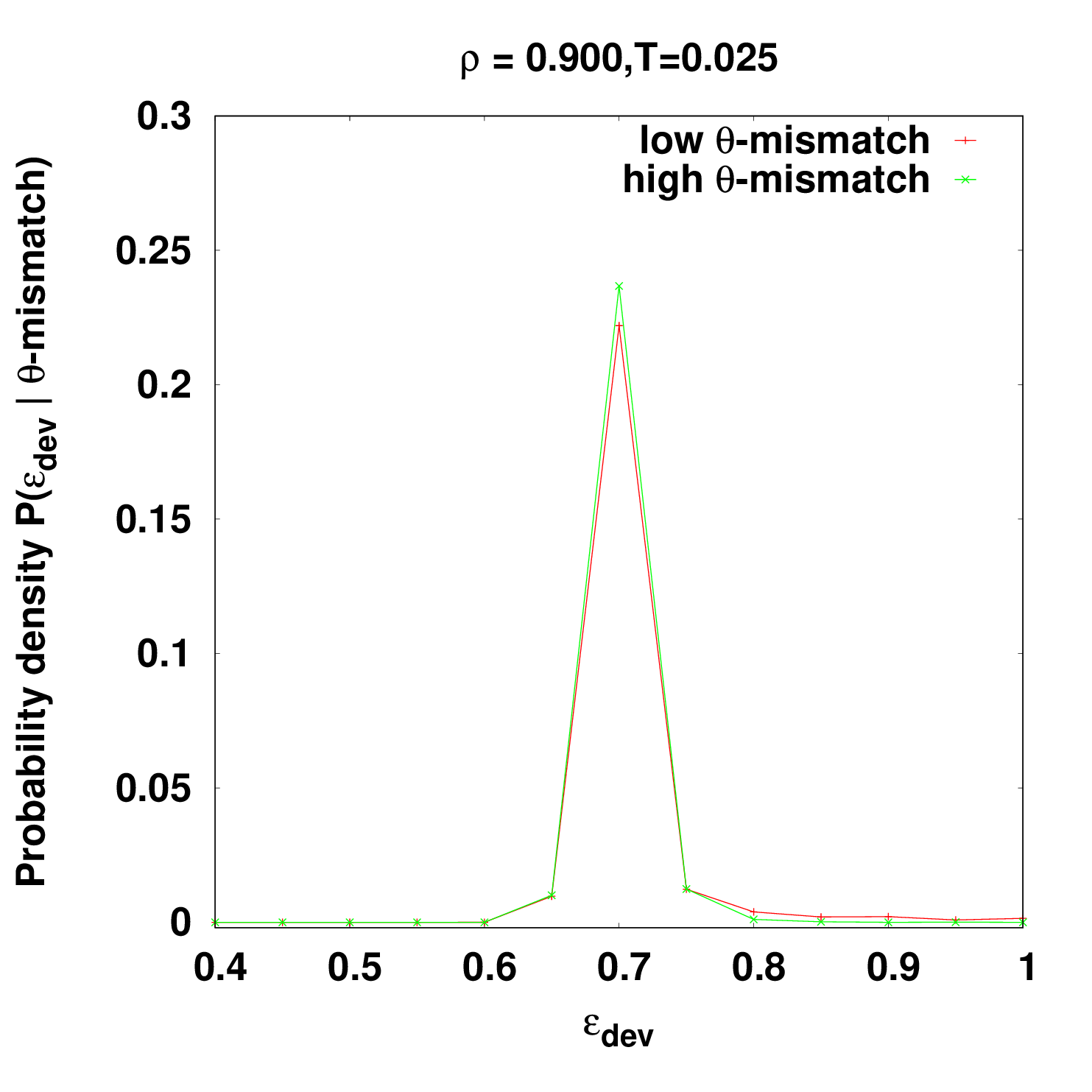}
    \caption{$T=0.025$}
  \end{subfigure}\hfill
   \begin{subfigure}[b]{0.32\textwidth}
    \includegraphics[width=\linewidth]{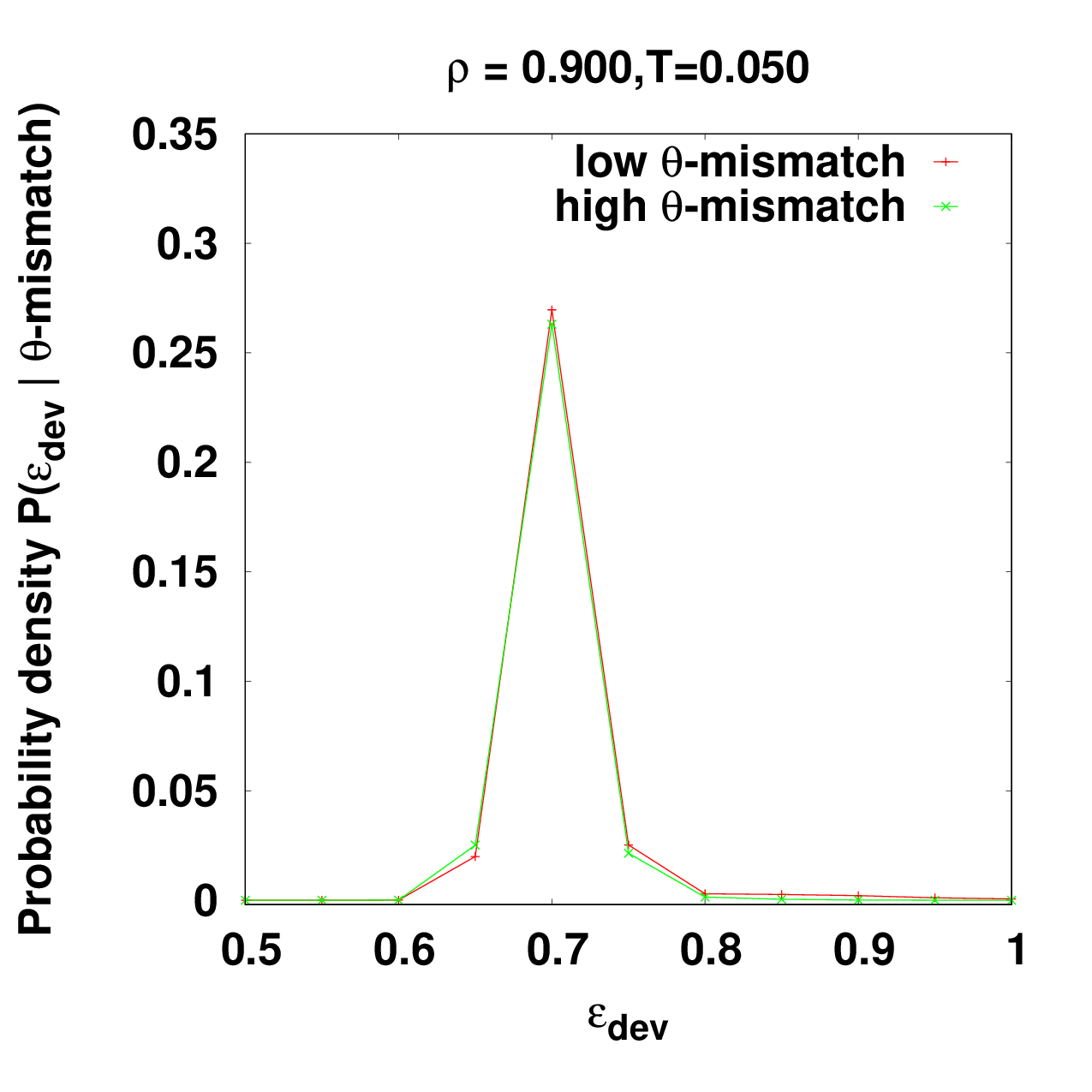}
    \caption{$T=0.050$}
  \end{subfigure}
  \vspace{0.5em}
  \begin{subfigure}[b]{0.32\textwidth}
    \includegraphics[width=\linewidth]{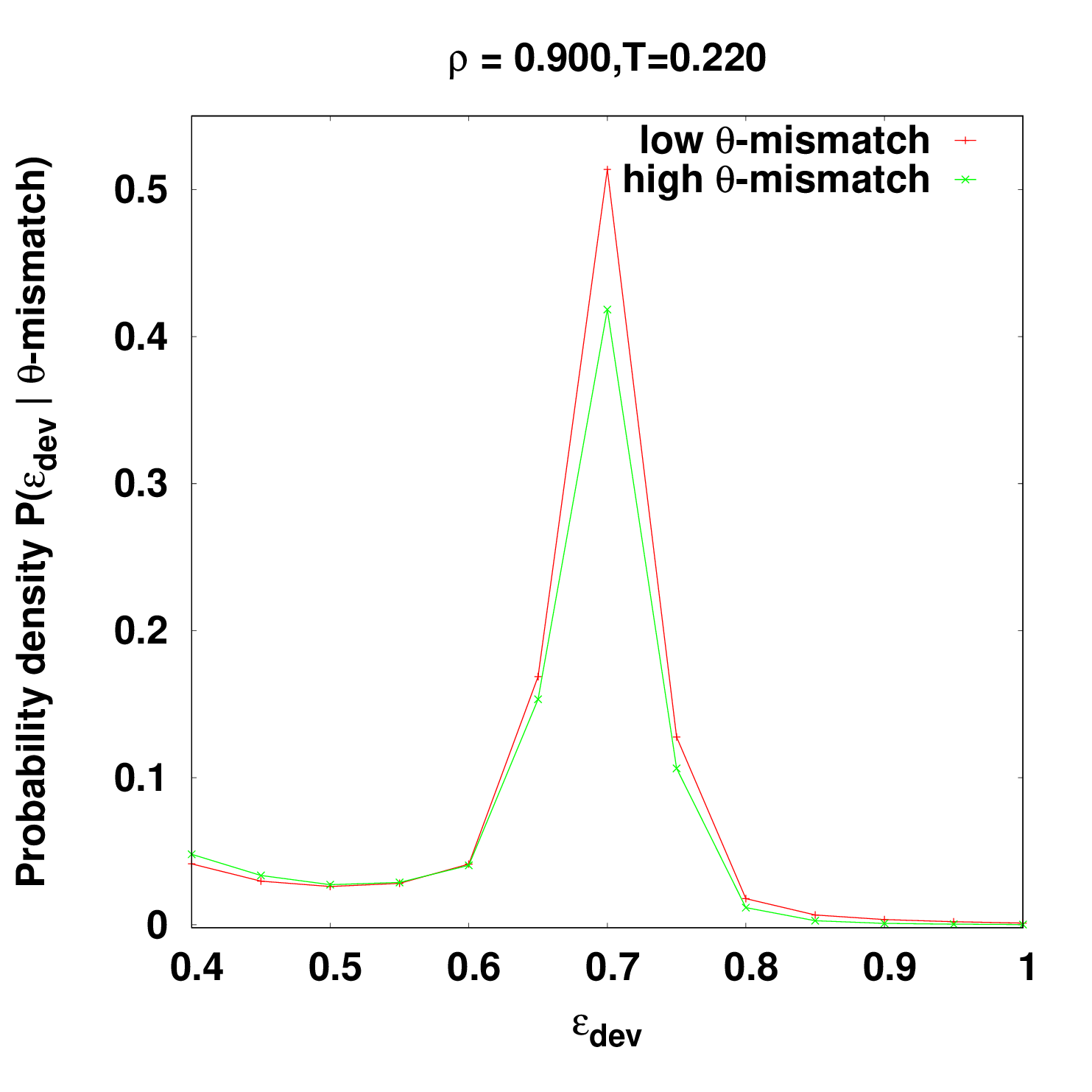}
    \caption{$T=0.220$}
  \end{subfigure}\hfill
  \begin{subfigure}[b]{0.32\textwidth}
    \includegraphics[width=\linewidth]{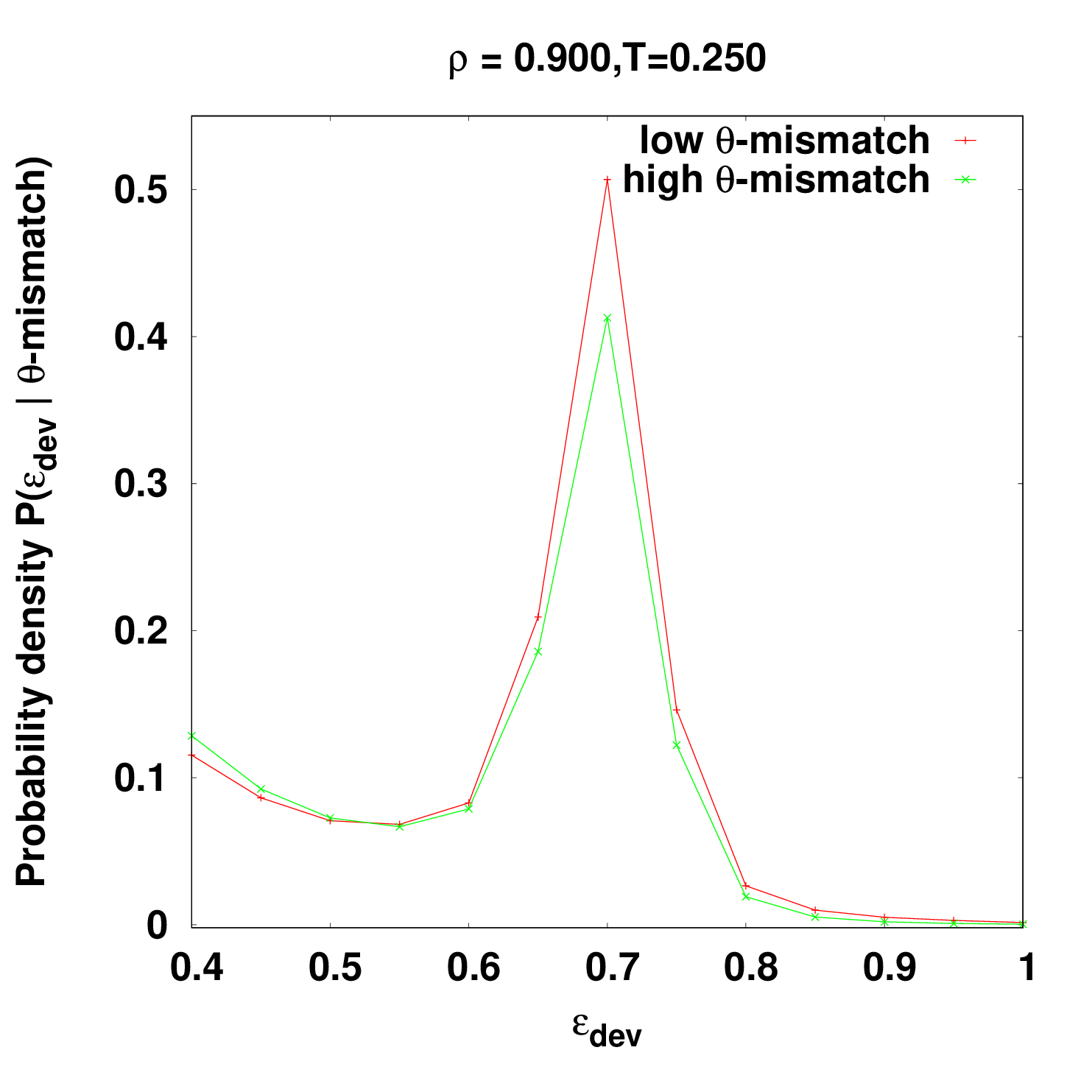}
    \caption{$T=0.250$}
  \end{subfigure}\hfill
  \begin{subfigure}[b]{0.32\textwidth}
    \includegraphics[width=\linewidth]{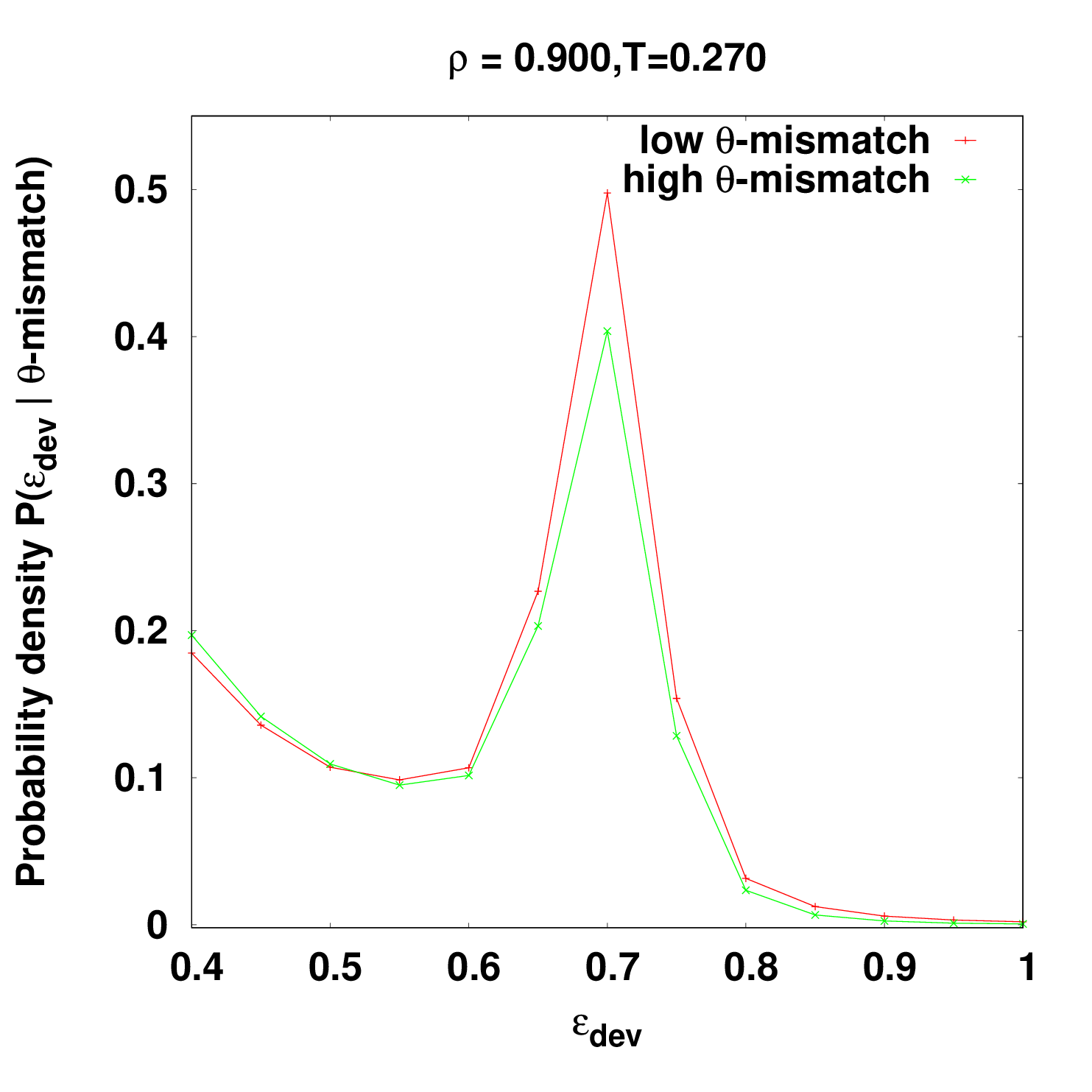}
    \caption{$T=0.270$}
  \end{subfigure}
  \caption{Conditional distributions $P(\varepsilon_{\rm dev}\mid\theta)$ at $\rho=0.900$ (red: lowest 10\% $\Delta\theta$, green: highest 10\% $\Delta\theta$), shown for temperatures spanning the deep glass, viscoelastic, supercooled and normal-liquid regimes.  In the deep glass ($T\le0.050$) high-mismatch sites (green) contribute relatively more deviatoric strain (enhanced moderate/large-$\varepsilon_{\rm dev}$ shoulder), signalling rare, frustration-driven microplastic events.  As the system softens (viscoelastic window, $0.100\le T\le0.160$) the two curves cross: the low-mismatch population (red) develops the larger primary peak (near $\varepsilon_{\rm dev}\!\approx\!0.7$) and remains dominant through the supercooled and normal-liquid regimes, although the red/green difference steadily decreases at high $T$.  The crossing identifies the temperature at which vibrational steric fluctuations give way to collective, elastic-like nonaffine strain.}

  \label{fig:thetaa2}
\end{figure*}

\begin{figure*}[ht!]
  \centering
  \begin{subfigure}[b]{0.32\textwidth}
    \includegraphics[width=\linewidth]{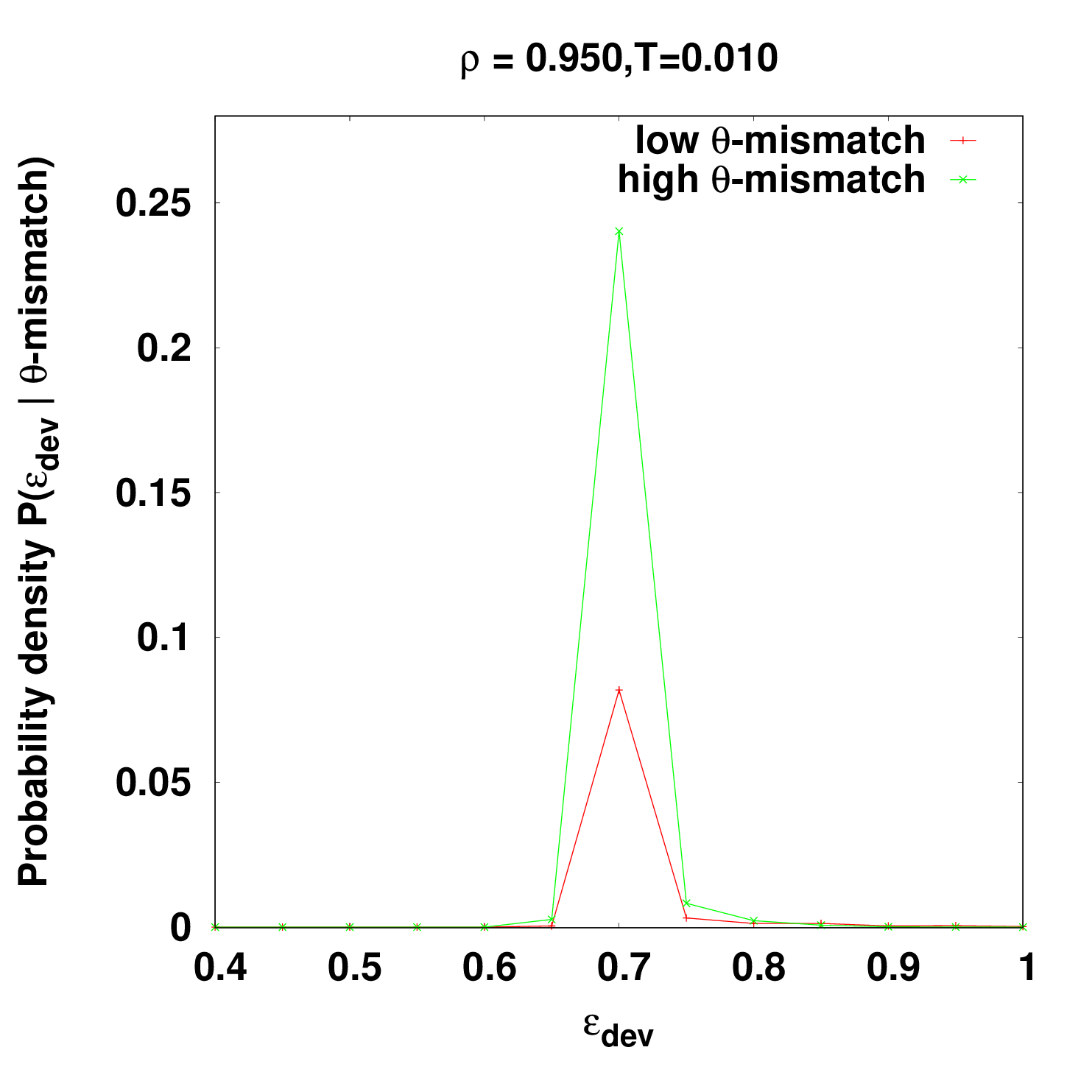}
    \caption{$T=0.010$}
  \end{subfigure}\hfill
  \begin{subfigure}[b]{0.32\textwidth}
    \includegraphics[width=\linewidth]{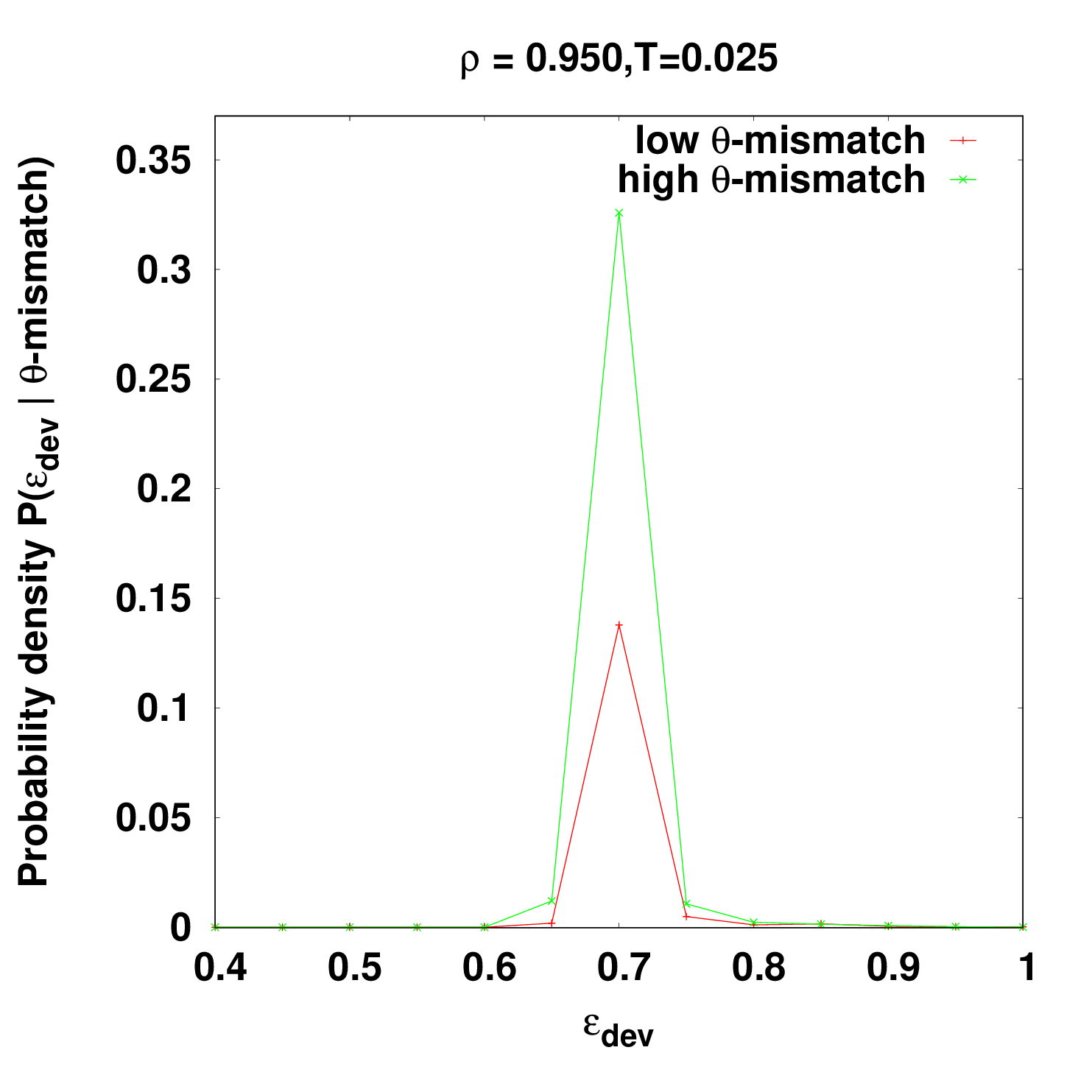}
    \caption{$T=0.025$}
  \end{subfigure}\hfill
   \begin{subfigure}[b]{0.32\textwidth}
    \includegraphics[width=\linewidth]{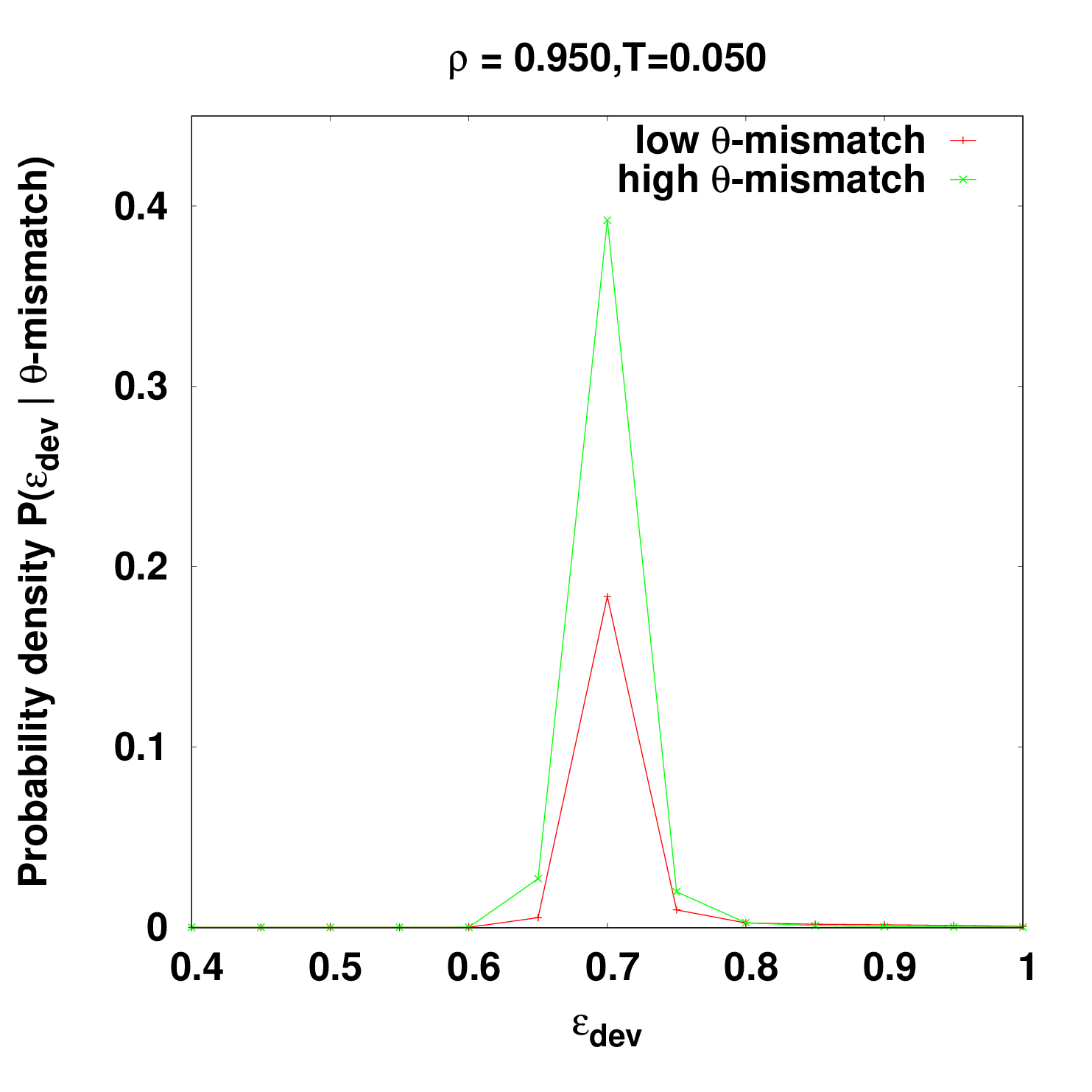}
    \caption{$T=0.050$}
  \end{subfigure}
 \vspace{0.5em}
  
   \begin{subfigure}[b]{0.32\textwidth}
    \includegraphics[width=\linewidth]{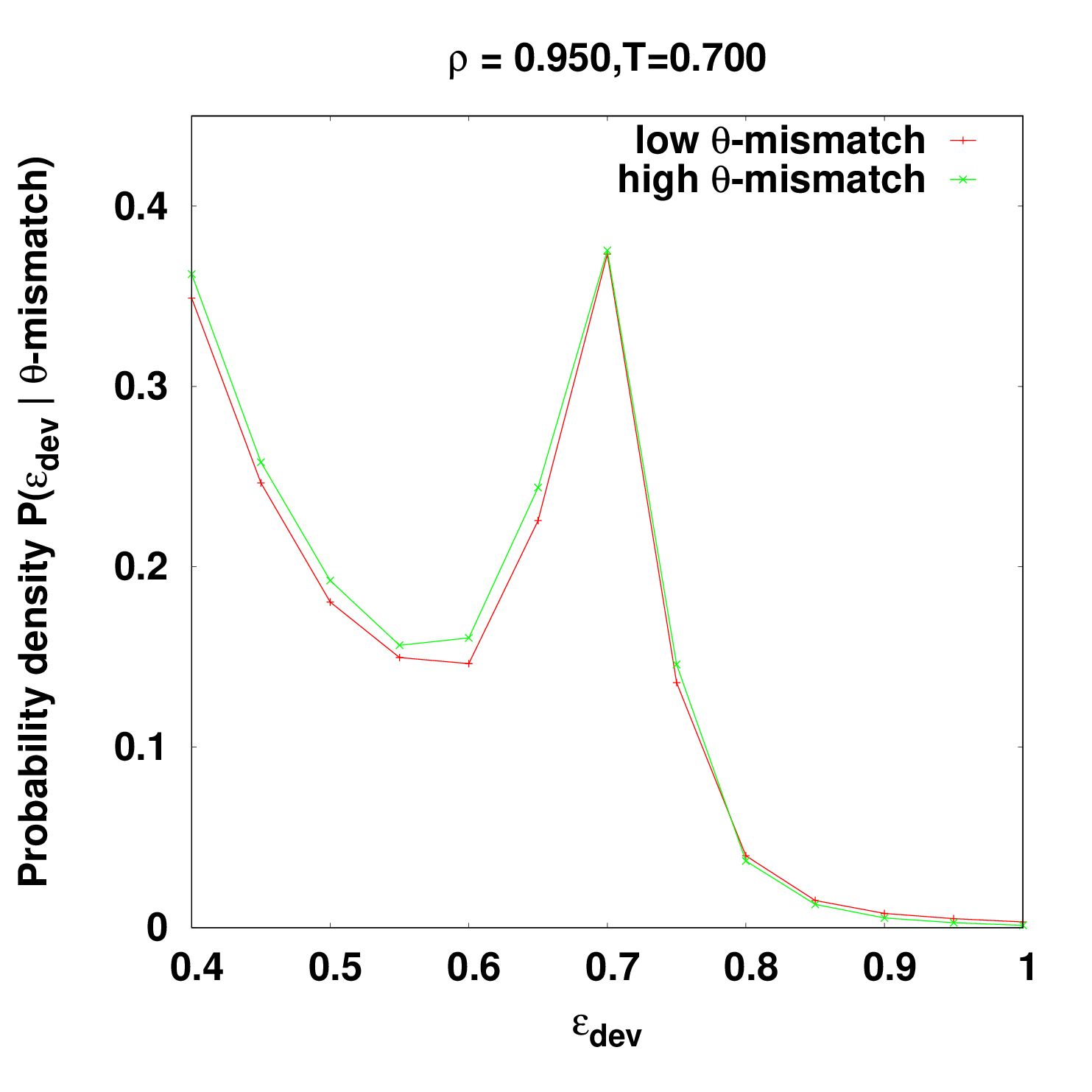}
    \caption{$T=0.700$}
  \end{subfigure}\hfill
  \begin{subfigure}[b]{0.32\textwidth}
    \includegraphics[width=\linewidth]{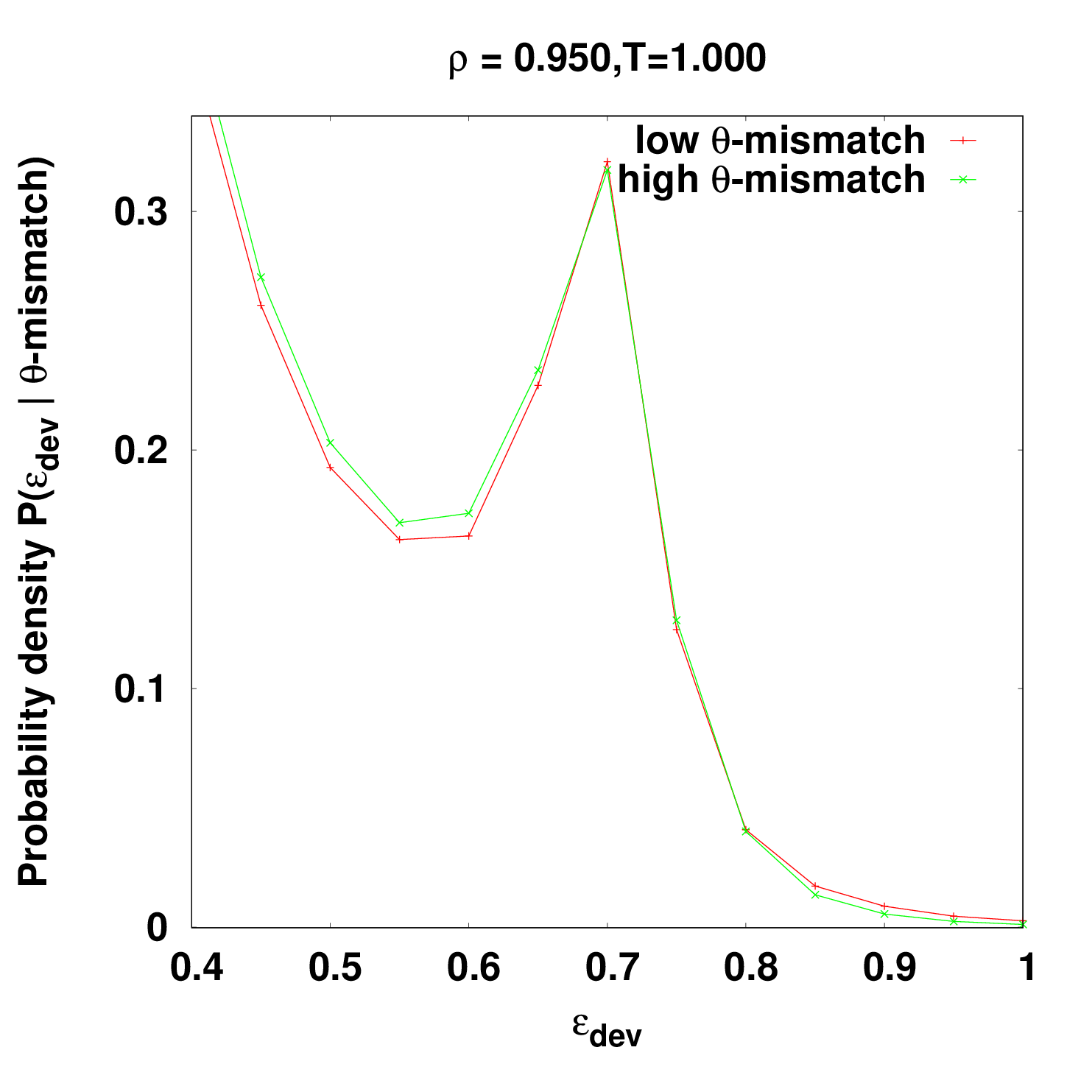}
    \caption{$T=1.000$}
  \end{subfigure}\hfill
  \begin{subfigure}[b]{0.32\textwidth}
    \includegraphics[width=\linewidth]{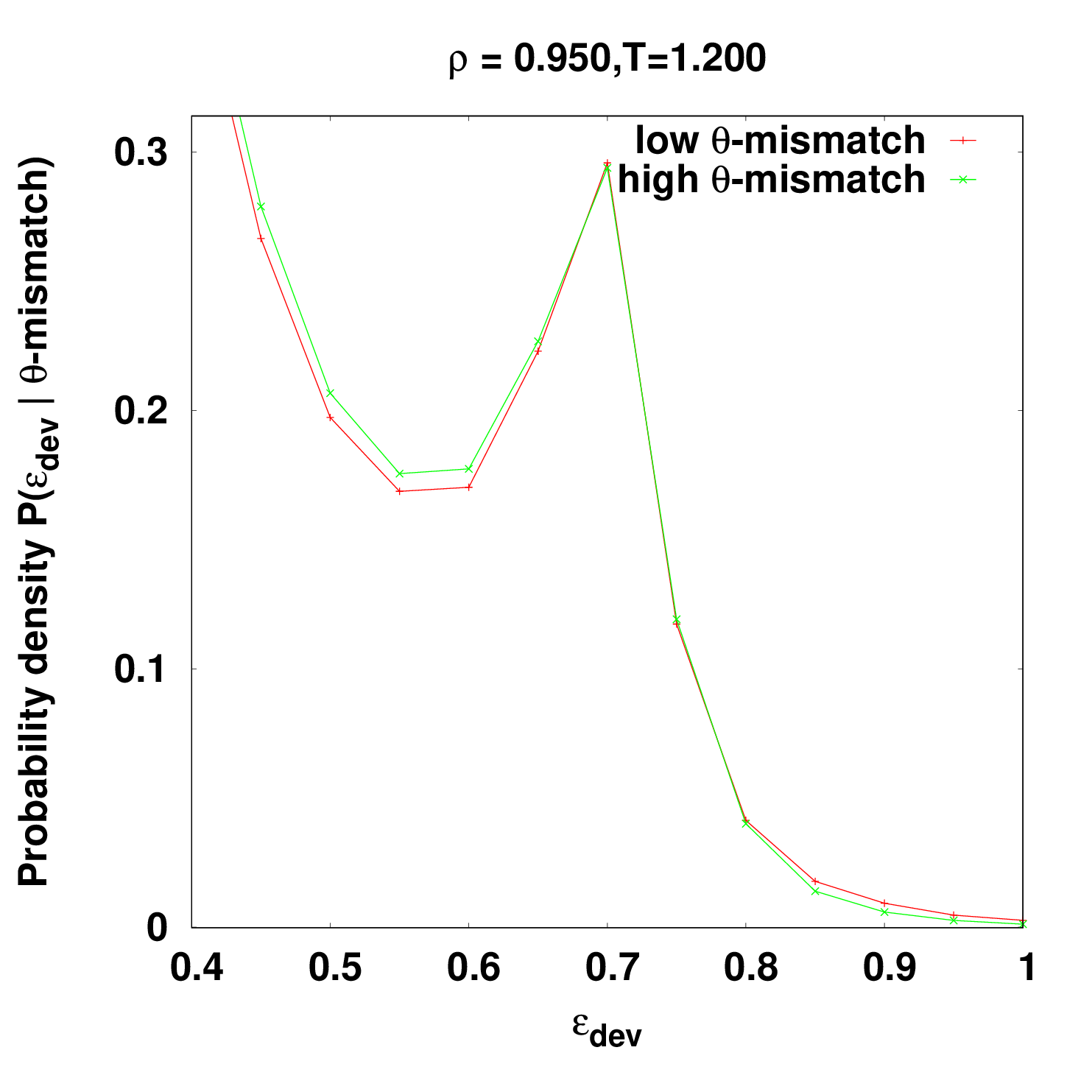}
    \caption{$T=1.200$}
  \end{subfigure}
 \caption{Conditional probability density $P(\varepsilon_{\rm dev}\mid\theta)$ for the 10\% lowest- and highest-$\Delta\theta$ particles at $\rho=0.950$ (red: low $\Delta\theta$; green: high $\Delta\theta$). Curves are shown over a temperature sweep that spans the deep solid, viscoelastic/intermittent window and the supercooled/normal liquid regimes (key markers: $T_{\rm s}=0.110$, $T_g=0.343$, $T_{\rm onset}\approx 0.679$). In the deeply glassy solid ($T\le T_{\rm s}$) the high-$\Delta\theta$ population carries the larger $\varepsilon_{\rm dev}$ peak ($\approx 0.7$), indicating that frustrated contacts host the dominant nonaffine response. As $T$ rises into the viscoelastic and supercooled regimes the two distributions broaden and their separation shrinks; by $T\sim 0.7$ the red and green curves are nearly coincident, and above $T_{\rm onset}$ they overlap, showing that steric mismatch ceases to be a strong predictor of deviatoric response once cages sufficiently soften.}
\label{fig:thetaa5}
\end{figure*}

\begin{figure*}[ht!]
  \centering
  \begin{subfigure}[b]{0.32\textwidth}
    \includegraphics[width=\linewidth]{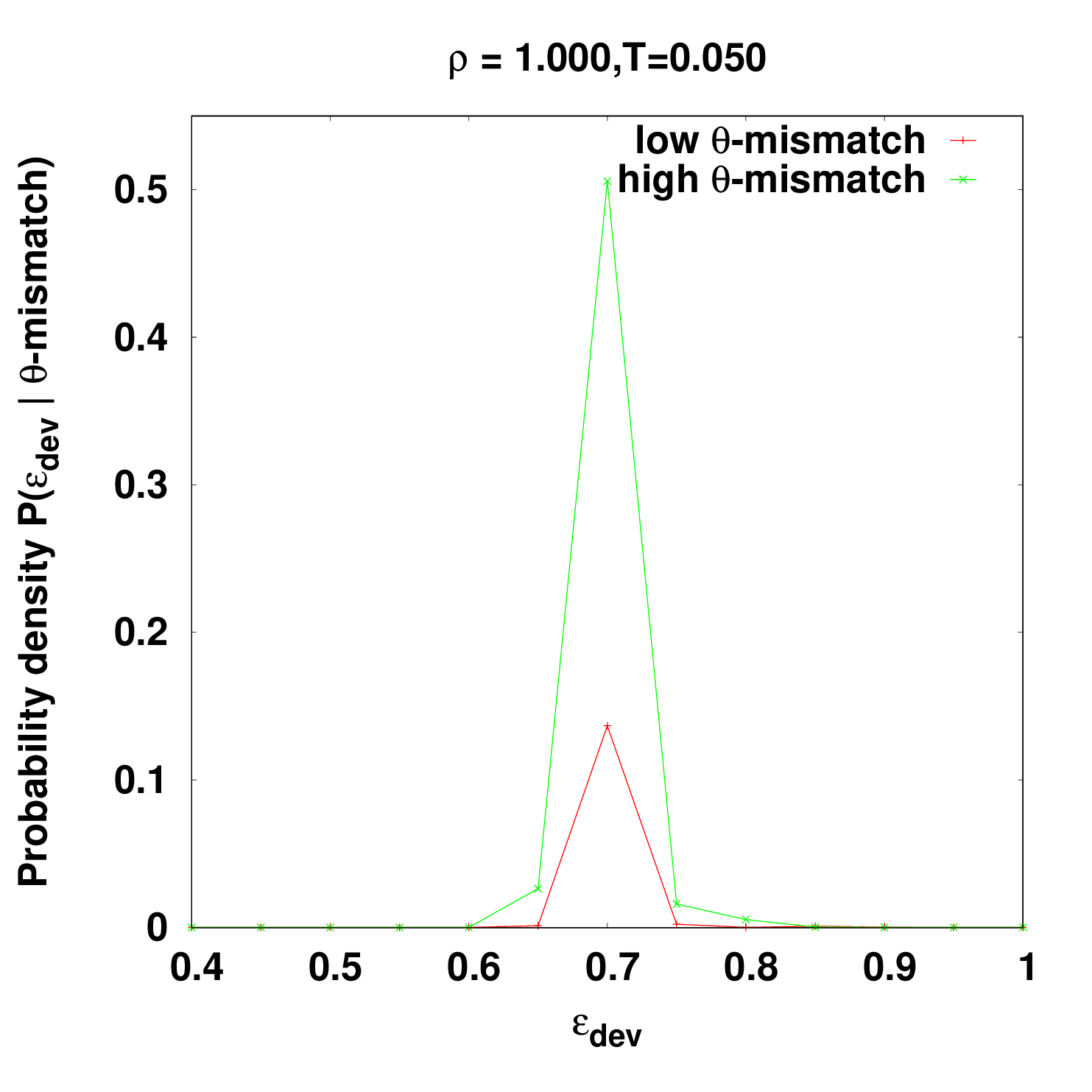}
    \caption{$T=0.050$}
  \end{subfigure}\hfill
   \begin{subfigure}[b]{0.32\textwidth}
    \includegraphics[width=\linewidth]{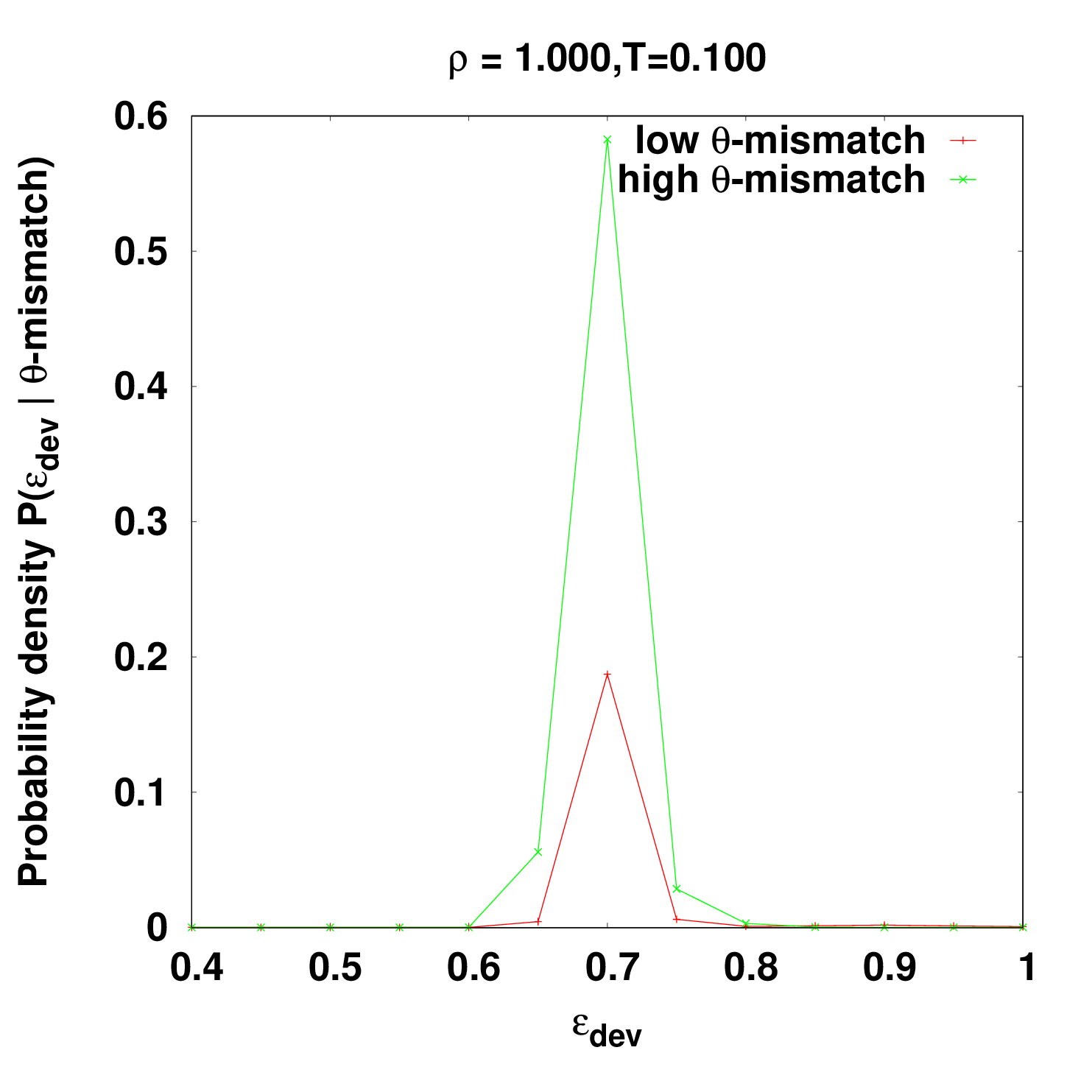}
    \caption{$T=0.100$}
  \end{subfigure}
   \begin{subfigure}[b]{0.32\textwidth}
    \includegraphics[width=\linewidth]{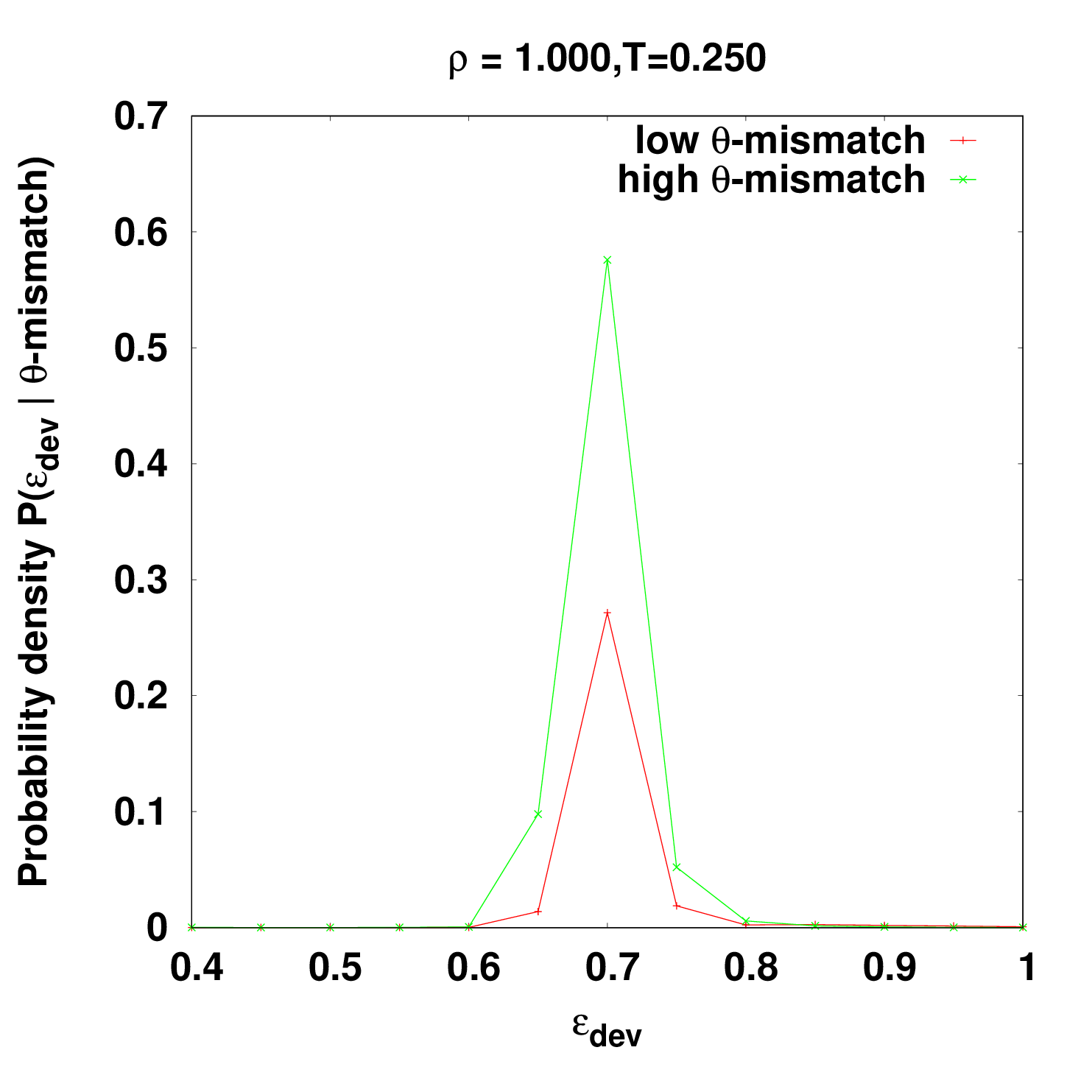}
    \caption{$T=0.250$}
  \end{subfigure}

 \vspace{0.5em}
   \begin{subfigure}[b]{0.32\textwidth}
    \includegraphics[width=\linewidth]{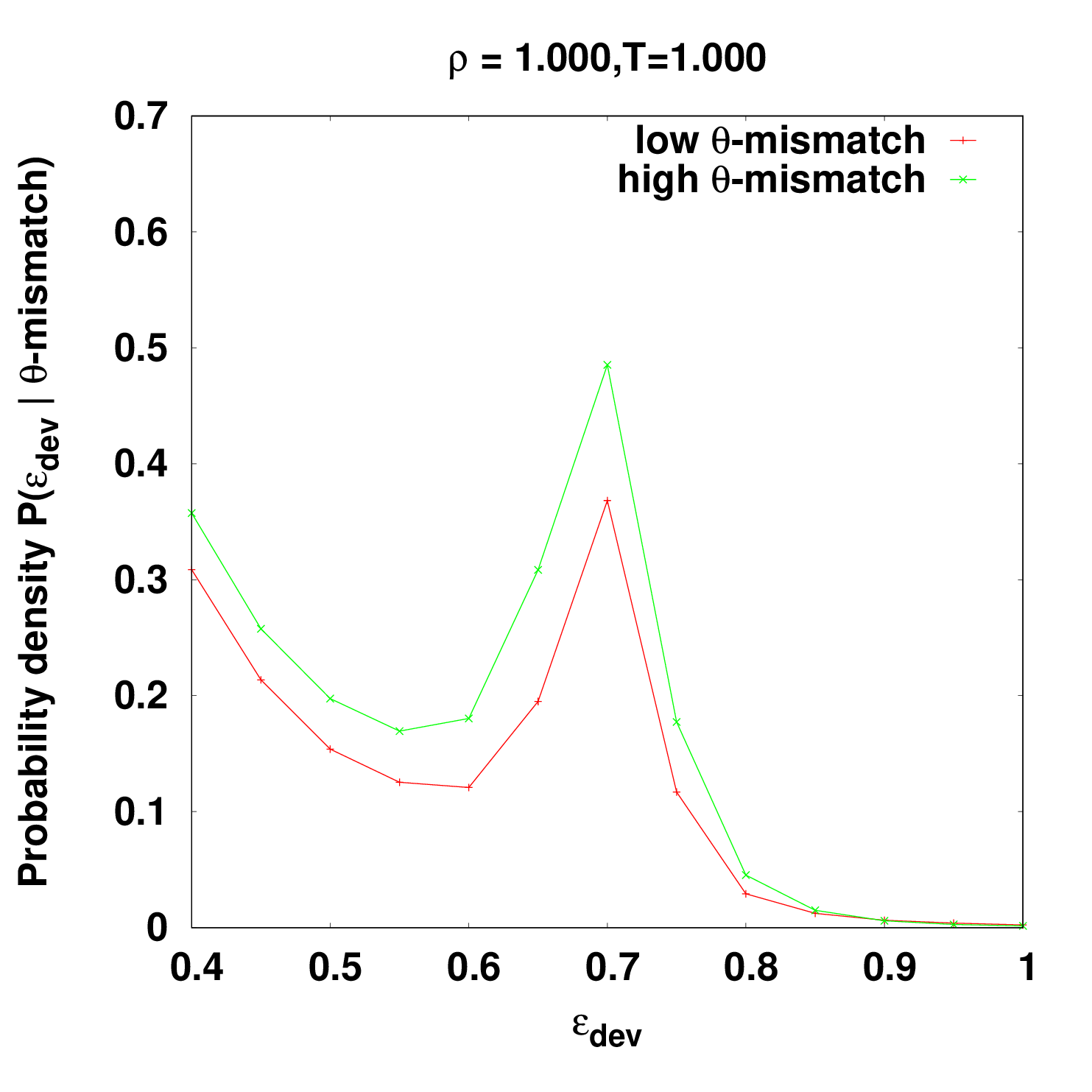}
    \caption{$T=1.000$}
  \end{subfigure}\hfill
   \begin{subfigure}[b]{0.32\textwidth}
    \includegraphics[width=\linewidth]{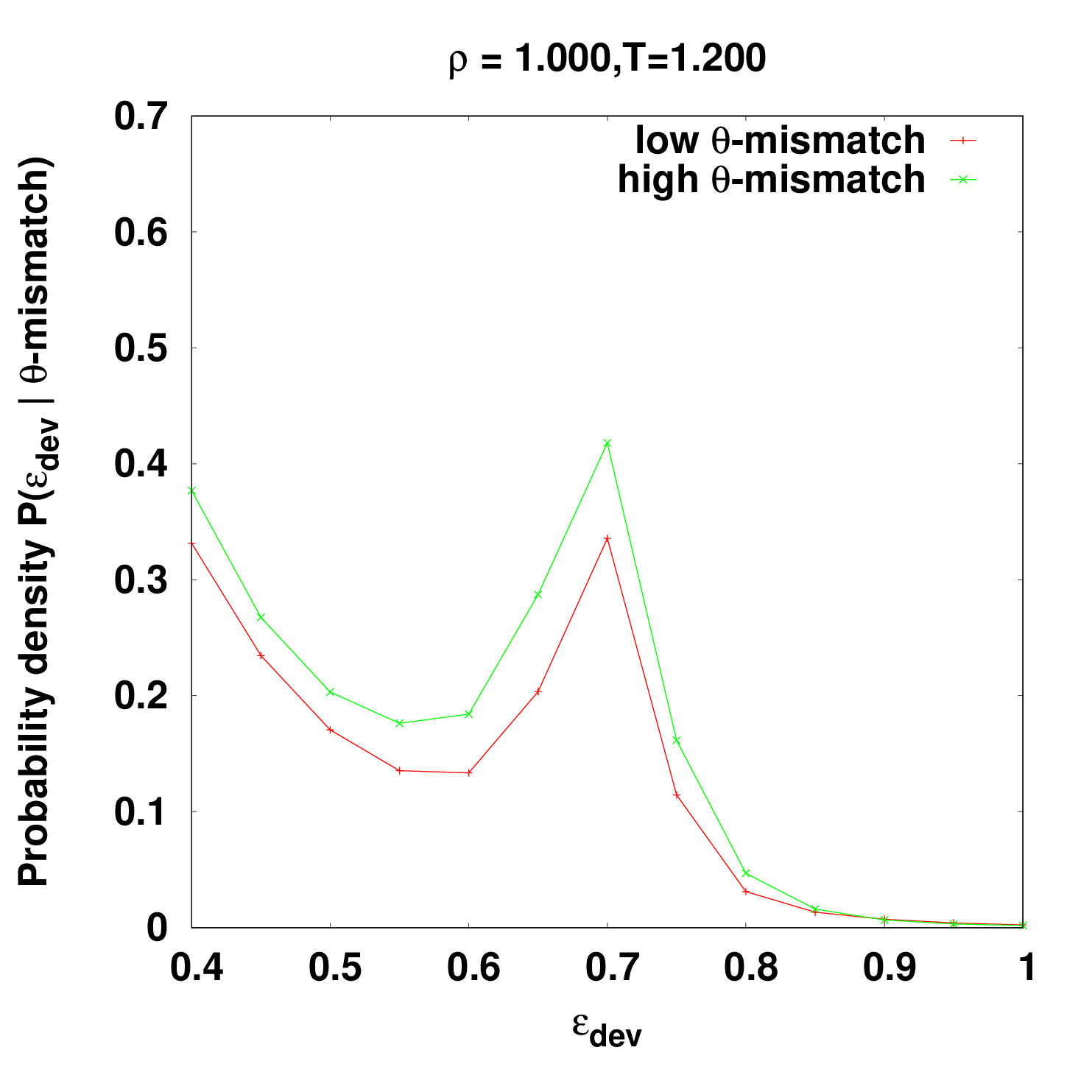}
    \caption{$T=1.200$}
  \end{subfigure}\hfill
  \begin{subfigure}[b]{0.32\textwidth}
    \includegraphics[width=\linewidth]{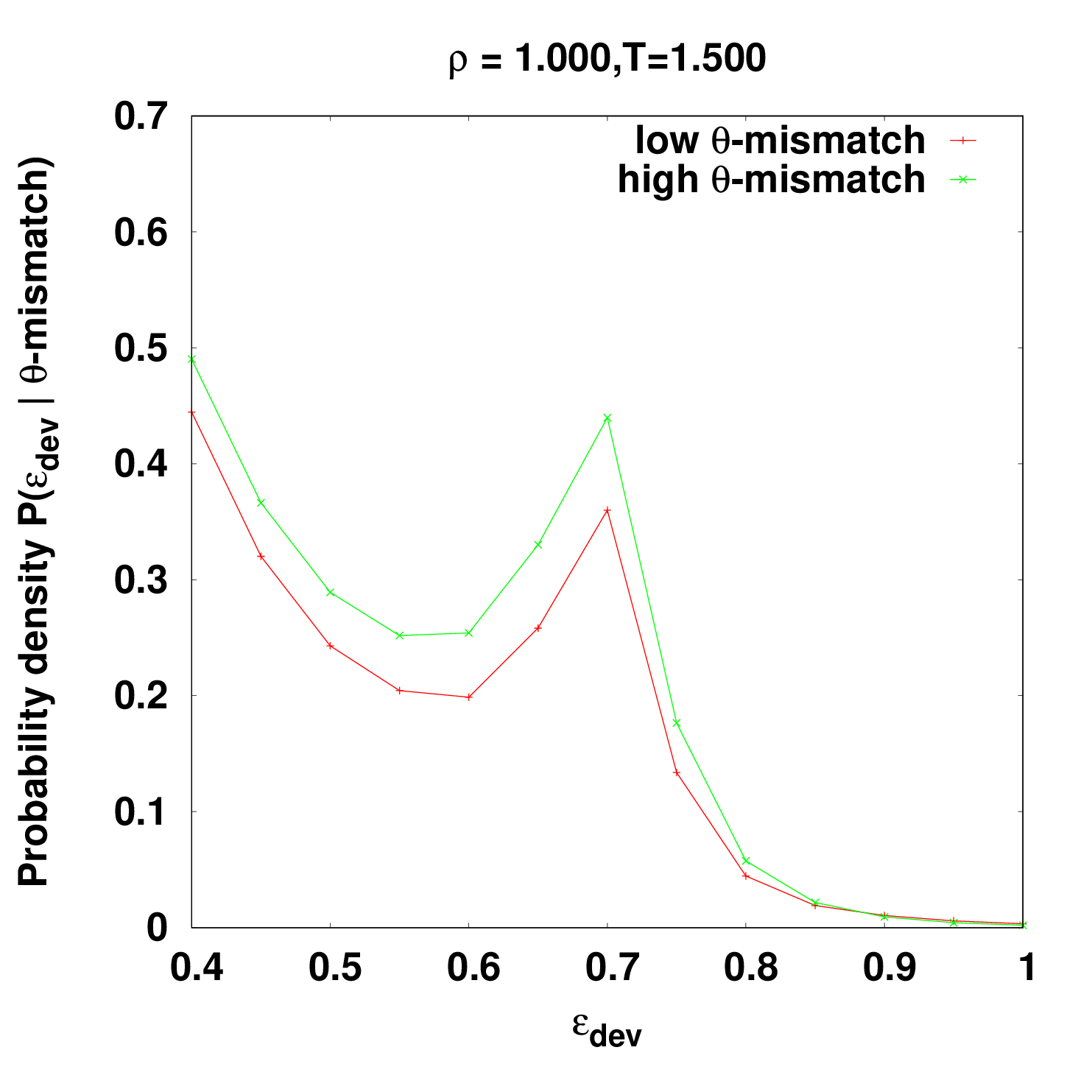}
    \caption{$T=1.500$}
  \end{subfigure}
 \caption{Conditional probability density $P(\varepsilon_{\rm dev}\mid\theta)$ for the 10\% lowest- and highest-$\Delta\theta$ particles at $\rho=1.000$ (red: low $\Delta\theta$; green: high $\Delta\theta$). Curves span temperatures across the solid, viscoelastic, supercooled and normal-liquid regimes (key markers: $T_{\rm s}=0.251$, $T_g=0.596$, $T_{\rm onset}\approx 1.483$). In the deeply glassy solid ($T\le T_{\rm s}$) the high-$\Delta\theta$ population carries the larger $\varepsilon_{\rm dev}$ peak (near $\varepsilon_{\rm dev}\approx 0.7$); this separation persists and even increases through the viscoelastic window. After crossing $T_g$ both distributions shift to higher strains but the green peak remains above the red one, and only in the high-$T$ normal-liquid regime does the gap appreciably narrow. These trends show that at the highest density studied structurally frustrated (high-$\Delta\theta$) sites remain the principal hosts of nonaffine, deviatoric response up to near-liquid conditions.}

  \label{fig:thetaa11}
\end{figure*}

\section{Thermal Deviatoric Strain Response for Low- and High-\(\theta\)-Mismatch Populations}
\label{sec:deviatoric_theta}

The thermal nonaffine metric characterizes the magnitude of local motion that cannot be represented by an affine transformation. To further resolve the character of this response, we next examine its deviatoric component, which quantifies the local shear-like distortion generated by thermal fluctuations. We therefore compare the conditional probability densities of the thermal deviatoric strain for the low- and high-\(\theta\)-mismatch populations.

\subsection{Deviatoric Strain Distributions at \(\rho=0.900\)}
\label{sec:dev_0900}

Fig.~\ref{fig:thetaa2} presents the conditional probability densities
\[
P\!\left(\varepsilon_{\rm dev}\mid\theta_i\leq\theta_{10\%}\right)
\qquad\text{and}\qquad
P\!\left(\varepsilon_{\rm dev}\mid\theta_i\geq\theta_{90\%}\right),
\]
corresponding to the \(10\%\) lowest- and highest-\(\theta\)-mismatch particles, respectively. The distributions are shown over thirteen temperatures spanning the deeply arrested regime (\(T\leq0.050\)), the intermediate viscoelastic regime (\(0.100\leq T\leq0.160\)), the supercooled regime (\(0.180\leq T\leq0.270\)), and the high-temperature liquid regime (\(T=0.350,\,0.500\)). A pronounced temperature-dependent inversion between the two mismatch populations emerges as the system leaves the deeply arrested regime.

At the lowest temperatures, \(T\leq0.050\), the high-\(\theta\)-mismatch distribution lies systematically above the low-\(\theta\)-mismatch distribution. Thus, although the system is in the mechanically locked regime, the strongly frustrated sites retain a slightly larger deviatoric thermal response. This indicates that a small population of localized micro-rearrangements or residual shear-like distortions remains preferentially associated with regions of large geometric mismatch. The ordering changes upon entering the viscoelastic regime. For \(0.100\leq T\leq0.160\), as the temperature increases beyond the solidification line (\(T_{\rm ML}\approx0.025\)) toward the thermodynamic glass-transition temperature (\(T_g\approx0.168\)), the two conditional distributions cross, and the low-\(\theta\)-mismatch population becomes dominant at the principal peak of \(P(\varepsilon_{\rm dev})\). The onset of this inversion indicates that once the cages become thermally softened, the sites that remain close to their steric reference geometry can accommodate a larger population of moderate shear-like thermal distortions than the already strongly mismatched sites.

The same ordering persists on further heating into the supercooled regime, \(0.180\leq T\leq0.270\). The low-\(\theta\) population remains above the high-\(\theta\) population around the principal peak, located near \(\varepsilon_{\rm dev}\approx0.7\), while both distributions broaden with increasing temperature. Thus, the inversion is not restricted to the immediate vicinity of the solidification crossover but extends well into the supercooled regime, where cooperative and anharmonic thermal distortions increasingly involve particles that were comparatively well aligned with their local steric reference. In the normal-liquid regime, represented here by \(T=0.350\) and \(0.500\), the low-\(\theta\) distribution continues to show a systematic enhancement around the main peak, although the difference from the high-\(\theta\) population becomes progressively smaller as thermal motion becomes more rapid and less spatially selective. The inversion therefore survives into the liquid-like regime, even though its contrast is substantially reduced.

At \(\rho=0.900\), the thermal deviatoric response consequently exhibits a clear change in its structural selectivity across temperature. Large-\(\theta\) sites carry the larger deviatoric response only in the deepest mechanically locked regime; once the system enters the viscoelastic regime, the ordering reverses and low-\(\theta\) sites dominate the principal deviatoric-strain response. This inverted ordering persists through the supercooled and liquid-like regimes, revealing a density-specific crossover from localized distortional activity at geometrically frustrated sites to a broader, predominantly low-mismatch population of thermally generated shear-like distortions.
\subsection{Deviatoric Strain Distributions at \(\rho=0.950\)}

Fig.~\ref{fig:thetaa5} presents the conditional probability densities of the deviatoric strain \(\varepsilon_{\rm dev}\) for the 10\% lowest‐mismatch (red) and 10\% highest‐mismatch (green) particles at \(\rho=0.950\) over a broad temperature sweep.  In the deeply glassy regime (\(T \le T_{\rm s}=0.110\)), the high‐mismatch population exhibits a markedly larger peak strain; its green curve dominates around \(\varepsilon_{\rm dev}\approx0.70\) signaling that severely misaligned local motifs are far more likely to undergo nonaffine deformations even when the system is essentially solid.  As we heat into the intermediate viscoelastic window (\(T_{\rm s} < T < T_{\rm g}=0.343\)), both distributions remain bimodal and well separated, but the height and area difference between the red and green peaks diminishes, reflecting the growing ability of low‐mismatch particles to participate in cooperative rearrangements.  Upon crossing the glass‐transition (\(T \ge T_{\rm g}\)) into the supercooled liquid (\(T_{\rm g} \le T \lesssim T_{\rm onset}=0.679\)), the two curves collapse toward one another: by \(T\approx0.700\) the red and green peaks are nearly coincident, indicating that once cages soften sufficiently, low‐ and high‐mismatch particles share a common nonaffine‐strain environment.  In the high‐temperature normal liquid (\(T > T_{\rm onset}\)), any residual separation vanishes and the conditional distributions fully overlap, showing that in the unjammed regime the prior steric mismatch has no bearing on the subsequent deviatoric response.

\subsection{Deviatoric Strain Distributions at \(\rho=1.000\)}
Fig.~\ref{fig:thetaa11} shows the conditional probability densities of the deviatoric strain \(\varepsilon_{\rm dev}\) for the 10\% lowest‐mismatch (red) and 10\% highest‐mismatch (green) particles at \(\rho=1.000\) over a wide temperature range.  In the deeply glassy regime (\(T \le T_{\rm s}=0.251\)), the high‐mismatch population exhibits a substantially larger peak strain; its green curve dominates around \(\varepsilon_{\rm dev}\approx0.70\), indicating that the most frustrated contacts bear the brunt of nonaffine deformations even in a jammed solid.  As temperature increases through the viscoelastic window (\(T_{\rm s}<T<T_{\rm g}=0.596\)), this separation persists and even grows slightly: low‐mismatch bonds remain remarkably resilient to shear, whereas high‐mismatch sites continue to deform most.  Crossing the glass transition into the supercooled liquid (\(T\ge T_{\rm g}\)), both distributions shift to higher \(\varepsilon_{\rm dev}\), but the green peak stays well above the red one.  Only upon reaching the high‐\(T\) normal‐liquid regime (\(T\gtrsim T_{\rm onset}=1.483\)) does the gap narrow appreciably, yet a clear offset between the two populations remains.  Thus, at the maximal packing fraction studied, the distinction between “weak” (high‐mismatch) and “strong” (low‐mismatch) sites survives all the way into the liquid.

\section{Comparative Summary of Deviatoric-Strain Trends Across Density}
\label{sec:dev_comparative}
Comparing the thermal deviatoric-strain response across the three densities reveals a systematic but density-dependent relation between local steric mismatch and thermal mechanical activity. In the deeply arrested regime, high-\(\Delta\theta\) particles exhibit larger \(\varepsilon_{\rm dev}\) than low-\(\Delta\theta\) particles at all three densities, consistent with the preferential localization of residual or rare thermally induced distortions in geometrically frustrated regions. Upon heating, however, the subsequent evolution depends strongly on density. At \(\rho=0.900\), the two populations undergo a clear inversion shortly above the solidification line, with low-\(\Delta\theta\) particles becoming dominant in the principal deviatoric-strain peak. This behavior indicates that the relatively loose contact network at the lowest density allows thermally generated shear-like distortions to spread from initially frustrated sites to the broader low-mismatch population. At the intermediate density \(\rho=0.950\), the distinction between the two populations decreases progressively with temperature: the high-mismatch population retains a larger response over the low- and intermediate-temperature regimes, but the separation gradually closes and the two distributions approach one another near the liquid onset without a clear inversion. At the highest density, \(\rho=1.000\), the high-\(\Delta\theta\) population remains dominant in deviatoric strain throughout the explored temperature range, with the separation becoming appreciably reduced only at the highest temperatures. Thus, increasing density not only shifts the characteristic temperature scales to higher values but also suppresses the transfer of thermal deviatoric response from highly frustrated sites to low-mismatch regions. The resulting density dependence demonstrates that packing fraction controls not only the extent of thermal mechanical activity but also which local steric environments preferentially accommodate shear-like thermal distortions.

\section{Thermal Volumetric Strain Response for Low- and High-\(\theta\)-Mismatch Populations}
\label{sec:volumetric_theta}

The thermal deviatoric strain characterizes the shear-like component of the local thermal response. We next examine the complementary volumetric component,
\(\varepsilon_{\rm vol}\), which characterizes local dilation and compression and therefore probes how thermal fluctuations modify the local packing and free-volume environment. We compare the volumetric response of the lowest- and highest-\(\theta\)-mismatch populations to determine whether the local steric geometry
identified by \(\Delta\theta\) is also reflected in the dilatational component of the thermal mechanical response.

\subsection{Volumetric Strain Distributions at \(\rho=0.900\)}
\label{sec:volumetric_0900}

\begin{figure*}[ht!]
  \centering
  \begin{subfigure}[b]{0.32\textwidth}
    \includegraphics[width=\linewidth]{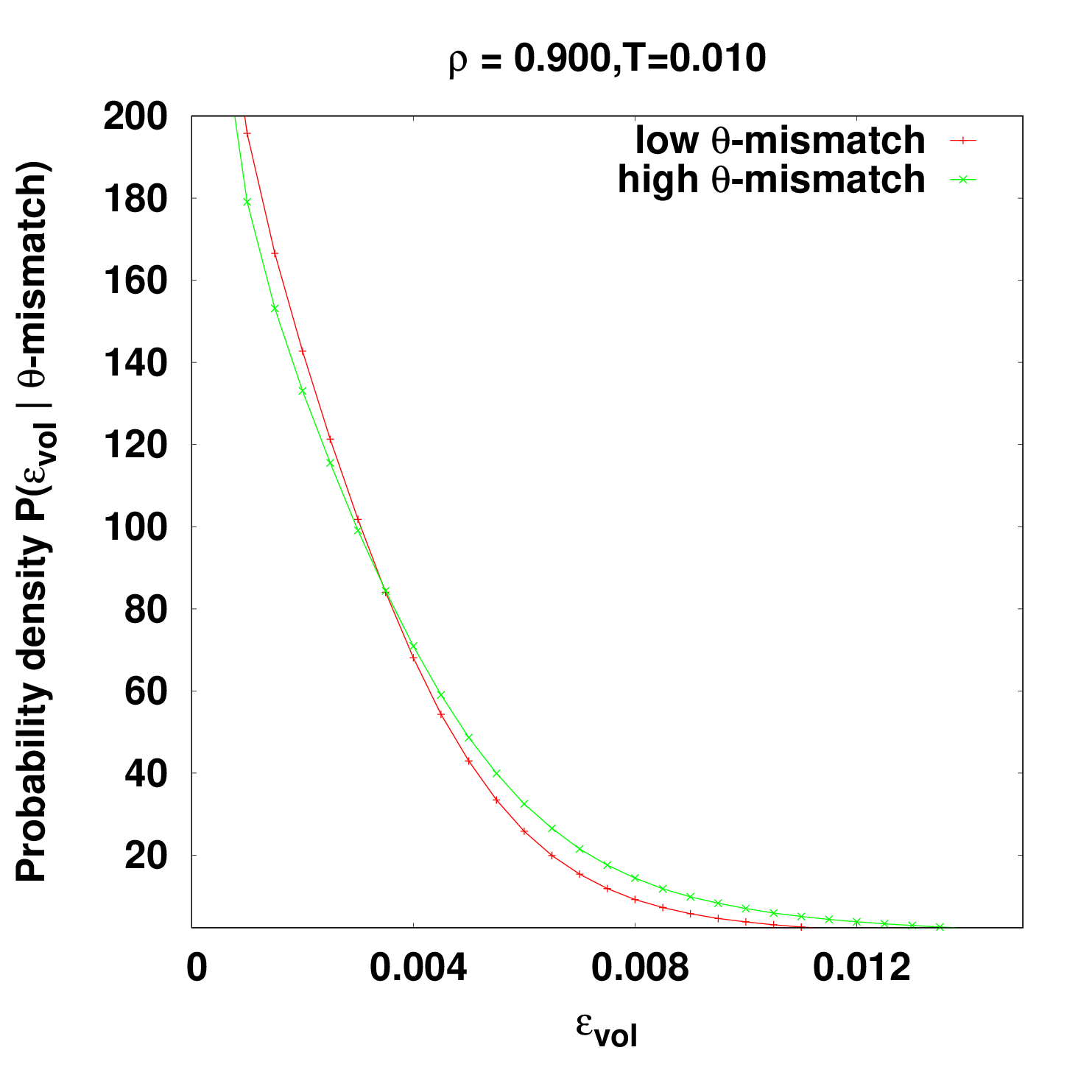}
    \caption{$T=0.010$}
  \end{subfigure}\hfill
  \begin{subfigure}[b]{0.32\textwidth}
    \includegraphics[width=\linewidth]{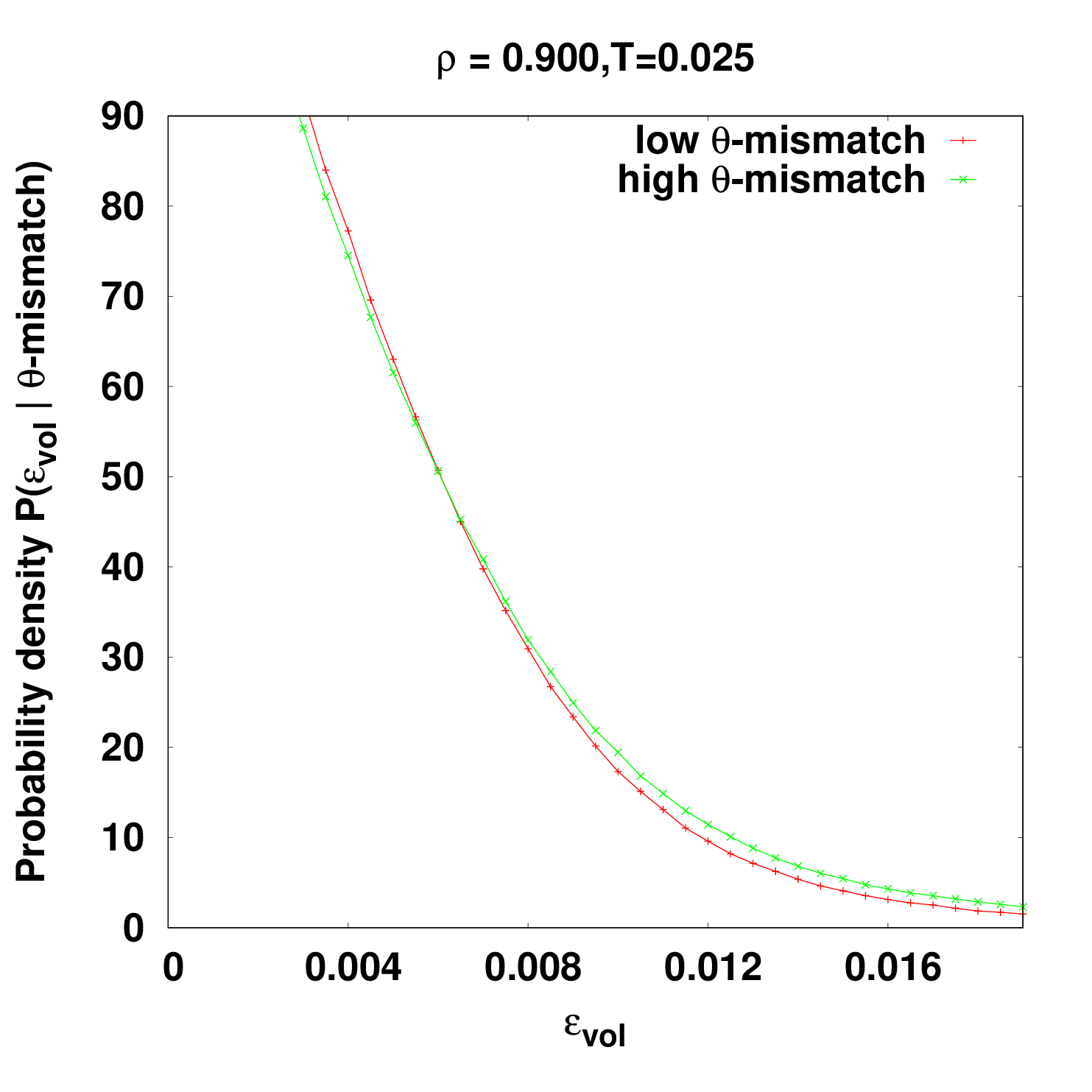}
    \caption{$T=0.025$}
  \end{subfigure}\hfill
   \begin{subfigure}[b]{0.32\textwidth}
    \includegraphics[width=\linewidth]{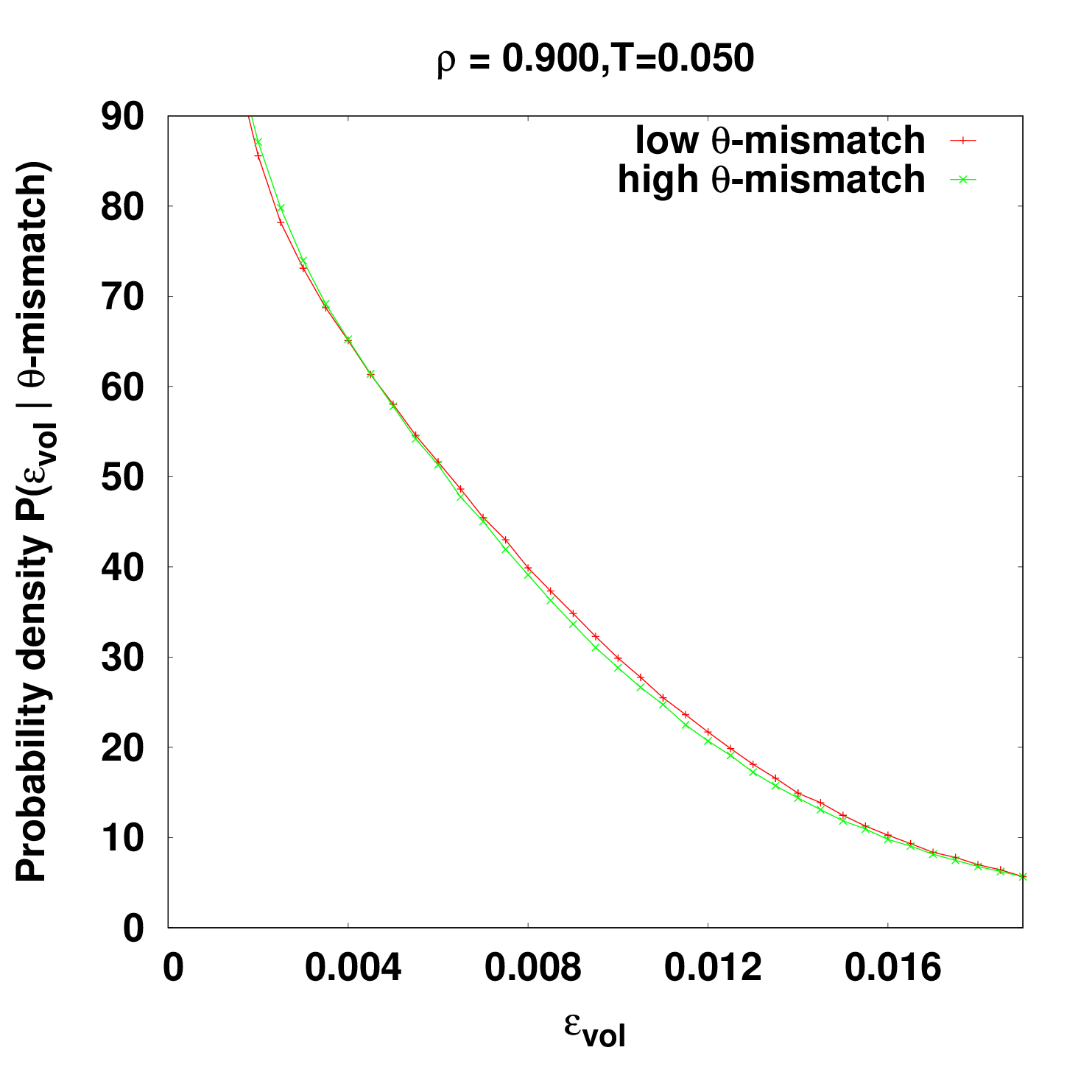}
    \caption{$T=0.050$}
  \end{subfigure}
  \vspace{0.5em}
  \begin{subfigure}[b]{0.32\textwidth}
    \includegraphics[width=\linewidth]{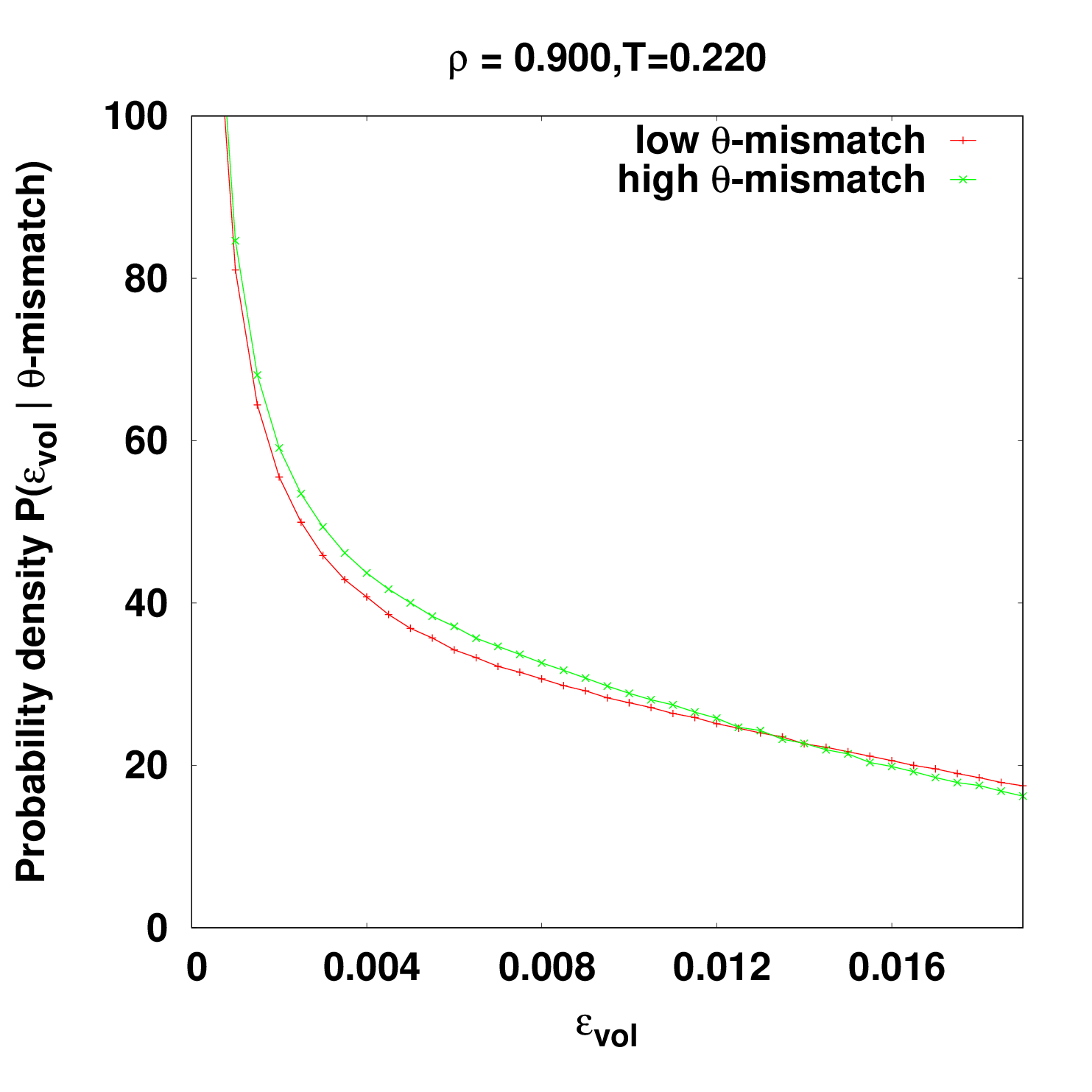}
    \caption{$T=0.220$}
  \end{subfigure}\hfill
  \begin{subfigure}[b]{0.32\textwidth}
    \includegraphics[width=\linewidth]{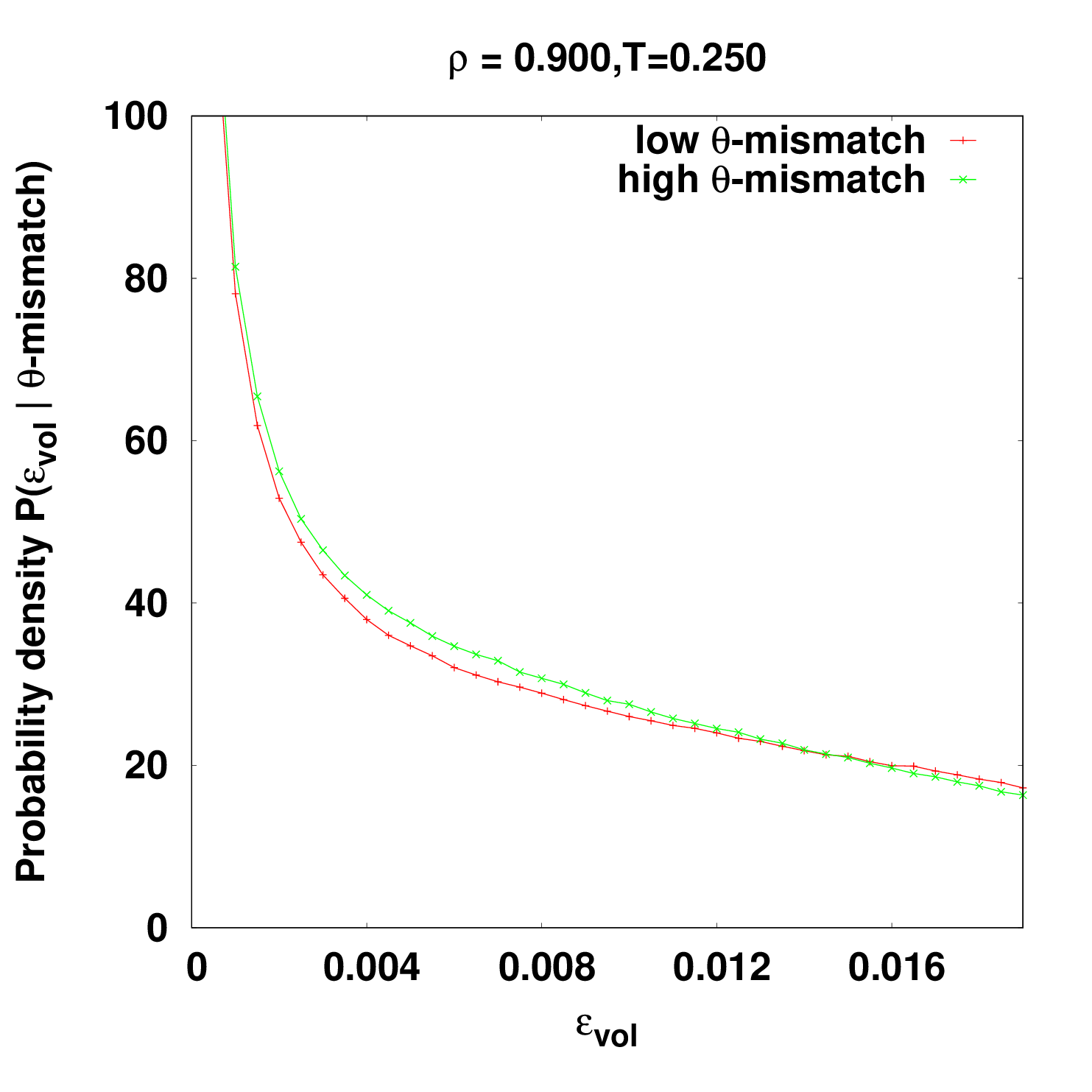}
    \caption{$T=0.250$}
  \end{subfigure}\hfill
  \begin{subfigure}[b]{0.32\textwidth}
    \includegraphics[width=\linewidth]{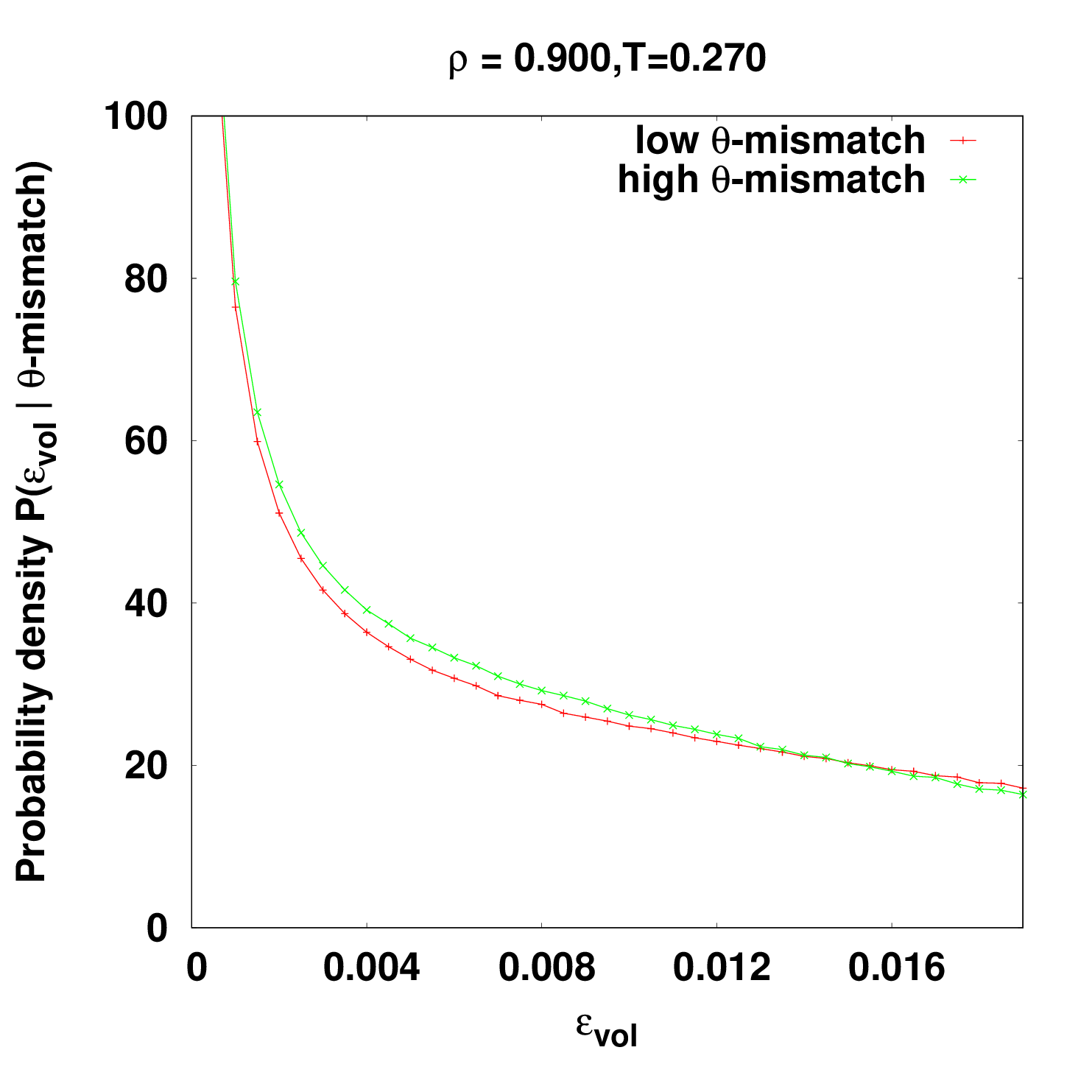}
    \caption{$T=0.270$}
  \end{subfigure}
\caption{Conditional probability density $P(\varepsilon_{\rm vol}\mid\theta)$ for the 10\% lowest- and highest-$\Delta\theta$ particles at \(\rho=0.900\) (red: low $\Delta\theta$; green: high $\Delta\theta$). Curves span temperatures across the deep solid, viscoelastic/intermittent, supercooled and normal-liquid regimes. In the deep solid the low-mismatch population dominates the small-$\varepsilon_{\rm vol}$ region while the high-mismatch population supplies the large-$\varepsilon_{\rm vol}$ tail, producing a clear crossover strain. In the intermediate (viscoelastic) window this small-strain ordering inverts (green $>$ red), and by the supercooled/normal-liquid regimes both distributions broaden and progressively merge as thermal activation erases the steric memory. These trends indicate that low-$\Delta\theta$ sites form a tight, low-dilation backbone in the solid whereas high-$\Delta\theta$ sites host larger local dilation; the inversion signals increasing participation of formerly rigid sites as cages soften.}
\label{fig:thetaa50}
\end{figure*}

Fig.~\ref{fig:thetaa50} presents the conditional probability densities
\[
P\!\left(\varepsilon_{\rm vol}\mid\theta_i\leq\theta_{10\%}\right)
\qquad\text{and}\qquad
P\!\left(\varepsilon_{\rm vol}\mid\theta_i\geq\theta_{90\%}\right),
\]
for the \(10\%\) lowest- and highest-\(\theta\)-mismatch particles, respectively, at \(\rho=0.900\). The distributions span the full temperature range, from the deeply arrested regime through the viscoelastic and supercooled regimes to the high-temperature liquid. At the lowest temperatures, the two populations exhibit a clear separation in their volumetric response. The low-\(\Delta\theta\) population has greater weight at small \(\varepsilon_{\rm vol}\), whereas the high-\(\Delta\theta\) population contributes more strongly to the large-\(\varepsilon_{\rm vol}\) tail, with the two distributions crossing at an intermediate strain. Thus, particles whose local bond geometry remains close to the steric reference predominantly contribute to the small-dilation response, while geometrically frustrated regions are preferentially represented among the larger local volumetric excursions.
With increasing temperature into the intermediate viscoelastic regime, the ordering in the small-strain region changes. The high-\(\Delta\theta\) distribution becomes larger than the low-\(\Delta\theta\) distribution at small \(\varepsilon_{\rm vol}\), while the low-mismatch population becomes comparable to or exceeds the high-mismatch population at larger volumetric strain. The crossover therefore persists, but its relative weight is reversed from that observed in the deepest glass. This evolution indicates that thermal activation increasingly involves the initially low-mismatch regions, allowing sites that were tightly constrained at low temperature to participate in collective local dilation as the cages soften.
Upon further heating into the supercooled and normal-liquid regimes, both conditional distributions broaden and progressively overlap. The distinction between the two populations at small volumetric strain becomes less pronounced, and their large-strain tails become increasingly similar, although a residual offset at intermediate strains can persist over part of the temperature range. Thus, at \(\rho=0.900\), the volumetric response evolves from a clearly mismatch-dependent distribution at low temperature, through a crossover and inversion in the intermediate regime, toward increasingly similar responses when thermal motion dominates.

\subsection{Volumetric Strain Distributions at \(\rho=0.950\)}
\label{sec:volumetric_0950}

\begin{figure*}[ht!]
  \centering
  \begin{subfigure}[b]{0.32\textwidth}
    \includegraphics[width=\linewidth]{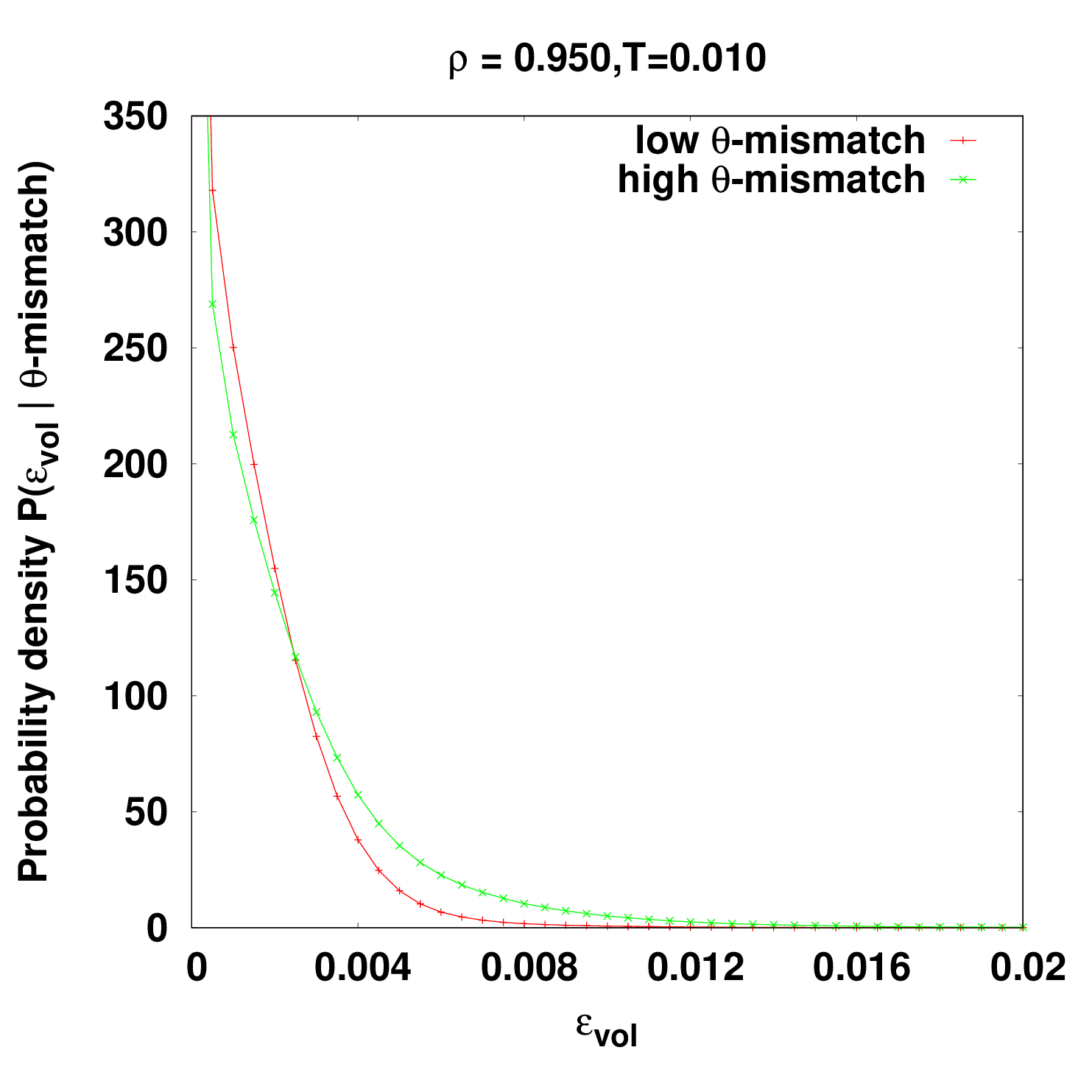}
    \caption{$T=0.010$}
  \end{subfigure}\hfill
  \begin{subfigure}[b]{0.32\textwidth}
    \includegraphics[width=\linewidth]{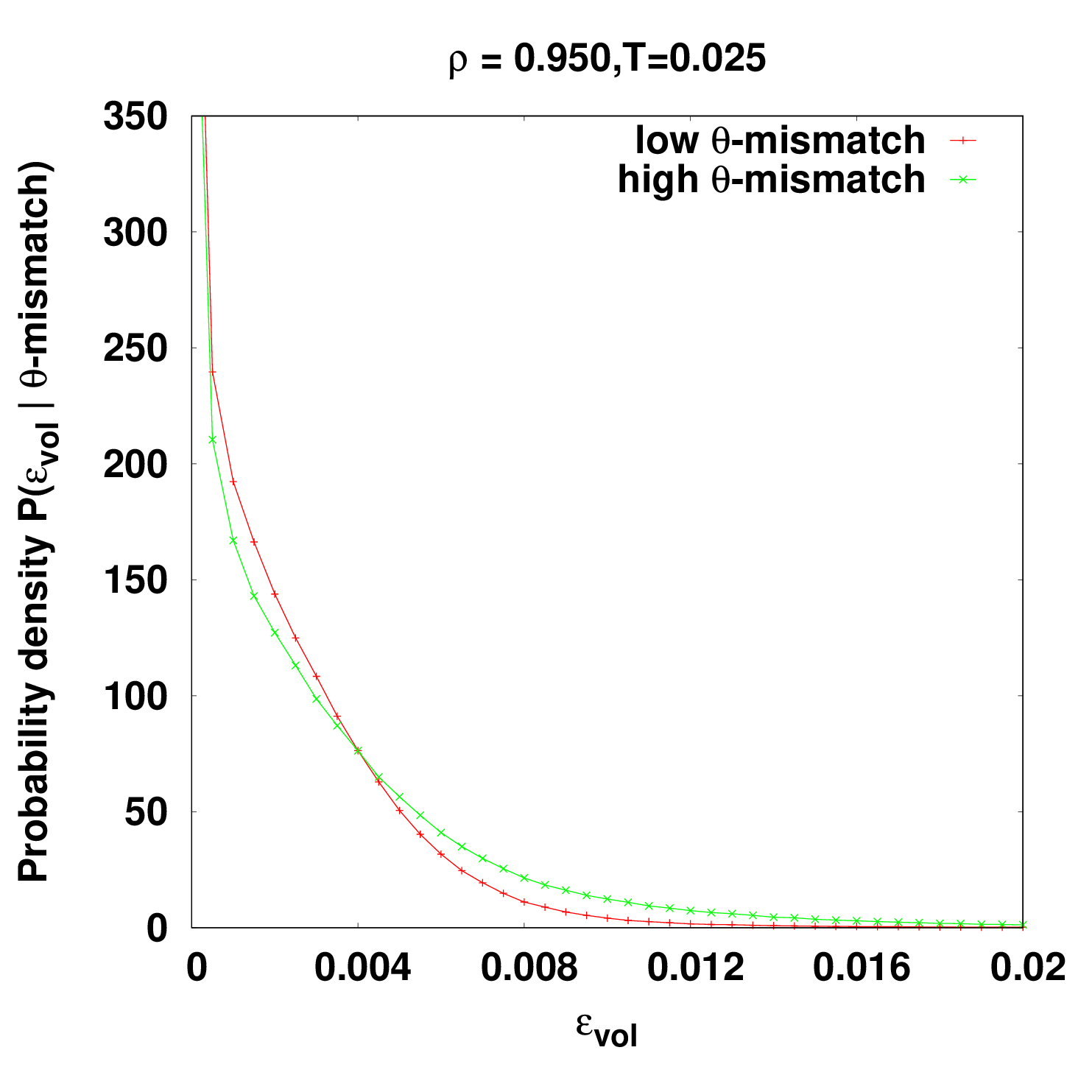}
    \caption{$T=0.025$}
  \end{subfigure}\hfill
   \begin{subfigure}[b]{0.32\textwidth}
    \includegraphics[width=\linewidth]{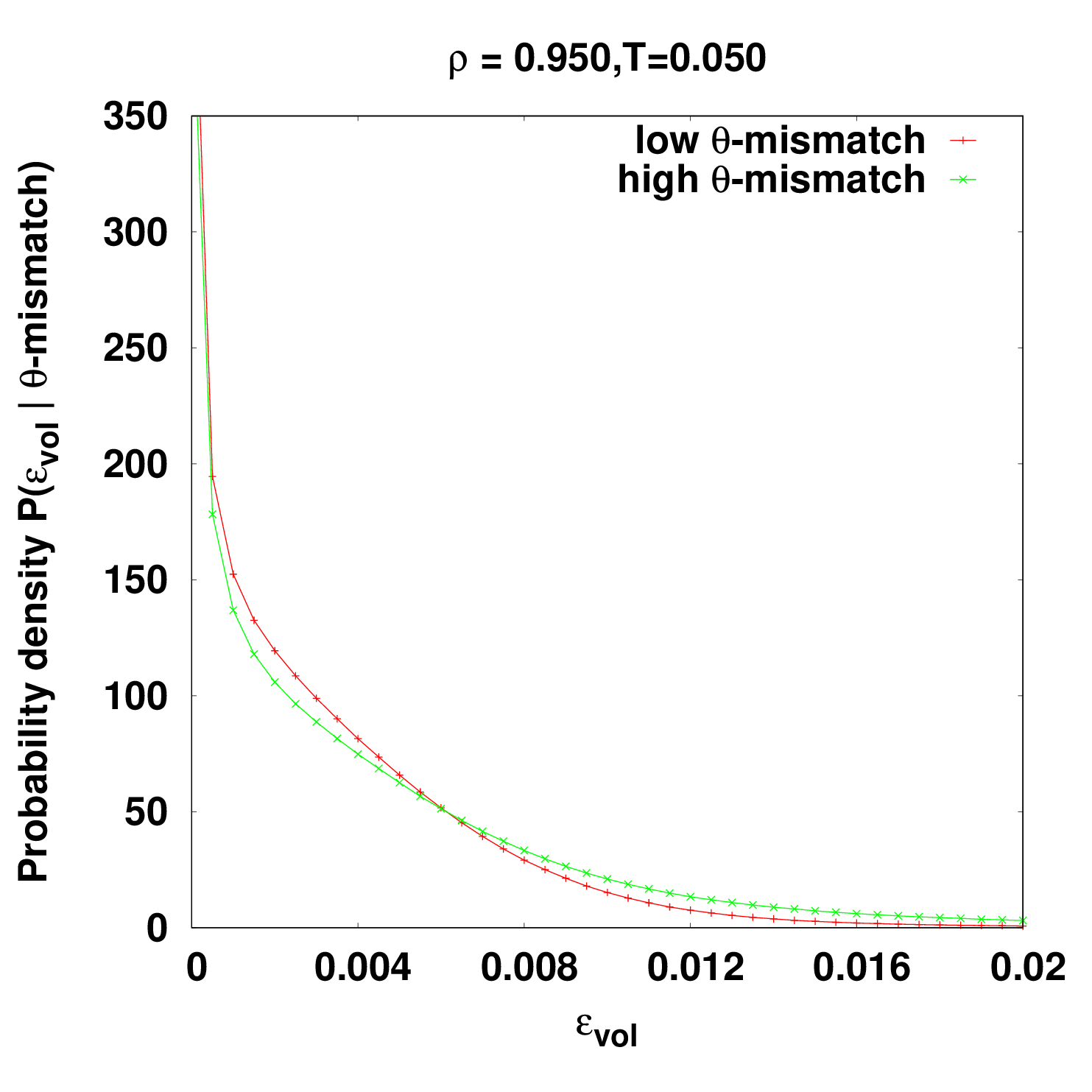}
    \caption{$T=0.050$}
  \end{subfigure}
 \vspace{0.5em}
  \begin{subfigure}[b]{0.32\textwidth}
    \includegraphics[width=\linewidth]{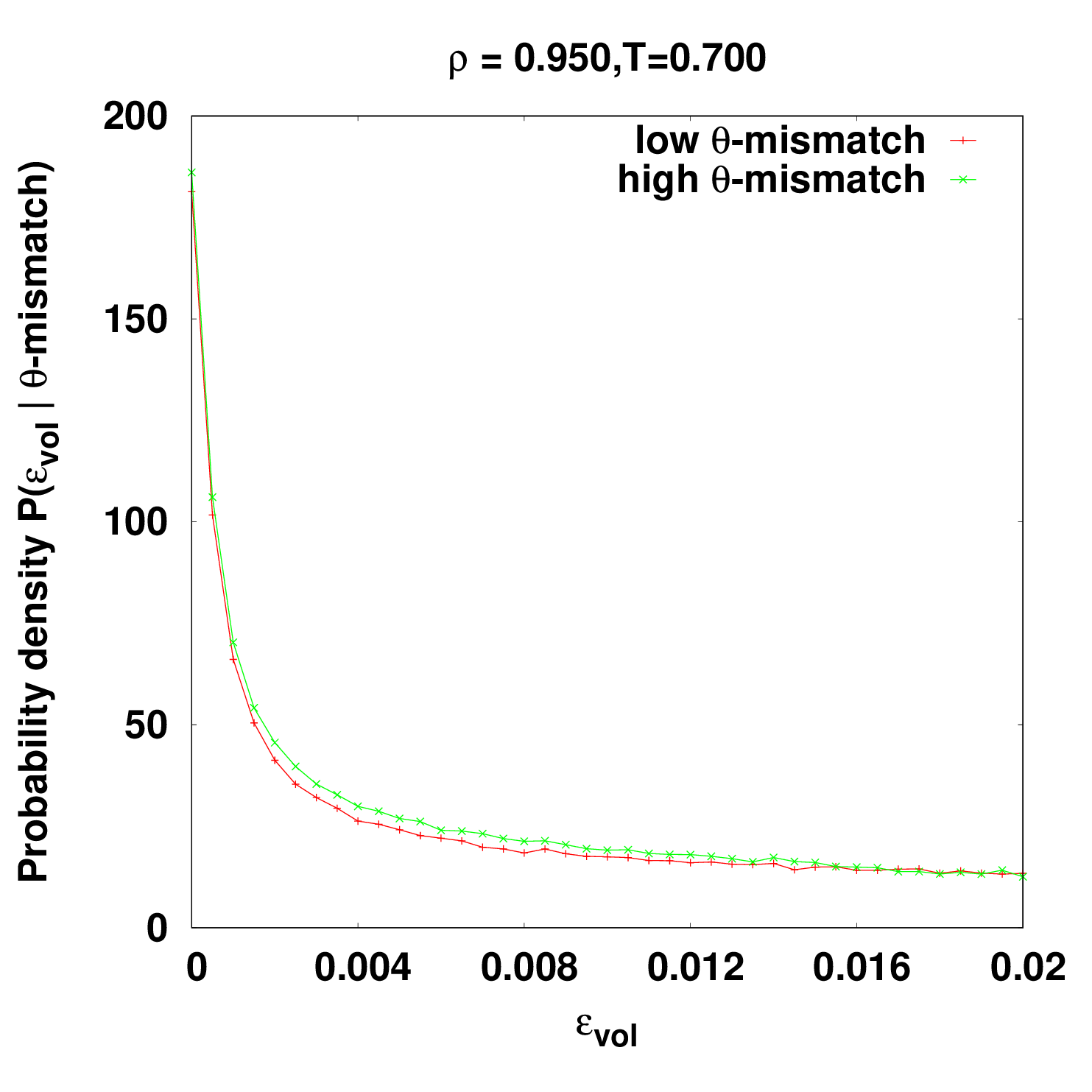}
    \caption{$T=0.700$}
  \end{subfigure}\hfill
  \begin{subfigure}[b]{0.32\textwidth}
    \includegraphics[width=\linewidth]{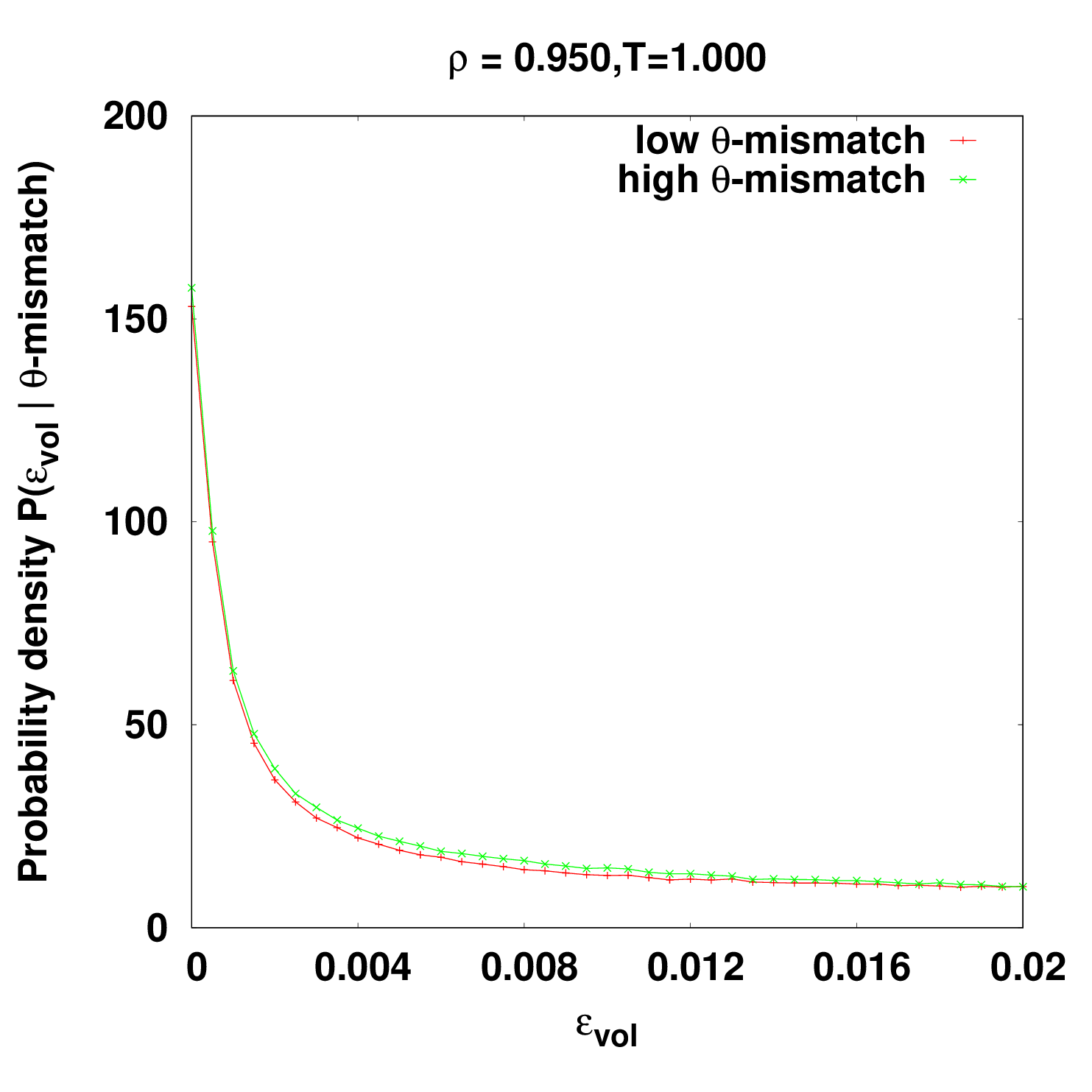}
    \caption{$T=1.000$}
  \end{subfigure}\hfill
  \begin{subfigure}[b]{0.32\textwidth}
    \includegraphics[width=\linewidth]{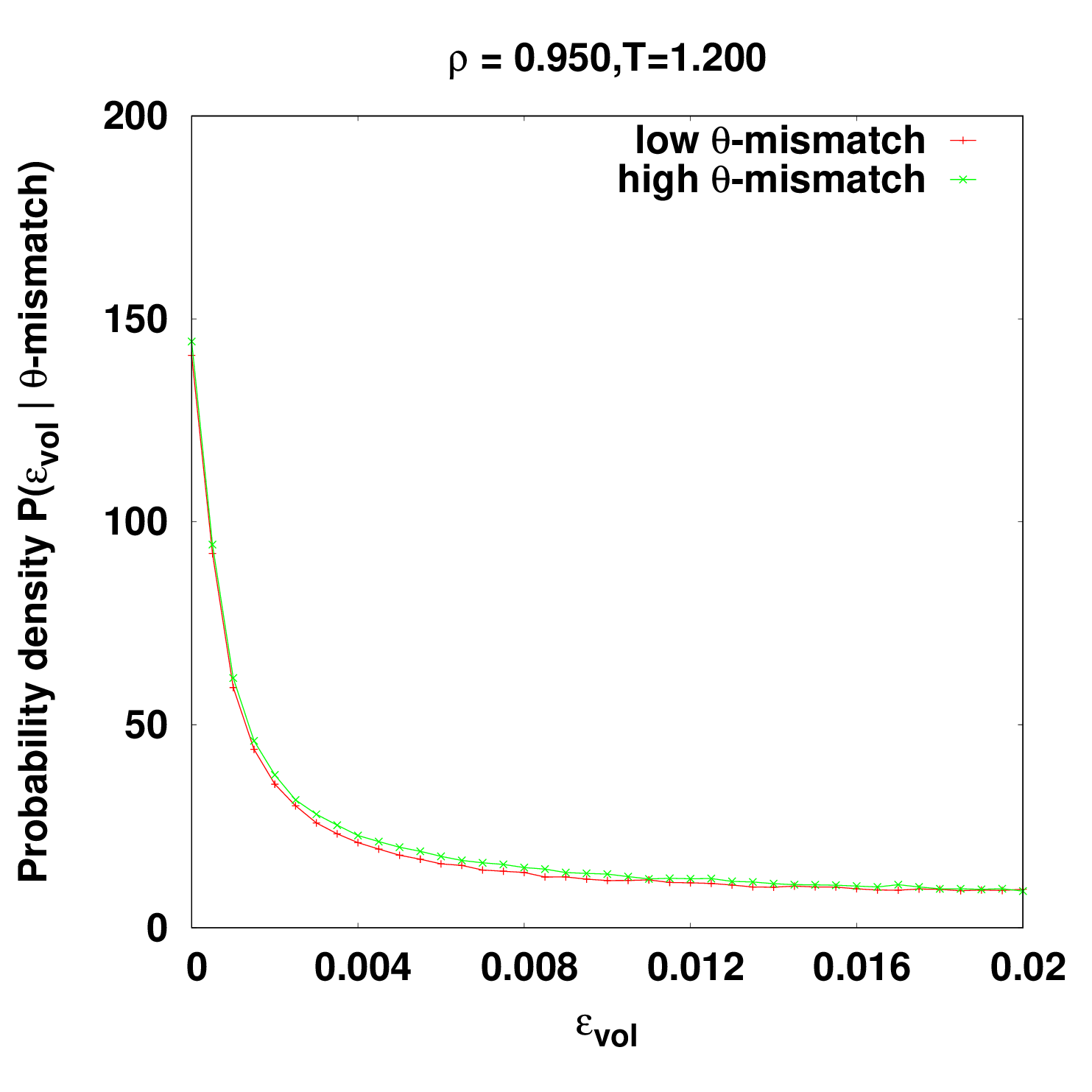}
    \caption{$T=1.200$}
  \end{subfigure}
 \caption{Conditional probability density $P(\varepsilon_{\rm vol}\mid\theta)$ for the 10\% lowest- and highest-$\Delta\theta$ particles at $\rho=0.950$ (red: low $\Delta\theta$; green: high $\Delta\theta$), shown over a temperature sweep from the deep glass into the normal liquid. In the solid ($T\lesssim0.110$) the PDFs are clearly separated and cross: low-$\Delta\theta$ particles dominate the near-zero $\varepsilon_{\rm vol}$ region while high-$\Delta\theta$ particles supply the large-$\varepsilon_{\rm vol}$ tail, indicating that dilational events concentrate on geometrically frustrated sites. In the intermediate (viscoelastic) window the separation weakens and a small inversion can appear at large $\varepsilon_{\rm vol}$ as cages soften; by the supercooled/normal-liquid regimes the two PDFs broaden and collapse, showing that steric mismatch ceases to strongly bias volumetric fluctuations at high $T$.}

  \label{fig:thetaa6}
\end{figure*}

Fig.~\ref{fig:thetaa6} shows the conditional volumetric-strain distributions for the \(10\%\) lowest- and highest-\(\theta\)-mismatch particles at \(\rho=0.950\). In the low-temperature regime, the two populations are clearly separated and display a well-defined crossover. The low-\(\Delta\theta\)
distribution is larger at very small volumetric strain, indicating a greater probability of negligible local volume change, whereas the high-\(\Delta\theta\)
population is enhanced at larger \(\varepsilon_{\rm vol}\), showing that geometrically frustrated sites are more strongly represented in the large-volume-change tail. This separation demonstrates that local steric frustration is directly reflected in the volumetric response of the jammed state.
As the system is heated into the intermediate viscoelastic regime, the separation between the two conditional distributions decreases markedly. A weak inversion develops on the large-\(\varepsilon_{\rm vol}\) side, where the low-mismatch population can become slightly more prominent than the high-mismatch population. Compared with the stronger inversion observed at \(\rho=0.900\), this effect is weaker, indicating that increasing density maintains a stronger distinction between the two structural populations over a broader temperature range. The evolution is consistent with progressively softening cages and increasing participation of low-mismatch particles in collective volumetric fluctuations.

At still higher temperatures, spanning the supercooled and normal-liquid regimes, the two distributions broaden and progressively merge. By the highest
temperatures they are essentially overlapping, showing that thermal mobility has largely erased the dependence of the volumetric response on the initial steric mismatch. Thus, at \(\rho=0.950\), the solid-state mismatch-dependent crossover remains visible, but its contrast and the associated inversion are substantially reduced as compared with the lowest density.

\subsection{Volumetric Strain Distributions at \(\rho=1.000\)}
\label{sec:volumetric_1000}

\begin{figure*}[ht!]
  \centering
 \begin{subfigure}[b]{0.32\textwidth}
    \includegraphics[width=\linewidth]{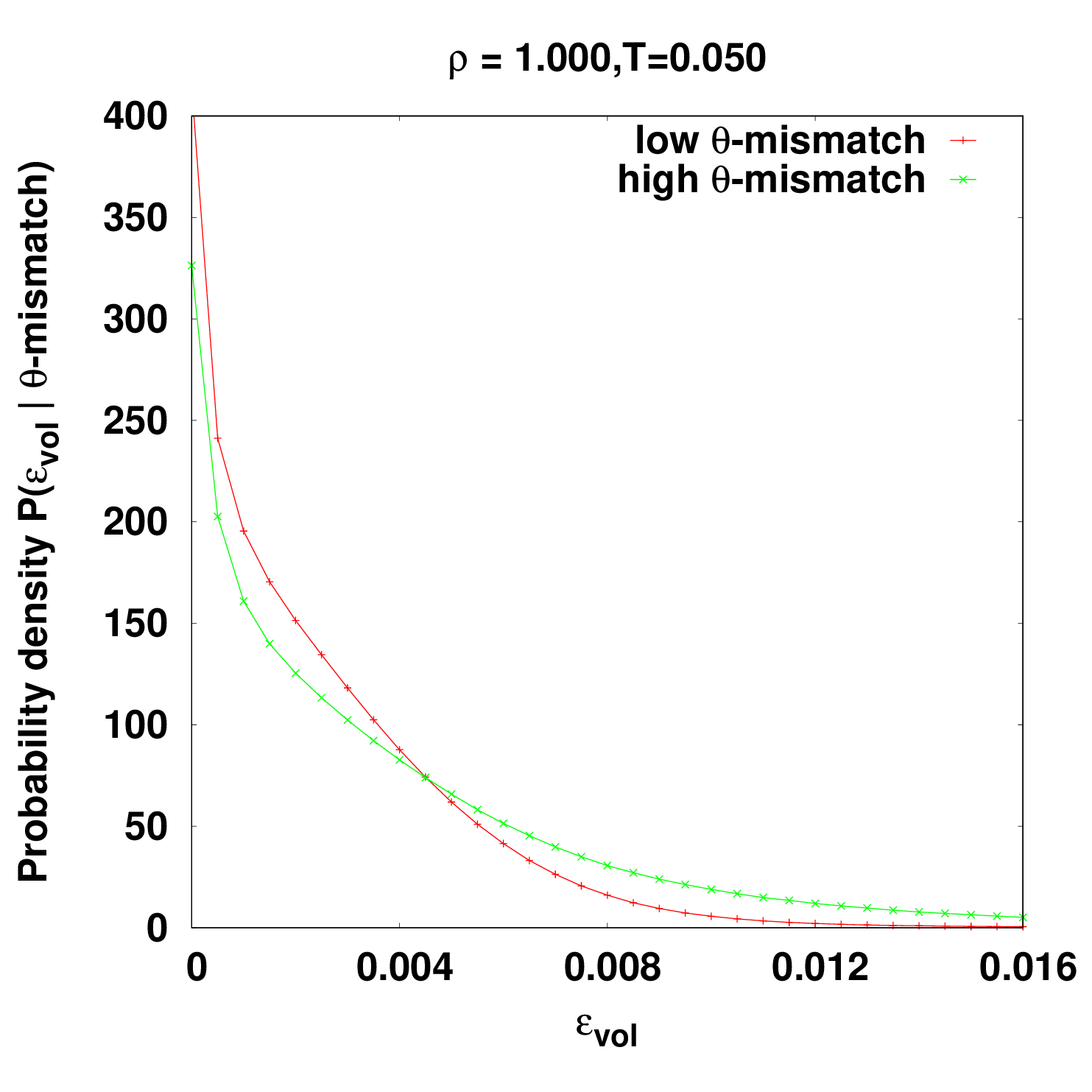}
    \caption{$T=0.050$}
  \end{subfigure}\hfill
   \begin{subfigure}[b]{0.32\textwidth}
    \includegraphics[width=\linewidth]{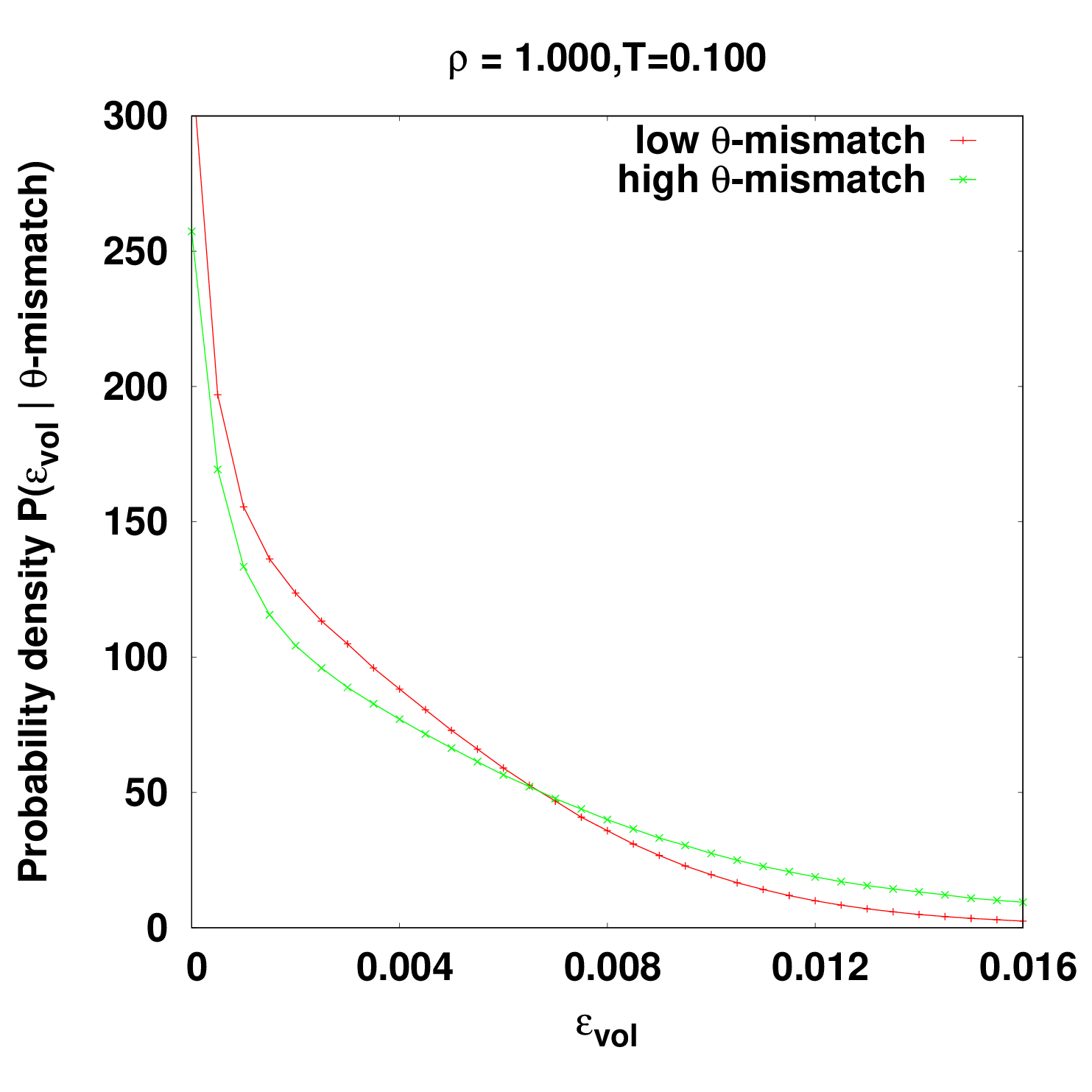}
    \caption{$T=0.100$}
  \end{subfigure}
  \begin{subfigure}[b]{0.32\textwidth}
    \includegraphics[width=\linewidth]{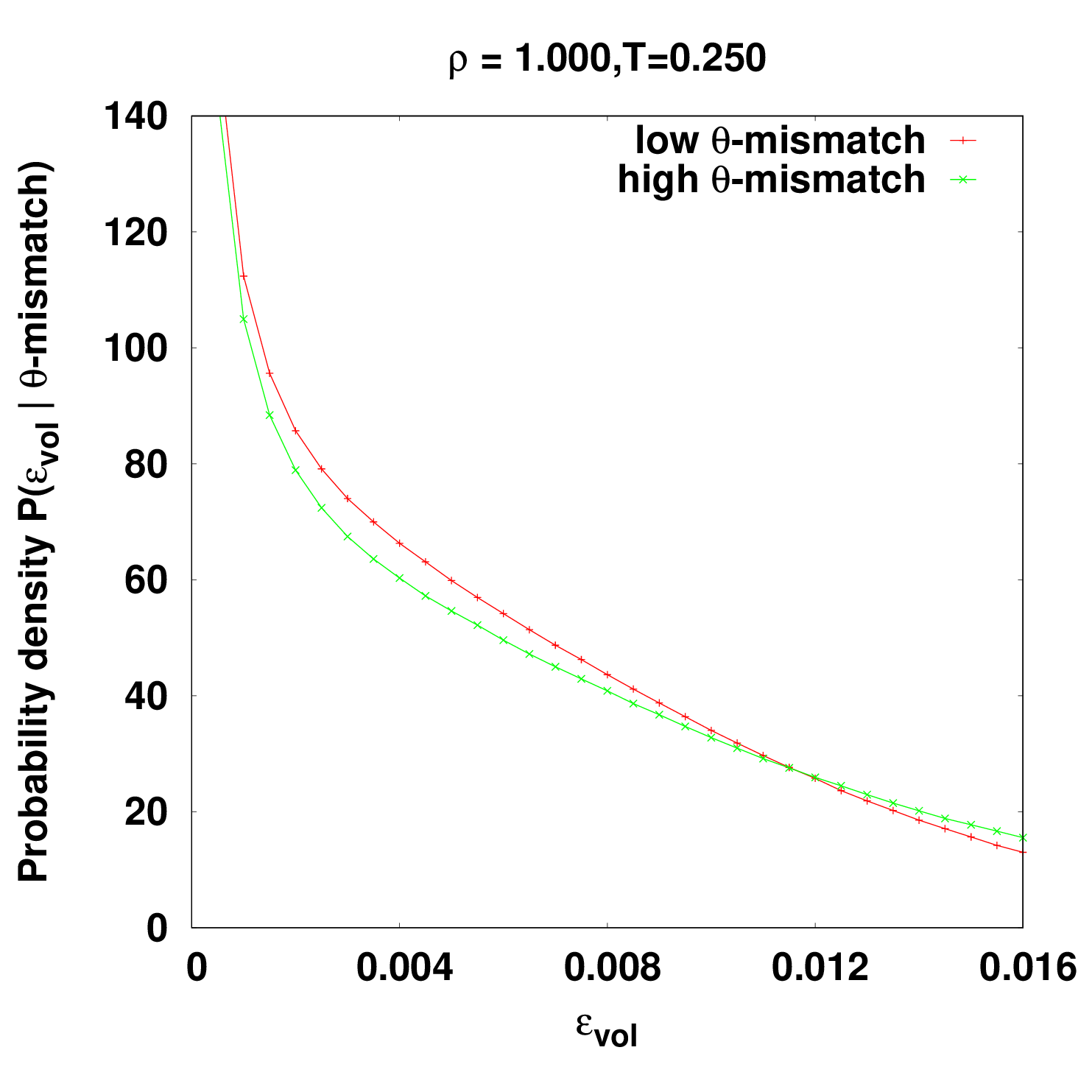}
    \caption{$T=0.250$}
  \end{subfigure}\hfill
\vspace{0.5em}
   \begin{subfigure}[b]{0.32\textwidth}
    \includegraphics[width=\linewidth]{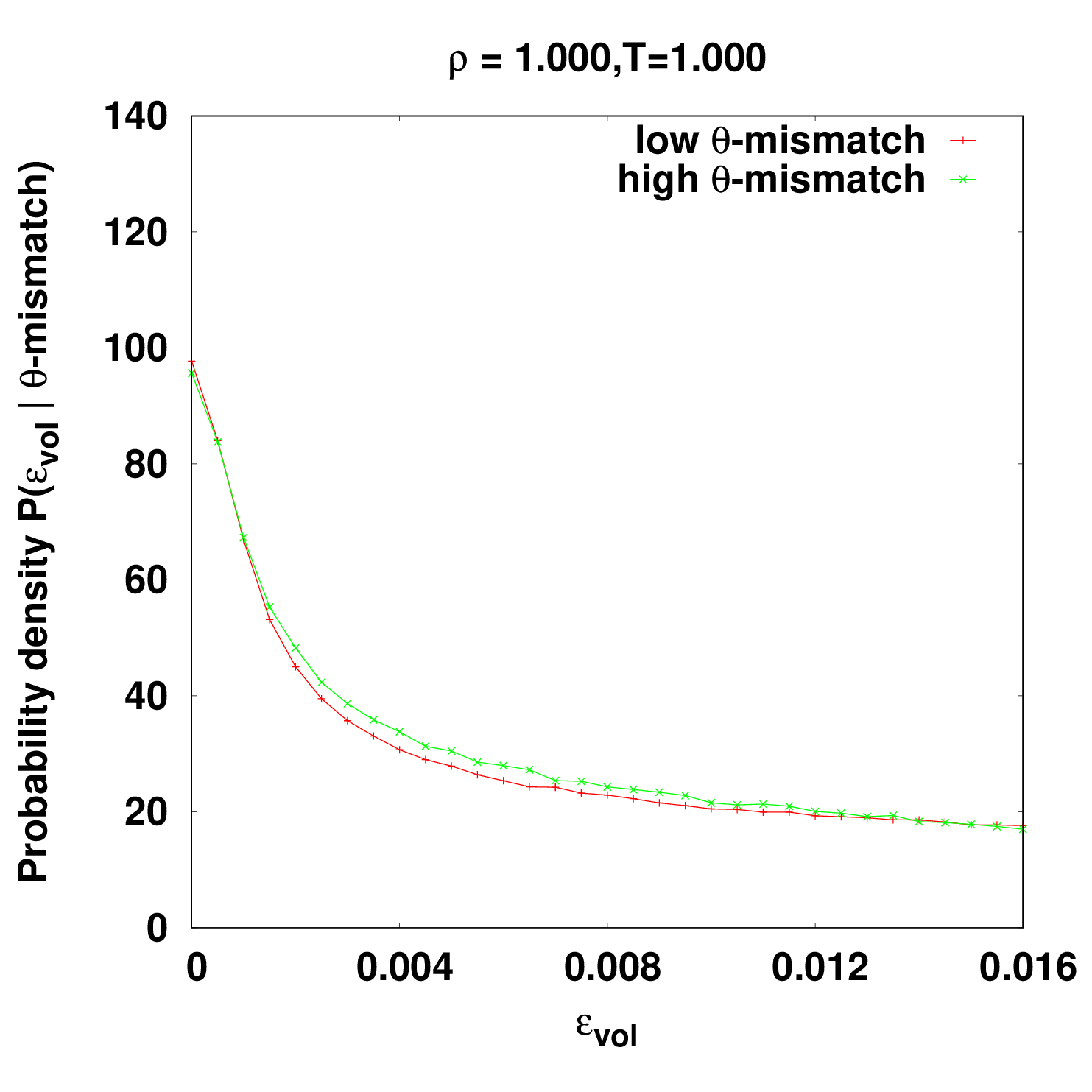}
    \caption{$T=1.000$}
  \end{subfigure}\hfill
   \begin{subfigure}[b]{0.32\textwidth}
    \includegraphics[width=\linewidth]{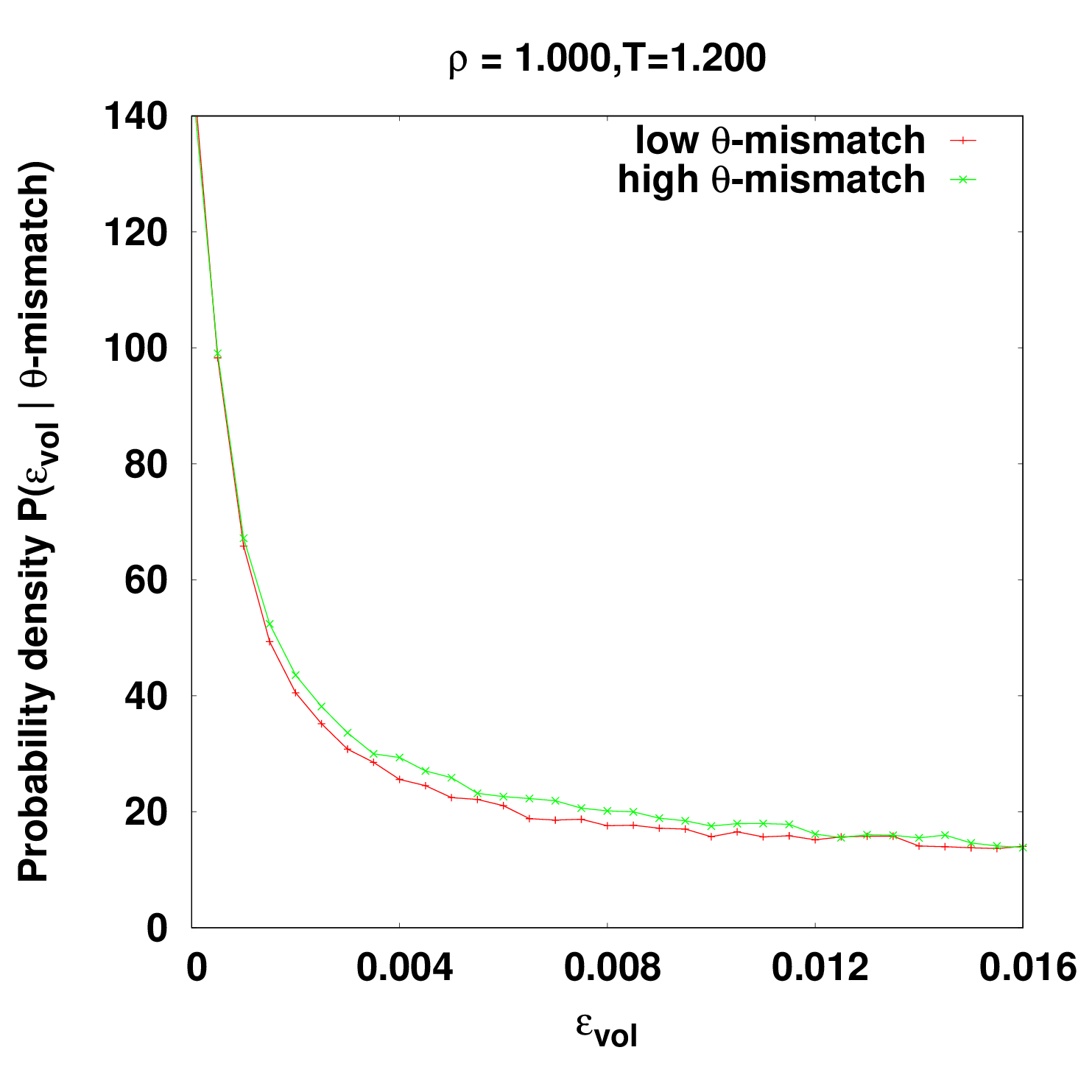}
    \caption{$T=1.200$}
  \end{subfigure}\hfill
  \begin{subfigure}[b]{0.32\textwidth}
    \includegraphics[width=\linewidth]{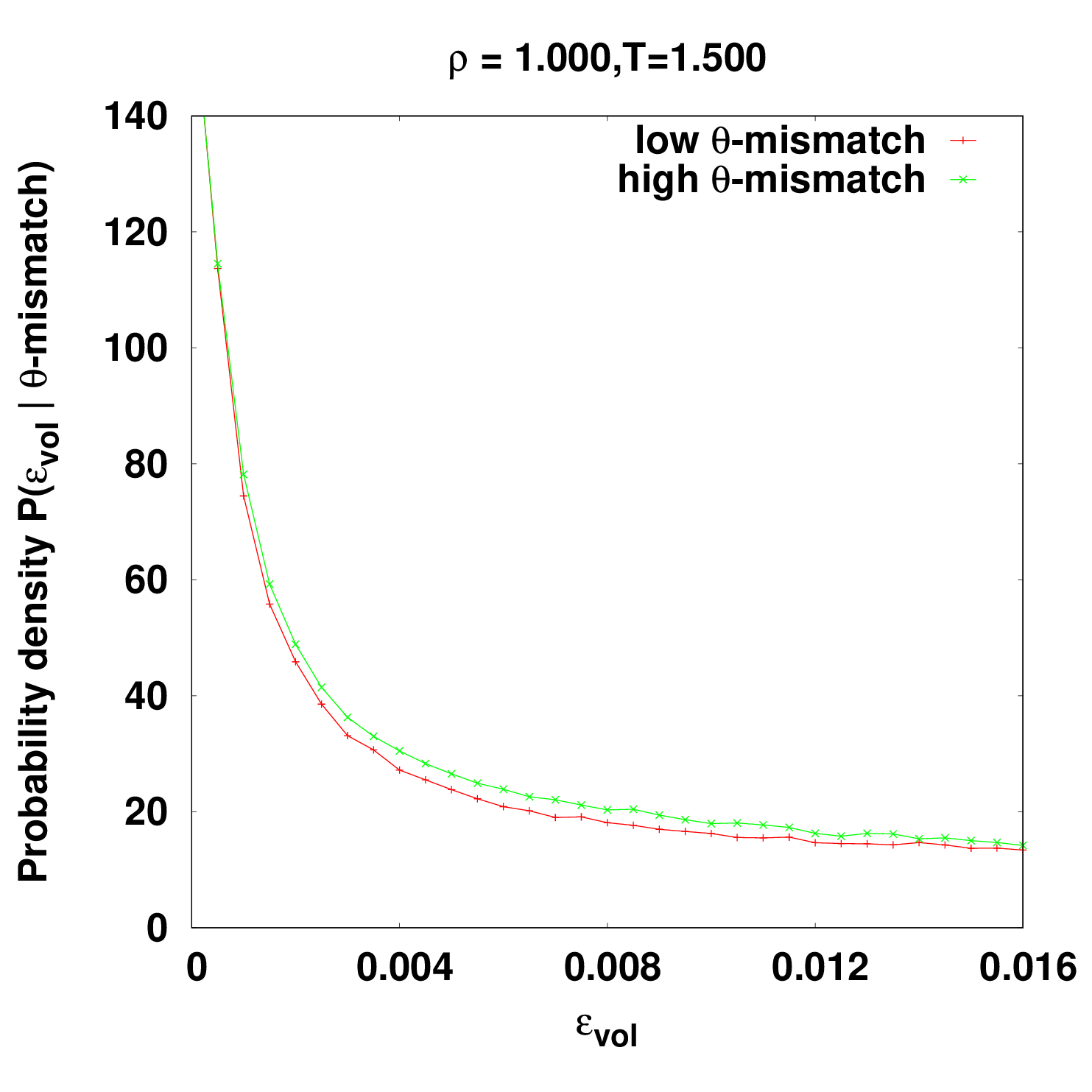}
    \caption{$T=1.500$}
  \end{subfigure}
 \caption{Conditional probability density $P(\varepsilon_{\rm vol}\mid\theta)$ for the 10\% lowest- (red) and 10\% highest- (green) $\Delta\theta$ particles at $\rho=1.000$. Curves span the temperature range from the deep glass to the normal liquid; key markers are $T_{\rm s}\approx0.251$, $T_g\approx0.596$, and $T_{\rm onset}\approx1.483$. In the jammed solid ($T\lesssim T_{\rm s}$) the PDFs are clearly separated and cross: low-mismatch particles dominate the near-zero $\varepsilon_{\rm vol}$ region while high-mismatch particles produce an enhanced large-$\varepsilon_{\rm vol}$ tail, indicating that frustrated contacts undergo larger local dilations when they yield. The crossover remains but weakens through the viscoelastic window, and by the supercooled/normal-liquid regimes the two distributions broaden and largely collapse as thermal motion homogenizes volumetric response.}
\label{fig:vol_dev_1000}

\label{fig:thetaa8}
\end{figure*}

Fig.~\ref{fig:vol_dev_1000} presents the conditional probability density of \(\varepsilon_{\rm vol}\) for the \(10\%\) lowest- and highest-\(\theta\)-mismatch
particles at \(\rho=1.000\). In the deeply arrested regime, the two populations again display a pronounced crossover. At small volumetric strain, the
low-\(\Delta\theta\) distribution lies above the high-\(\Delta\theta\) distribution, whereas the high-\(\Delta\theta\) population dominates the
large-\(\varepsilon_{\rm vol}\) tail. The result shows that even under strong steric confinement, local geometric frustration is associated with a greater probability of large volume-changing thermal distortions.

On entering the intermediate viscoelastic regime, the crossover remains present but its magnitude decreases substantially. A weak inversion of the relative peak heights appears at some temperatures, with the low-mismatch distribution becoming slightly larger at moderate volumetric strain. The effect is nevertheless much smaller than the separation observed in the deeply arrested regime. This reduced contrast indicates that the strong confinement at \(\rho=1.000\) preserves the structural distinction in the volumetric response while thermal activation progressively allows both populations to participate.

Upon further heating through the supercooled and normal-liquid regimes, the two conditional distributions broaden and progressively converge. At high temperature the distinction between low- and high-mismatch particles becomes small, indicating that thermal motion increasingly homogenizes the local volumetric response. The overall evolution therefore remains the same in qualitative form as at lower densities: a pronounced solid-state separation, followed by progressive convergence with temperature, but the solid-state mismatch dependence is more strongly retained under the tighter steric
confinement.

\section{Comparative Summary of Volumetric-Strain Trends Across Density}
\label{sec:volumetric_comparative}

A comparison of the volumetric-strain response at \(\rho=0.900\), \(0.950\), and \(1.000\) reveals a common structural signature together with a systematic dependence on density. In the deeply arrested regime at all three state points, the low-\(\Delta\theta\) population has greater weight at small volumetric strain, whereas the high-\(\Delta\theta\) population is preferentially represented in the large-strain tail, producing a clear crossover between the two conditional distributions. This establishes a robust connection between local steric frustration and thermally generated volume changes: regions with larger geometric mismatch are more strongly associated with large local volumetric excursions.

The subsequent temperature evolution depends on density. At \(\rho=0.900\), the mismatch-dependent separation changes most strongly upon heating, with an inversion of the small-strain ordering in the intermediate viscoelastic regime and a noticeable persistence of this effect into the
higher-temperature regime before the two populations ultimately merge. At \(\rho=0.950\), the solid-state crossover remains clear, but the subsequent inversion is substantially weaker and the two distributions converge more rapidly with increasing temperature. At \(\rho=1.000\), the solid-state separation and the enhancement of the high-\(\Delta\theta\) contribution to the large-strain tail remain pronounced, while the post-solid inversion is weak and the two populations progressively converge only upon further heating. Thus, increasing density strengthens the mismatch-dependent volumetric response in the arrested regime and suppresses the extent to which the ordering is inverted in the intermediate regime.

Across all three densities, the common sequence is therefore a well-separated volumetric response in the deeply arrested regime, a weakened and progressively redistributed response in the intermediate viscoelastic regime, and increasingly overlapping distributions in the liquid-like regime. The persistence of the solid-state crossover across density, followed by its systematic attenuation with temperature, shows that thermal volumetric strain carries a robust structural fingerprint of steric mismatch while remaining sensitive to the degree of steric confinement. In this sense, the volumetric component complements the thermal nonaffine and deviatoric responses by directly revealing how geometric frustration is reflected in local dilation and compression during thermally driven rearrangements.

\begin{figure*}[htb!]
  \centering
  \begin{subfigure}[b]{0.32\textwidth}
    \centering
    \textbf{\small $\rho=0.900,\; T=0.010$}\vspace{4pt}\\
    \includegraphics[width=\linewidth]{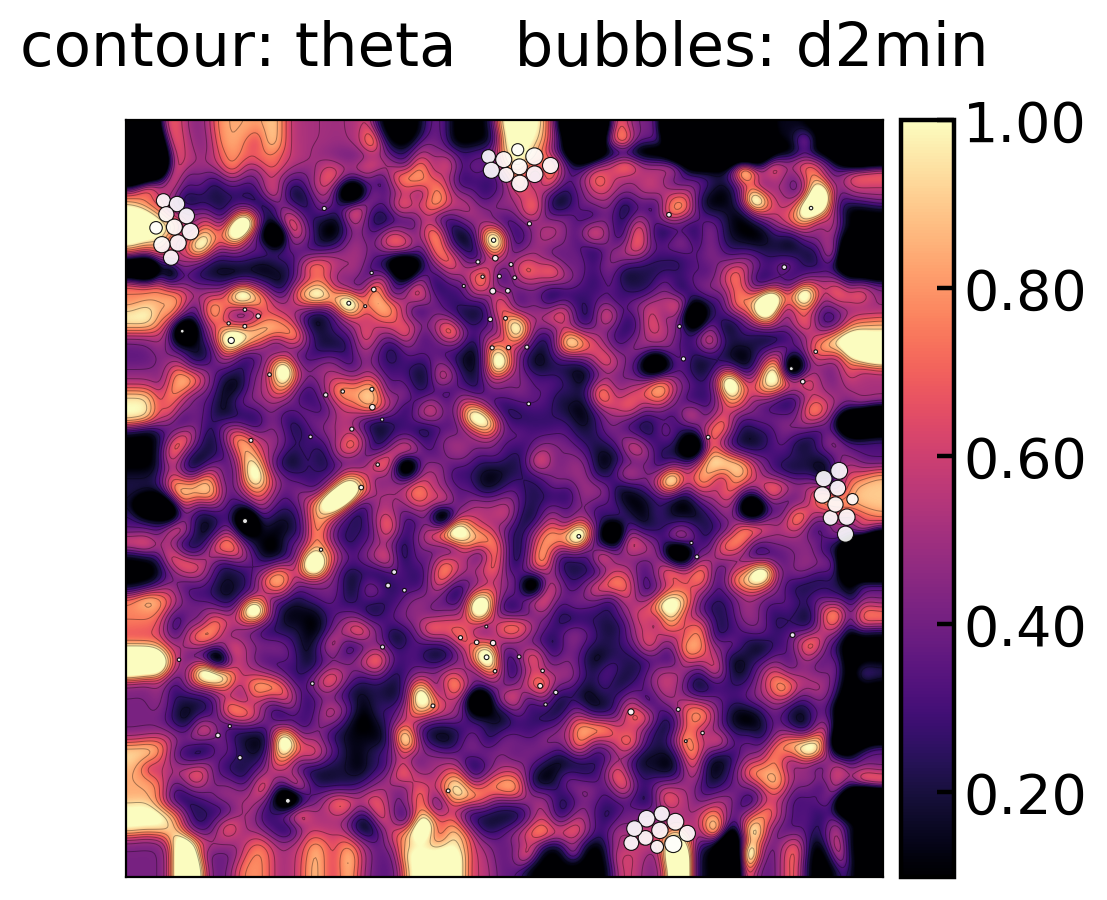}
  \end{subfigure}\hfill
  \begin{subfigure}[b]{0.32\textwidth}
    \centering
    \textbf{\small $\rho=0.900,\; T=0.010$}\vspace{4pt}\\
    \includegraphics[width=\linewidth]{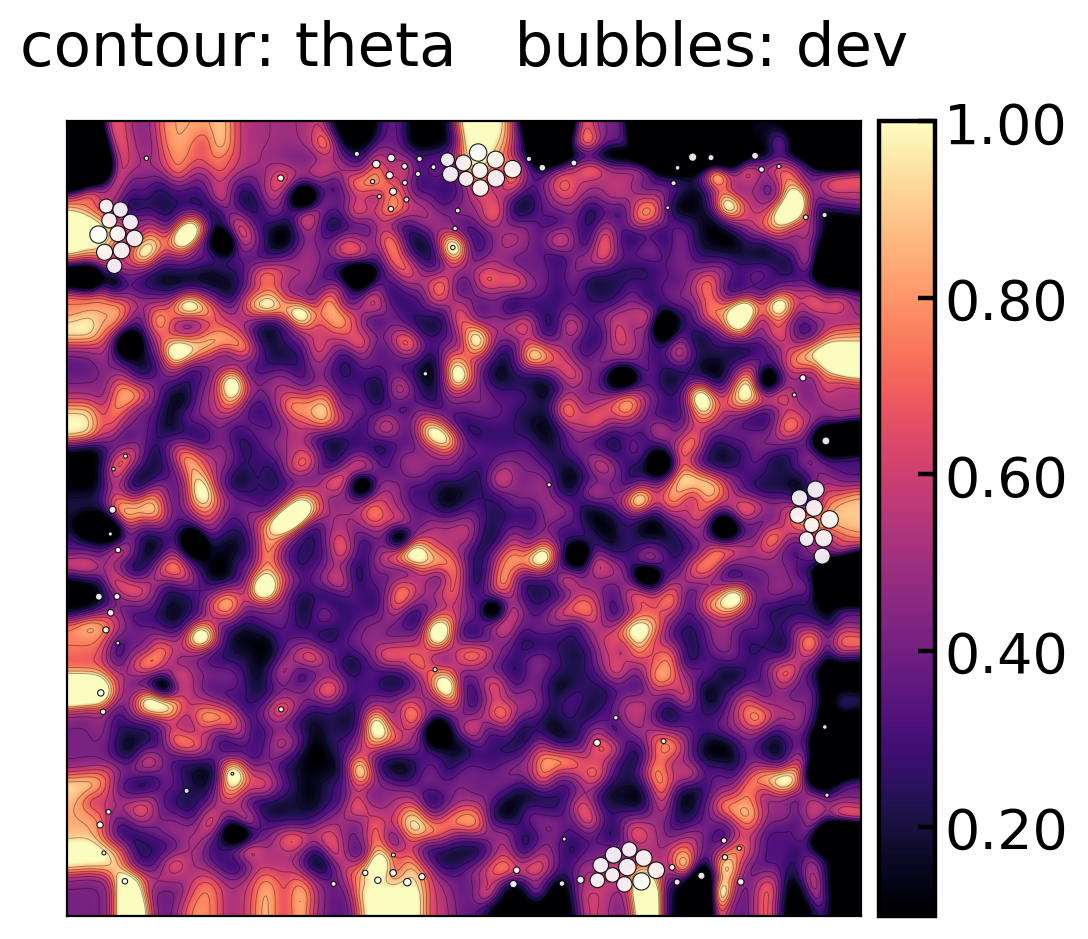}
  \end{subfigure}\hfill
  \begin{subfigure}[b]{0.32\textwidth}
    \centering
    \textbf{\small $\rho=0.900,\; T=0.010$}\vspace{4pt}\\
    \includegraphics[width=\linewidth]{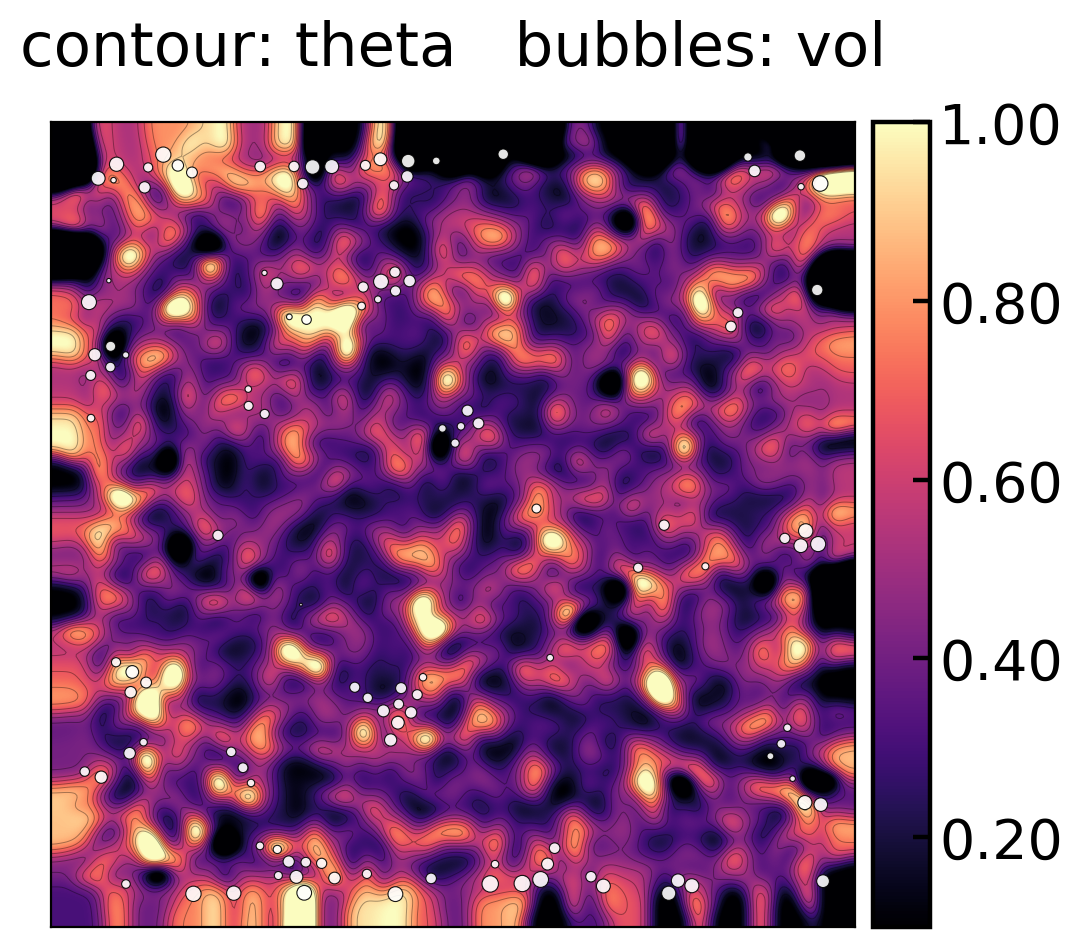}
  \end{subfigure}

  \vspace{8pt}

  \begin{subfigure}[b]{0.32\textwidth}
    \centering
    \textbf{\small $\rho=0.950,\; T=0.100$}\vspace{4pt}\\
    \includegraphics[width=\linewidth]{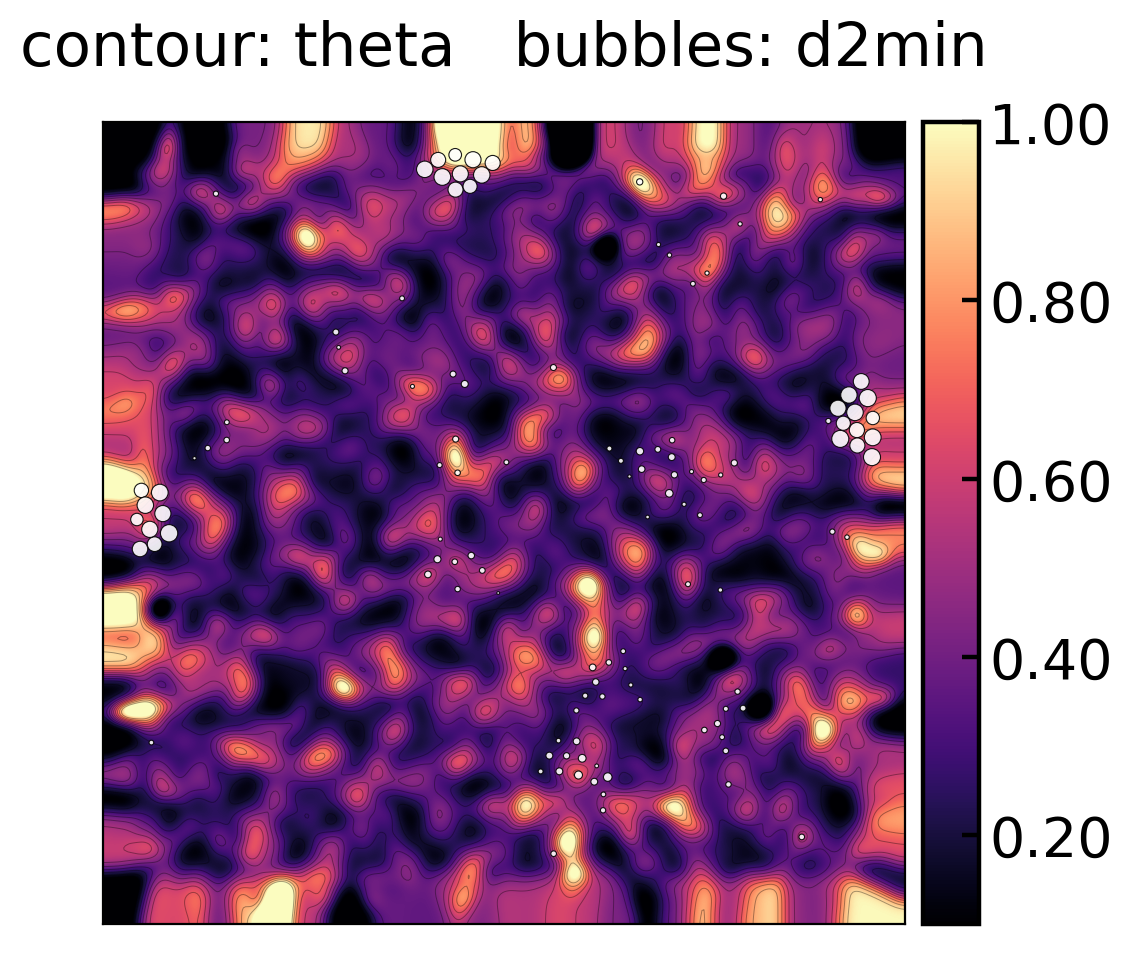}
  \end{subfigure}\hfill
  \begin{subfigure}[b]{0.32\textwidth}
    \centering
    \textbf{\small $\rho=0.950,\; T=0.100$}\vspace{4pt}\\
    \includegraphics[width=\linewidth]{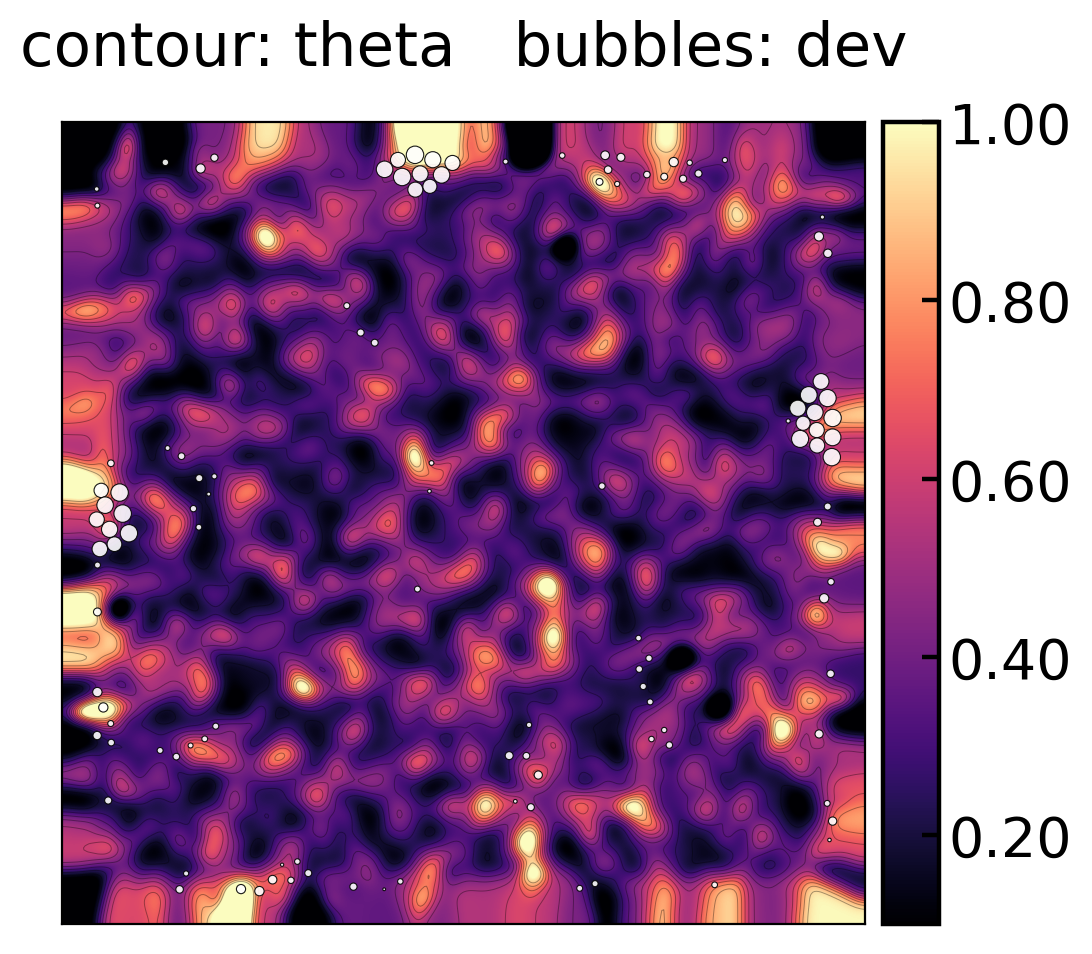}
  \end{subfigure}\hfill
  \begin{subfigure}[b]{0.32\textwidth}
    \centering
    \textbf{\small $\rho=0.950,\; T=0.100$}\vspace{4pt}\\
    \includegraphics[width=\linewidth]{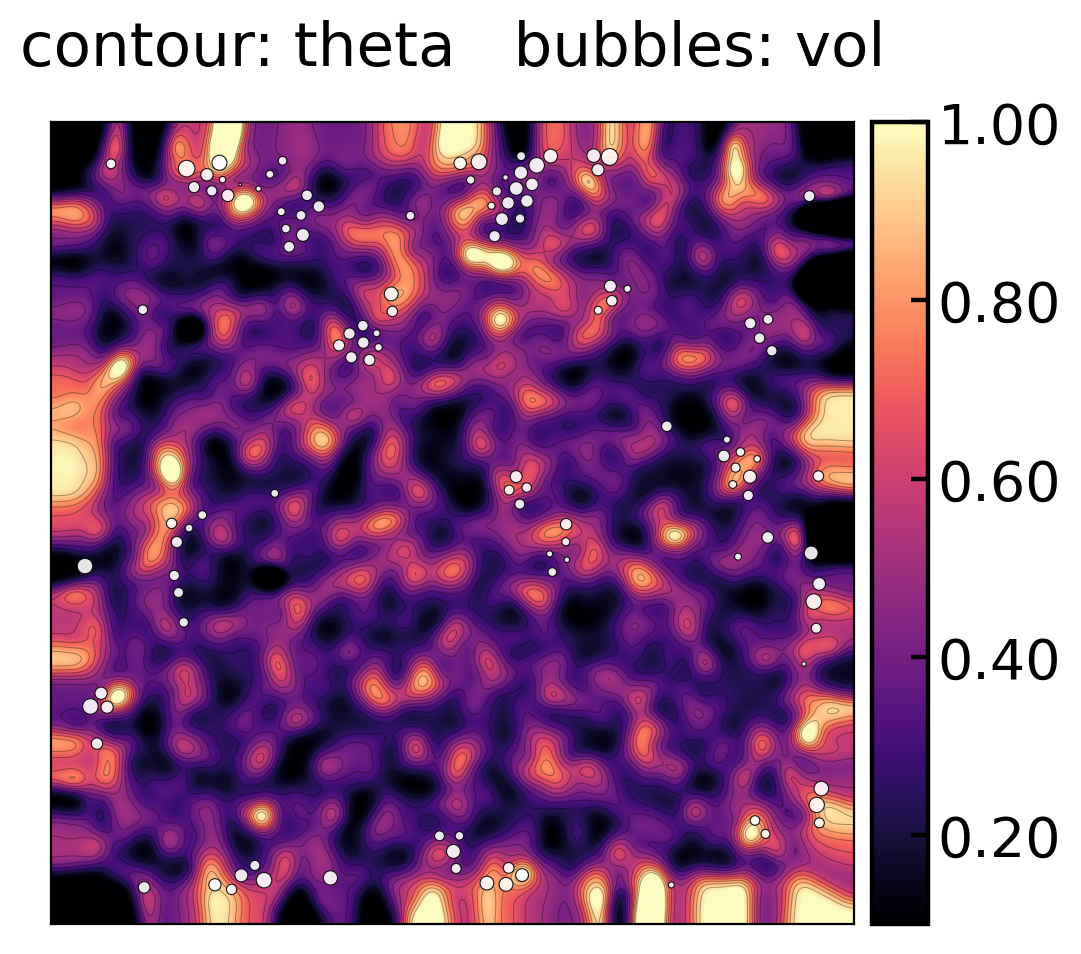}
  \end{subfigure}

  \vspace{8pt}

  \begin{subfigure}[b]{0.32\textwidth}
    \centering
    \textbf{\small $\rho=1.000,\; T=0.550$}\vspace{4pt}\\
    \includegraphics[width=\linewidth]{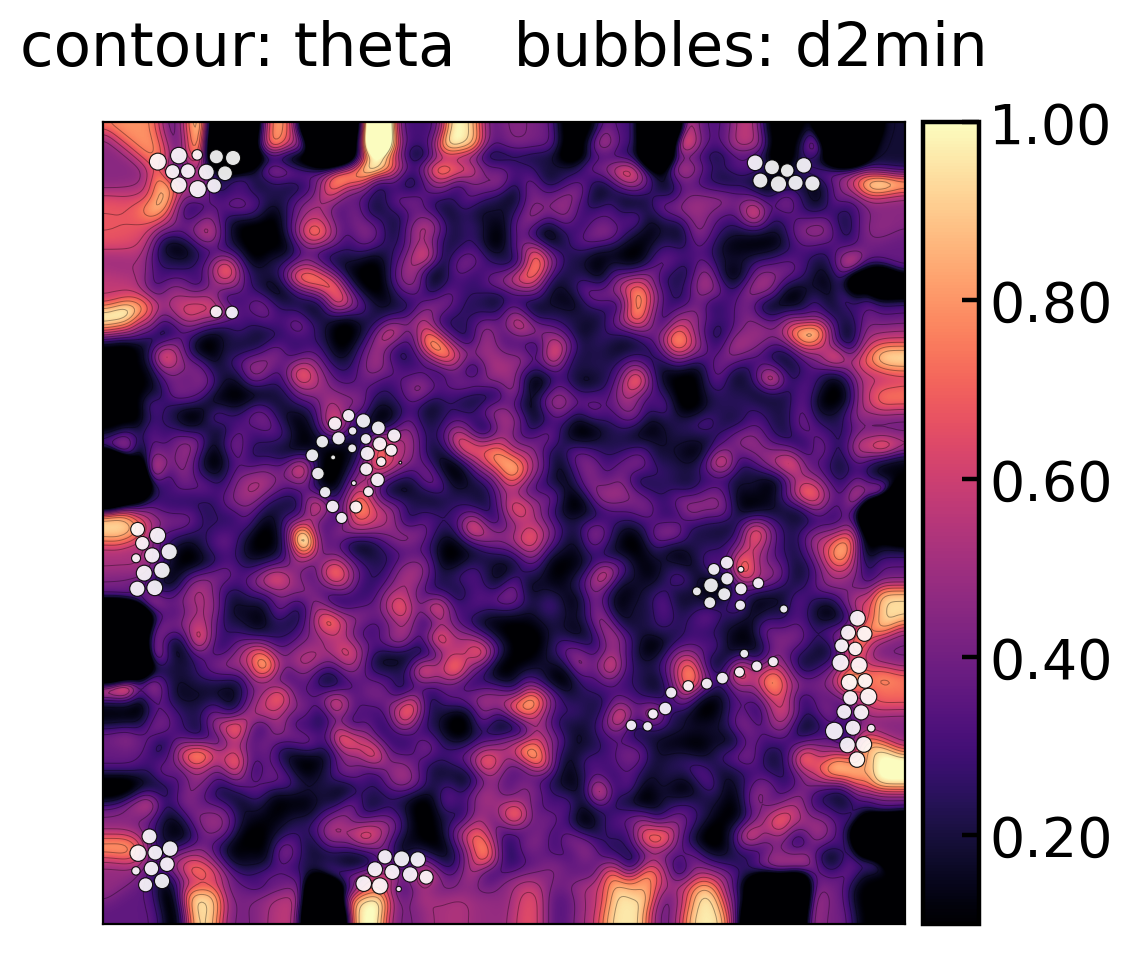}
  \end{subfigure}\hfill
  \begin{subfigure}[b]{0.32\textwidth}
    \centering
    \textbf{\small $\rho=1.000,\; T=0.550$}\vspace{4pt}\\
    \includegraphics[width=\linewidth]{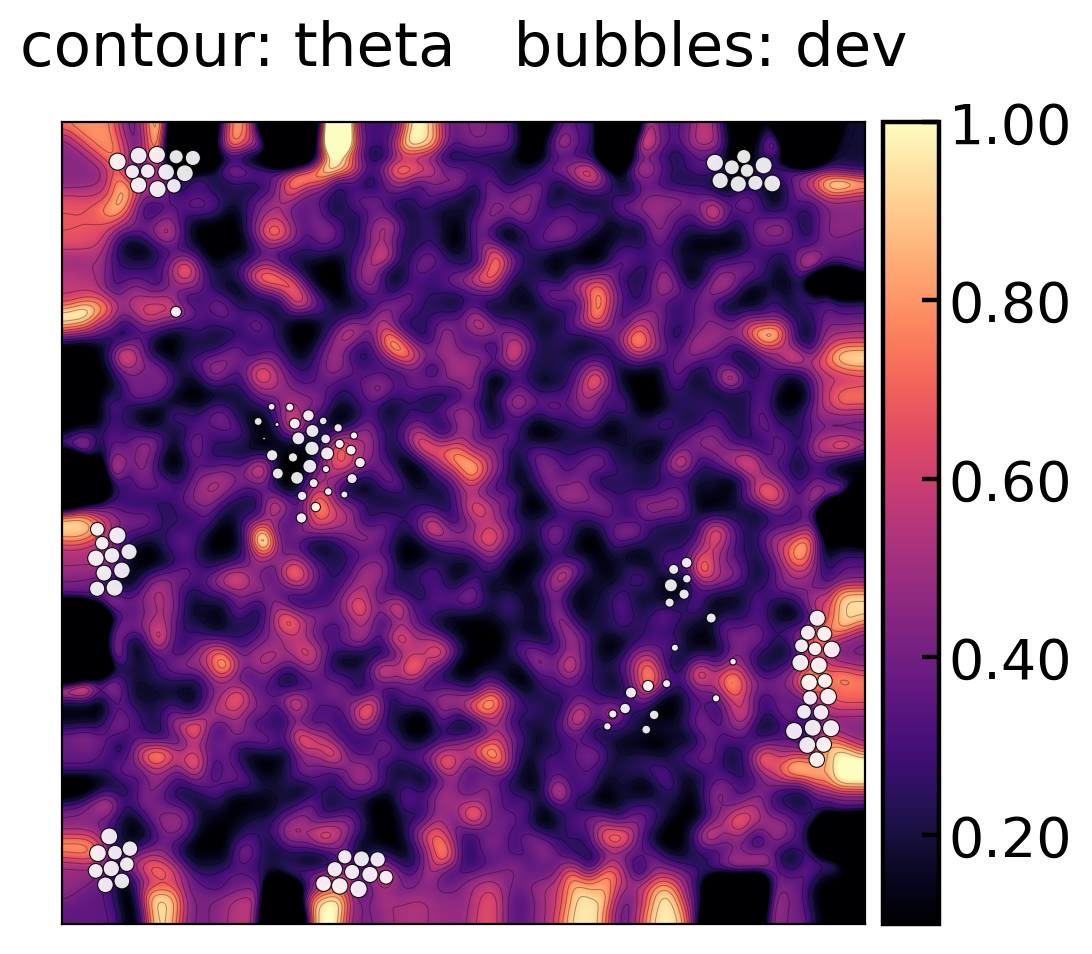}
  \end{subfigure}\hfill
  \begin{subfigure}[b]{0.32\textwidth}
    \centering
    \textbf{\small $\rho=1.000,\; T=0.550$}\vspace{4pt}\\
    \includegraphics[width=\linewidth]{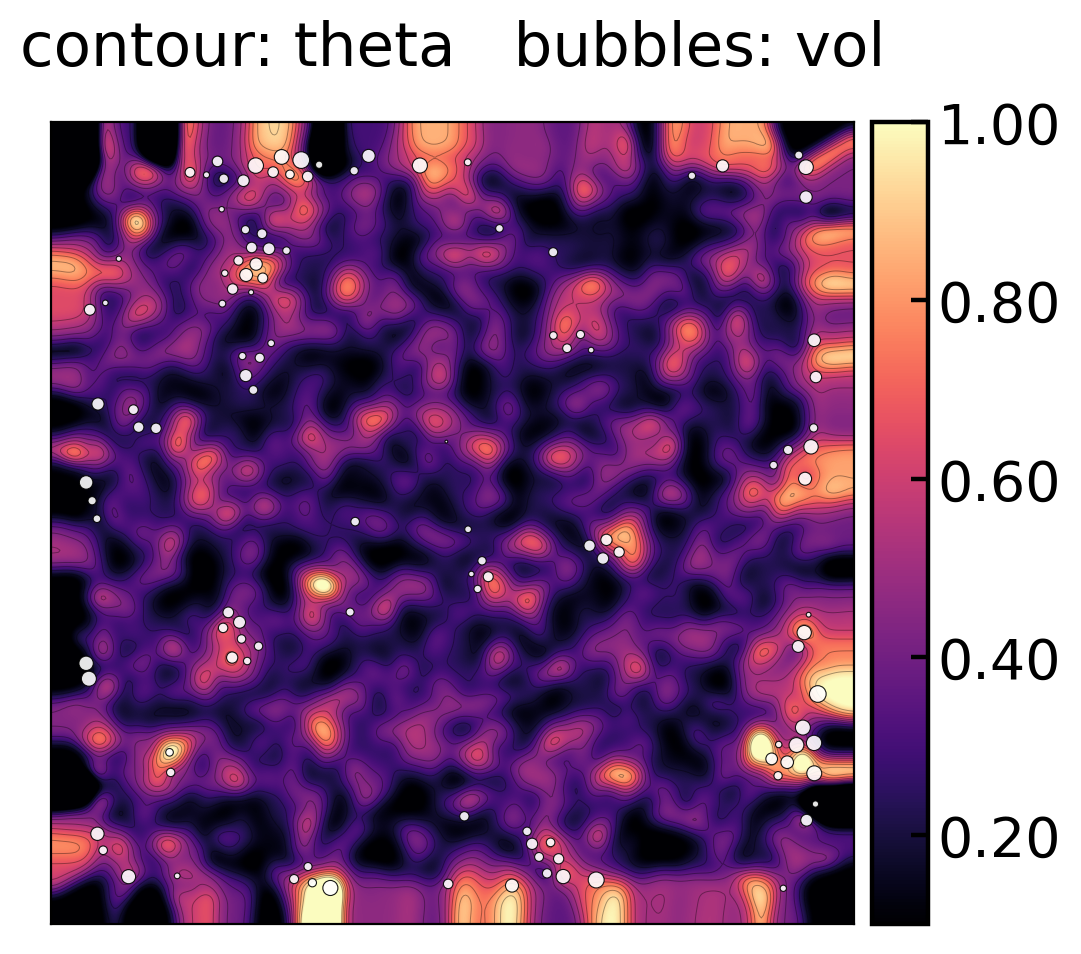}
  \end{subfigure}

 \caption{Spatial maps of local elastic activity overlaid on the contour of the orientational-mismatch field $\theta$. 
Each subpanel title indicates the global packing fraction (density) and temperature for that dataset. 
Columns (left to right) display particle-wise $D^{2}_{\min}$, the deviatoric strain, and the volumetric strain, respectively. 
The background shading is the $\theta$ mismatch field; the same color scale is used across panels to permit direct comparison of magnitudes. 
Filled bubbles mark a selected subset of particles: bubble area is proportional to the plotted elastic metric (larger bubbles indicate larger local elastic/plastic response), and bubble outlines are drawn for visual clarity. 
Bubbles show the particles in the top decile of the chosen elastic metric, i.e. those whose metric value exceeds the frame-specific 90th percentile (values $\geq$ the 90th percentile of the per-frame distribution). 
For $N=2000$ this corresponds to roughly the 200 particles with largest metric values. 
These panels illustrate that regions of large orientational mismatch $\theta$ spatially coincide with enhanced local elastic activity, consistent with the conditional distributions discussed in the text.}
\label{fig:theta_elastic_maps}

\label{fig:theta_elastic_maps}

\end{figure*}


\section{Spatial Correlation Between Steric Mismatch and Thermal Mechanical Activity}
\label{sec:spatial_correlation}

The conditional distributions presented in the preceding sections establish a statistical relation between local steric mismatch and the thermal mechanical response. We now examine this relation directly in real space by superimposing the spatially resolved thermal mechanical fields on the local steric-mismatch field. Fig.~\ref{fig:theta_elastic_maps} shows representative configurations at three state points, with the rows corresponding to \(\rho=0.900\), \(0.950\), and \(1.000\), and the columns showing, from left to right, the thermal nonaffine metric, thermal deviatoric strain, and thermal volumetric strain. In each panel, the background contour represents the spatial distribution of the steric mismatch \(\Delta\theta\).

To identify the regions carrying the strongest thermal mechanical response, we mark the particles belonging to the upper \(10\%\) of the corresponding thermal metric in each frame. Specifically, a particle is selected when its metric value is greater than or equal to the frame-specific \(90^{\rm th}\) percentile. For \(N=2000\), this corresponds to approximately 200 particles per frame. The area of each bubble is proportional to the magnitude of the corresponding thermal mechanical metric, so that larger bubbles represent larger local response; the bubble outlines are included only for visual clarity. This representation allows the spatial location and relative magnitude of the strongest thermal mechanical responses to be compared directly with the underlying steric-mismatch field.

In the deeply arrested state at \(\rho=0.900\) and \(T=0.010\), the thermal nonaffine, deviatoric, and volumetric maps contain only sparse regions of strong local response, consistent with the mechanically locked character of this regime. Nevertheless, the selected high-response particles are not distributed uniformly across the system. They preferentially occur within or close to regions of enhanced \(\Delta\theta\), indicating that the residual thermal mechanical activity is spatially associated with geometrically frustrated environments even when the overall response is strongly suppressed. The same spatial correspondence is visible across the three mechanical measures, although the detailed arrangement of the high-response particles differs because the three quantities probe different components of the local thermal response.

At the intermediate state point, \(\rho=0.950\) and \(T=0.100\), the thermal mechanical fields become more spatially extended. The upper-tail particles of the thermal nonaffine metric and thermal deviatoric and volumetric strains form localized clusters that are frequently found in regions of enhanced steric mismatch. The correspondence is again not one-to-one: a region of large \(\Delta\theta\) does not imply that every particle within that region belongs to the upper tail of a given mechanical metric, and high-response particles can also occur near regions of moderate mismatch. Nevertheless, the repeated spatial association demonstrates that the structural selectivity observed in the conditional probability distributions is reflected in the real-space organization of the thermal response.

The same relation persists at the highest density shown, \(\rho=1.000\) and \(T=0.550\), despite the stronger steric confinement. The particles selected from the upper tails of the thermal nonaffine, deviatoric, and volumetric responses again occur preferentially in localized regions of enhanced \(\Delta\theta\) or in their immediate surroundings. The persistence of this association at the highest density is consistent with the conditional distributions, where the distinction between low- and high-\(\Delta\theta\) populations remains identifiable over a broader temperature range. Thus, increased confinement does not remove the spatial connection between steric frustration and thermal mechanical activity.

An important feature of Fig.~\ref{fig:theta_elastic_maps} is that the correspondence is observed simultaneously for all three thermal mechanical descriptors. The thermal nonaffine metric identifies regions of large nonaffine motion, while the thermal deviatoric and volumetric strains resolve the shear-like and dilatational components of that response. Although these observables are not identical and need not select exactly the same particles, their upper-tail populations repeatedly appear in the same broad regions of enhanced steric mismatch. The spatial maps therefore complement the conditional distributions presented above: the latter establish the statistical dependence of the response on \(\Delta\theta\), whereas the present maps demonstrate that this dependence is also spatially organized.

Taken together, the maps provide a real-space view of the structure--response relation established throughout this work. Local steric mismatch does not uniquely determine the response of every particle; rather, regions of larger geometric frustration preferentially host enhanced thermal nonaffine motion and thermal strain. The persistence of this spatial association for the nonaffine, deviatoric, and volumetric components strengthens the conclusion that the steric geometry encoded by \(\Delta\theta\) is systematically coupled to thermally driven mechanical activity. At the same time, the progressively broader and less localized response upon heating is consistent with the weakening of the mismatch dependence found in the corresponding conditional distributions. The spatial maps thus provide a direct real-space complement to the statistical topology-mechanics mapping developed in the preceding sections.

\section{Conclusions}
\label{sec:conclusions}
Building on our previous development of thermal nonaffine metrics and thermal strain fields, in the present study we investigate how local steric geometry is related to the thermally driven mechanical response of an amorphous glass. Using the steric \(\theta\)-mismatch as a purely geometric measure of local frustration, we compared low- and high-mismatch populations with the thermal nonaffine response and its deviatoric and volumetric components over a broad range of temperatures and densities. The steric mismatch distribution exhibits well-defined temperature- and density-dependent structure, with low-mismatch motifs becoming increasingly sharp and prominent under stronger confinement, while thermal activation progressively broadens and redistributes the distinct structural populations.

The comparison with the thermal nonaffine response establishes a direct structural--mechanical connection. In the deeply mechanically locked regime, regions of larger \(\Delta\theta\) preferentially accommodate larger thermal nonaffine displacements, indicating that the rare residual rearrangements that survive deep in the arrested state are associated with geometrically frustrated environments. Upon entering the mechanically active viscoelastic regime, this distinction becomes more pronounced as the high-\(\Delta\theta\) population develops enhanced weight at larger nonaffine displacements. With further heating, the conditional distributions progressively broaden and converge, showing that the ability of the steric mismatch to distinguish regions of enhanced nonaffine activity weakens as thermal motion becomes increasingly dominant.

The thermal strain fields further resolve the character of this structural--mechanical relation. The deviatoric response follows the nonaffine activity in the deeply arrested regime, where high-\(\Delta\theta\) sites exhibit larger shear-like thermal distortions, while its subsequent evolution depends strongly on density. At the lowest density, the ordering reverses above the solidification line, with low-\(\Delta\theta\) sites becoming dominant in the principal deviatoric-strain response, whereas increasing density suppresses this inversion and preserves the preferential response of high-mismatch sites over a broader temperature range. The volumetric response exhibits a more robust solid-state signature: at all densities, low-\(\Delta\theta\) sites are preferentially associated with small volumetric strains, whereas high-\(\Delta\theta\) sites contribute more strongly to the large-strain tail, with the two populations exhibiting a clear crossover-like behavior in the solid and intermediate regimes. This mismatch-dependent crossover weakens progressively with increasing temperature as the conditional distributions broaden and approach one another, eventually becoming less distinguishable in the liquid state.

The real-space maps provide a complementary demonstration of these statistical correlations. By overlaying the spatially resolved thermal nonaffine, deviatoric, and volumetric responses on the local \(\Delta\theta\) field, we find that particles selected from the upper \(10\%\) of each thermal mechanical metric are preferentially localized within or near regions of enhanced steric mismatch. The bubble representation, in which bubble size reflects the magnitude of the corresponding thermal mechanical response, shows that the strongest nonaffine, deviatoric, and volumetric responses form localized spatial clusters rather than being distributed uniformly throughout the system. This spatial association persists across the representative densities and temperatures examined, although its contrast becomes progressively weaker as thermal activation increases. The maps therefore provide a real-space counterpart to the conditional distributions: the latter establish the statistical dependence of the thermal mechanical response on \(\Delta\theta\), while the former demonstrate that the same structure--response relation is spatially organized in the material.

Taken together, these results show that the steric mismatch provides a local geometric reference that is systematically coupled to thermally generated mechanical response. Geometrically frustrated regions preferentially host enhanced thermal nonaffine motion and, in the arrested regime, stronger deviatoric response and larger-amplitude volumetric distortions. The strength and form of this coupling depend on the mechanical component, temperature, and density, with stronger confinement preserving the structural signature over a broader temperature range and thermal activation progressively reducing the distinction between mismatch-selected populations. Thus, combining a purely geometric structural descriptor with thermal nonaffine and thermal strain measures provides a direct framework for connecting local steric frustration to thermally driven mechanical rearrangements without externally imposed deformation.

The present study is based on a two-dimensional polydisperse, short-range repulsive glass former and therefore does not by itself establish the universality of the observed mapping. Extensions to three-dimensional systems, different interaction potentials, and experimentally accessible amorphous materials will be important for determining the broader applicability of this structure--mechanics relation. More generally, the results demonstrate that thermal nonaffine and strain-based metrics, when analyzed together with local geometric descriptors, can resolve both the magnitude and spatial organization of thermally activated rearrangements in amorphous matter.
\begin{acknowledgments}
A.J. acknowledges support from the University Grants Commission (UGC), India,
through the Junior Research Fellowship and Senior Research Fellowship.
A.J. thanks S. Sengupta for valuable discussions. A.J. also acknowledges the use of the Horizon workstation and the Neo machine in the Department of Physics, Indian Institute of Technology Roorkee, and the Param Ganga supercomputer at the Indian Institute of Technology Roorkee.
\end{acknowledgments}
\bibliographystyle{apsrev4-1}
\bibliography{mybib}
\end{document}